\documentclass[12pt,preprint]{aastex}
\usepackage{graphicx,color}
\shorttitle{NEAR INFRARED BACKGROUND FLUCTUATIONS}
\shortauthors{FERNANDEZ, KOMATSU, ILIEV, & SHAPIRO}
\begin{document}
\title{%
 THE COSMIC NEAR INFRARED BACKGROUND II: FLUCTUATIONS
}%
\author{%
  Elizabeth R. Fernandez$^1$, Eiichiro Komatsu$^2$, Ilian
 T. Iliev$^{3,4}$, Paul R. Shapiro$^2$
}%
\affil{%
  $^1$Center for Astrophysics and Space Astronomy, University of Colorado,
  389 UCB, Boulder, CO 80309-0389\\
  $^2$Texas Cosmology Center and the Department of Astronomy, The University of Texas at Austin,
  1 University Station, C1400, Austin, TX 78712\\
  $^3$Astronomy Centre, Department of Physics \& Astronomy, Pevensey II
 Building, University of Sussex, Falmer, Brighton BN1 9QH, United Kingdom\\
  $^4$Universit\"at Z\"urich, Institut f\"ur Theoretische Physik,
 Winterthurerstrasse 190, CH-8057 Z\"urich, Switzerland
}%
\email{%
  $^1$elizabeth.fernandez@colorado.edu
}%
%%%%%%%%%%%%%%%%%%%%%%%%%%%%%%%%%%%%%%%%%%%%%%%%%%%%%%%%%%%%%%%%%%%%%%
\begin{abstract}
The Near Infrared Background (NIRB) is one of a few methods that can be
 used to observe the redshifted light from early stars at a
 redshift of six and above, and thus it is imperative to understand the
 significance of any detection or non-detection of the
 NIRB. Fluctuations of the NIRB can provide information on the first
 structures, such as halos and their surrounding ionized regions in the
 Inter  Galactic Medium (IGM).  We combine, for the first time, $N$-body
 simulations, radiative transfer code, and analytic calculations of
 luminosity of early structures to predict the angular power spectrum
 ($C_l$) of  fluctuations in the NIRB.
 We study, in detail, the effects of various
 assumptions about the stellar mass, the initial mass spectrum of stars,
 metallicity, the star formation efficiency ($f_*$), the escape fraction
 of ionizing 
 photons ($f_{\rm esc}$), and the star formation timescale ($t_{\rm
 SF}$), on the amplitude as well as the  shape of $C_l$. The power 
 spectrum of NIRB fluctuations  is 
 maximized when $f_*$ is the largest (as $C_l\propto f_*^2$) and 
 $f_{\rm esc}$ is the smallest (as more nebular emission is produced within
 halos). A significant uncertainty in the predicted amplitude of $C_l$
 exists due to our lack of knowledge of $t_{\rm SF}$ of
 these early populations of galaxies, which is equivalent to our lack of
 knowledge of the mass-to-light ratio of these sources. 
  We do not see a turnover in the NIRB angular power spectrum of the halo
 contribution, which was claimed to exist in the literature, and explain
 this as the effect of high levels of non-linear bias that was ignored
 in the previous calculations. This is partly due to our choice of the
 minimum mass of  halos contributing to NIRB ($\sim 2\times 10^9~M_\sun$), and a
 smaller minimum mass, which has a smaller non-linear bias,  may still
 exhibit a turn over. Therefore, our results suggest that both the
 amplitude and shape of the NIRB power spectrum provide important
 information regarding the nature of sources contributing to the cosmic
 reionization. 
 The angular power spectrum of the IGM, in most
 cases, is much smaller than the halo angular power spectrum,
 except when $f_{\rm esc}$ is close to unity, $t_{\rm SF}$ is longer, or
 the minimum  redshift at which the star formation is occurring is high.
 In addition, low levels of the observed mean background
 intensity tend to 
 rule out high values of $f_* \gtrsim 0.2$.
\end{abstract}  
\keywords{cosmology: theory --- diffuse radiation --- galaxies:
  high-redshift --- infrared: galaxies}
%%%%%%%%%%%%%%%%%%%%%%%%%%%%%%%%%%%%%%%%%%%%%%%%%%%%%%%%%%%%%%%%%%%%%%
\section{INTRODUCTION}
\label{sec:introduction}
We have few probes of the early universe and the first few generations of
stars.  We know that stars had to form early in order to pollute the
universe with metals and reionize the universe.  There is evidence that the
universe was reionized at around $z\sim 11$, 
such as from the Wilkinson Microwave Anisotropy Probe (WMAP) satellite
\citep{kogut/etal:2003,spergel/etal:2003,spergel/etal:2007,page/etal:2007,dunkley/etal:2008,komatsu/etal:2008}. Stars
are efficient producers of ionizing photons, so are likely candidates
for the bulk of reionization. These ultraviolet photons at redshifts
$6 \lesssim z \lesssim 30$ would be
redshifted into the near-infrared bands.  Therefore, it makes sense to
search for this remnant light in the near infrared bands to learn about
this early epoch of star formation and reionization \citep{santos/bromm/kamionkowski:2002,magliocchetti/salvaterra/ferrara:2003,salvaterra/ferrara:2003,cooray/etal:2004,cooray/yoshida:2004,kashlinsky/etal:2004,madau/silk:2005,fernandez/komatsu:2006}.
Observations of the Near Infrared Background (NIRB) may indicate that
there is an excess mean background above that normal galaxies can
account for
\citep{dwek/arendt:1998,gorjian/wright/chary:2000,kashlinsky/odenwald:2000,wright/reese:2000,wright:2001,cambresy/etal:2001,totani/etal:2001,matsumoto/etal:2005,kashlinsky:2005}.
In addition, there also appears to be a peak in the NIRB spectrum at 1--2
$\mu{\rm m}$, which could represent a Lyman-cutoff signature
\citep{bock/etal:2006}.  However, the interpretation of the current
observational data,  in particular accuracy of the subtraction of
Zodiacal light and foreground galaxies, is highly controversial \citep{thompson/etal:2007a,thompson/etal:2007b}.
Nevertheless, any detection or non-detection of this excess light could tell us
properties of early stars.   

In addition to the mean intensity, fluctuations in the NIRB can provide
an additional source of information about the first generations of
stars \citep{kashlinsky/odenwald:2000,kashlinsky/etal:2002,kashlinsky/etal:2004,kashlinsky/etal:2005,kash/etal:2007,kashlinsky/etal:2007,kashlinskyb/etal:2007b,kashlinsky:2005,magliocchetti/salvaterra/ferrara:2003,odenwald/etal:2003,cooray/etal:2004,matsumoto/etal:2005,thompson/etal:2007a,thompson/etal:2007b}.
Fluctuations are in general easier to study than the mean intensity
because an accurate determination of the zero point is not needed; thus,
they are less sensitive to the imperfect subtraction of Zodiacal
light. However, as the contribution to fluctuations from low redshift
populations, i.e., $z<6$, can confuse the signal from higher redshift
populations, the level of contamination from low redshift populations
must be estimated and subtracted carefully
\citep{sullivan/etal:2007,cooray/etal:2007,kashlinskyb/etal:2007b,thompson/etal:2007b,chary/cooray/sullivan:2008}.
Upcoming measurements with AKARI \citep{matsuhara/etal:2008} and CIBER
(the Cosmic Infrared Background Experiment) \citep{bock/etal:2006, cooray/etal:2009} may
be able to put a more solid constraint on what fraction of the NIRB is
from high redshift stars and galaxies.   

In the previous paper we have presented detailed theoretical calculations of the
spectrum and metallicity/initial-mass-spectrum dependence of the mean
intensity of NIRB (\citet{fernandez/komatsu:2006}, hereafter FK06). In this paper we
present calculations of the power spectrum and
metallicity/initial-mass-spectrum dependence of the NIRB fluctuations, as
well as dependence on the star formation efficiency and the escape
fraction of ionizing photons.
While the previous work in the literature
\citep{cooray/etal:2004,kashlinsky/etal:2004} relied solely on
simplified analytical arguments, we use, for the first time, large-scale 
cosmological simulations of cosmic reionization given in 
\citet{iliev/etal:2006,iliev/etal:2007,iliev/etal:2008}, coupled with the
analytical calculations given in FK06, to
predict the power spectrum of NIRB fluctuations. In this way we are able
to capture the contribution from ionized bubbles surrounding the halos,
which has been ignored completely in the previous work.

In \S~\ref{sec:simulations} we outline the simulations 
\citep{iliev/etal:2006,iliev/etal:2007,iliev/etal:2008}
and in
\S~\ref{sec:analytical} we explain the analytic formulas we use to obtain the luminosity of the halos and the
surrounding IGM.  In \S~\ref{sec:lpk} we present our calculation of the
luminosity-density power spectrum, $P_L(k)$. 
Predictions for  $P_L(k)$ and the angular power spectrum of NIRB
fluctuations, $C_l$, are presented in \S~\ref{sec:results}.
Various parameters' effects on the results are discussed in \S~\ref{sec:varysection}.
We compare our results to the previous literature in \S~\ref{sec:compare} and to
observations in \S~\ref{sec:obs}.  We take a look at the constraints from the
mean NIRB in \S~\ref{sec:mean}, and 
compute the fractional anisotropy, i.e., the ratio of the power
spectrum and the mean intensity squared, in \S~\ref{sec:fractional}. We
conclude in 
\S~\ref{sec:conclusions}. 
%%%%%%%%%%%%%%%%%%%%%%%%%%%%%%%%%%%%%%%%%%%%%%%%%%%%%%%%%%%%%%%%%%%%%%
\section{SIMULATION}
\label{sec:simulations} 
We use simulations from \citet{iliev/etal:2006,
iliev/etal:2007,iliev/etal:2008}, which are the first truly large scale
simulations to include radiative transfer, and  
are therefore ideal for predicting the distribution of luminosities from high redshift stellar populations.  Simulations provide the advantage of being able to 
simultaneously model the distribution of halos and the density of the IGM,
as well as the ionization front that propagates through the IGM.  We combine this $N$-body code with radiative transfer and analytic formulas
for luminosity to simulate their luminosity-density power
spectrum.  

The particular simulation that we use in this paper is the run 
``f250C'' in Table I of \citet{iliev/etal:2008}, which was 
run with the cosmological parameters given by the
WMAP 3-year results 
\citep{spergel/etal:2007}, ($\Omega_{\rm m}$, $\Omega_\Lambda$,
$\Omega_{\rm b}$,
$h$, $\sigma_8$, $n_{\rm s}$)=(0.24, 0.76, 0.042, 0.73, 0.74, 0.95). 
Aside from the cosmological parameters, the only free parameter in the
reionization simulation of this kind is the production rate of ionizing photons
 escaping into the IGM  per halo.
We shall come back to this parameter, called $f_\gamma/t_{\rm SF}$, in 
\S~\ref{sec:massspectrum}.

These simulations combine a high resolution $N$-body code \citep[PMFAST,
see][]{merz/etal:2005} with a
radiative transfer code \citep[$C^2$-Ray, see][]{mellema/etal:2006},
which is a conservative, causal ray-tracing radiative transfer code.  
The $C^2$-Ray code traces the ionization front by tracking 
photons using photon conservation.  The code allows for large time
steps and coarse grids without loss of accuracy.  

 The box size of the simulation is 100$~h^{-1}$ Mpc, which is large enough 
to sample the history, geometry, and statistical properties of reionization.  The number of particles is $1624^3$, and the density field
was sampled on a lattice of $3248^3$ cells.  The density field was then
binned to $203^3$ cells for the radiative transfer calculations. We
use the $203^3$ cells when we compute the radiation from the IGM in
\S~\ref{sec:PLIGM}. 
The minimum mass of the halos is $2.2\times 10^9~M_\sun$, which 
represents dwarf galaxies.  These halos have virial
  temperatures of $1.2 \times 10^4 \: {\rm K}$, $1.8 \times 10^4 \: {\rm K}$,
  and $2.6 \times 10^4 \: {\rm K}$ at $z=6$, $10$, and $15$ respectively.  For these halos
  the dominant cooling process is hydrogen atomic cooling.  It is important to sample these dwarf galaxies,
as they are far more numerous than larger galaxies and may provide 
most of the photons needed for reionization.   

Even though this simulation is a very powerful tool, it is important to
consider its limitations.  Halos slightly below the resolution of this simulation ($10^8$ to $10^9~$ $M_\sun$) may also be an important source for ionizing radiation.  \citet{iliev/etal:2007} also did a smaller box-size simulation [(35$~h^{-1}$ Mpc)$^3$] that resolves 
halos down to $10^8~M_\sun$, which includes halos that form stars as a
result of atomic cooling.  These smaller halos allow the ionization
fraction to reach 50\% at an earlier epoch than the simulations that
only resolved down to $2.2\times 10^9~M_\sun$.  
However,  \citet{iliev/etal:2007} found that the redshift in which
reionization was completed remained about the same for the 100$~h^{-1}$
Mpc and 35$~h^{-1}$ Mpc simulations, due to the ``self-regulation''
\citep[see][for details]{iliev/etal:2007}. 

The results discussed in this paper are based on the 
larger box size with halos resolved down to $2.2\times 10^9~M_\sun$.  It is possible that the 
smaller halos would affect the fluctuations in the NIRB from both the halos and the IGM.  Future simulations 
will allow both a larger box size along with a smaller minimum mass.  These
future simulations will be able to provide more robust 
predictions for the fluctuations in the NIRB if these smaller halos contribute to
the NIRB.  Simulations that resolve halos smaller than $10^8~M_\sun$ may not be needed, however: while these minihalos were likely the sites of the
truly first generation of stars, they may not be a significant source of
ionizing photons to reionize the universe, as UV photons in the Lyman
Werner bands dissociate molecular hydrogen, terminating star formation
in these small halos \citep{haiman/etal:1997,haiman/etal:2000,Mackacek:2003,Yoshida:2003,Johnson/etal:2008,ahn/etal:2008}.  However,
there is on-going discussion as to what the radiation
feedback actually does for the formation of second generation stars
\citep{wiseetal:2007,ahn/shapiro:2007,oshea:2008}.  While the Lyman Werner background from early star formation has a
primarily negative
feedback effect, other processes (e.g., cooling in supernova remnant shocks)
may mitigate
the suppression of H2 molecules \citep{ferrara/etal:1998, ricotti/etal:2001}.
%%%%%%%%%%%%%%%%%%%%%%%%%%%%%%%%%%%%%%%%%%%%%%%%%%%%%%%%%%%%%%%%%%%%%%
\section{ANALYTICAL CALCULATION}
\label{sec:analytical}
In this section we describe how we assign the luminosity 
to the halos and the IGM in the simulation. Note that our method is fully
analytical, and thus can be adopted to any other reionization simulations.
%%%%%%%%%%%%%%%%%%%%%%%%%%%%%%%%%%%%%%%%%%%%%%%%%%%%%%%%%%%%%%%%%%%%%%
\subsection{Luminosity of the halos}
\label{sec:Lhalo}
Luminosity within the halos is dominated by five radiative processes:
stellar (black-body) emission, and the nebular emission including
free-free, free-bound, and two-photon emission, as well as
any emission lines (here, Lyman-$\alpha$ is the most important one for
our study of NIRB).  The
luminosity of each component can be found analytically using the equations
in FK06.  Equations for the stars' luminosity, temperature,
  number of ionizing photons per second, and lifetime were based on
  equations from Table 3 of \citet{FK08}, which were fit from stellar
  models or fitting functions \citep{marigo/etal:2001, lejeune/schaerer:2001, schaerer:2002}.

First, let us define the ``volume emissivity.'' 
The volume emissivity (luminosity per comoving
volume per frequency), $p(\nu)$, is related to the ``emission
coefficient'' (luminosity per comoving volume per frequency per
steradian), 
$j_\nu$, in \citet{santos/bromm/kamionkowski:2002,cooray/etal:2004} by
$p(\nu)=4\pi j_\nu$. In other words, the luminosity is given by
\begin{equation}
 dL=p(\nu)d\nu dV=j_\nu d\Omega d\nu dV,
\label{eq:def_emissivity}
\end{equation}
where $dV$ is the {\it comoving} volume
element, and $d\Omega$ is the solid angle element. Integrating $p(\nu)$
over $\nu$, one obtains the ``comoving luminosity density,'' $\rho_L$,
as $dL=\rho_LdV$, where $\rho_L=\int p(\nu)d\nu$.

When the main-sequence lifetime of stars under consideration is shorter
than the time scale at which the star formation takes place, the volume
emissivity is given by 
a product of the star formation rate (the stellar mass density formed
per unit time), $\dot{\rho}_*(z)$, and the ratio of the
mass-weighted average of the total radiative energy (including stellar
emission and reprocessed light) emitted over the
stellar lifetime to the stellar rest-mass energy, $\langle \epsilon_\nu^\alpha
\rangle$  (see Eq.~(2) of FK06):
\begin{equation}
 p(\nu,z)=\sum_\alpha p_\alpha(\nu,z)=
\dot{\rho}_*(z)c^2\sum_\alpha\langle \epsilon^\alpha_\nu\rangle,
\label{eq:pnu}
\end{equation}
where
\begin{equation}
 \langle \epsilon_\nu^\alpha \rangle
= \frac{\int^{m_2}_{m_1} dm
   \left[L^\alpha_{\nu}(m)\tau(m)/(mc^2)\right]f(m)m}{\int^{m_2}_{m_1}
   dm f(m) m}.
\end{equation}
Here, $m$ is the stellar mass, $L_\nu^\alpha(m)$ is the time-averaged
luminosity per frequency of a given radiative process $\alpha$ (which
includes the stellar, 
free-free, free-bound, two-photon, and Lyman-$\alpha$ emission), 
$\tau(m)$ is the main sequence lifetime, and $f(m)$ is the initial mass
spectrum of stars under consideration (specified later in
\S~\ref{sec:massspectrum}).
Note that $\langle \epsilon_\nu^\alpha \rangle$ may also be interpreted
as a ratio of the total radiative energy within a unit frequency to the
total stellar rest-mass energy, 
\begin{equation}
 \langle \epsilon_\nu^\alpha \rangle= \frac{\int^{m_2}_{m_1} dm
   f(m)L^\alpha_{\nu}(m)\tau(m)}{\int^{m_2}_{m_1}
   dm f(m) mc^2}.
\end{equation}
Either way, $\langle \epsilon_\nu^\alpha \rangle$ is a convenient
quantity that tells us how much of the stellar rest-mass
energy is converted into the radiative energy  within a unit frequency
interval. 

 In FK06 we have shown that 
$\langle \epsilon_\nu^\alpha \rangle$ can be calculated robustly for a given
stellar population, i.e., $f(m)$, using the basic stellar physics and
radiative processes in the interstellar medium.
For the radiative processes and stellar populations we
consider in this paper, $\nu \langle \epsilon_\nu^\alpha \rangle\lesssim
10^{-3}$ (see Figure~2 of FK06).
From this one may obtain a quantity that is commonly used in the
literature, the luminosity {\it per stellar mass}, ${l}^\alpha_\nu$, as
\begin{equation}
 l^\alpha_\nu(z) = \frac{p_\alpha(\nu,z)}{\rho_*(z)}
= \frac{d\ln\rho_*(z)}{dt} 
\frac{\int^{m_2}_{m_1} dm
   f(m)L^\alpha_{\nu}(m)\tau(m)}{\int^{m_2}_{m_1} dm f(m) m}.
\end{equation}
In this expression one may identify 
$d\ln\rho_*(z)/dt$ as the inverse of the star formation timescale,
$t_{\rm SF}(z)$, i.e., $t_{\rm SF}(z)\equiv
[d\ln\rho_*(z)/dt]^{-1}$.\footnote{If one assumes that the star
formation is triggered by mergers of dark matter halos, then the star
formation timescale may be related to the halo merger rate, i.e., 
$t^{-1}_{\rm SF}(z)=[\int dM_h M_h(d^2n_h/dM_hdt)]/[\int dM_h
M_h(dn_h/dM_h)]$, 
where $dn_h/dM_h$ is the mass function of dark matter
halos. This approach was used by
\citet{santos/bromm/kamionkowski:2002,cooray/etal:2004,cooray/yoshida:2004}. In this paper
we shall use $t_{\rm SF}=20$~Myr as our fiducial value, to be consistent with
the value used by the simulation of \citet{iliev/etal:2008}. We also
study the effects of changing $t_{\rm SF}$ in \S~\ref{sec:tsf}.}
Therefore, we finally obtain
\begin{equation}
 l^\alpha_\nu(z) = \frac1{t_{\rm SF}(z)}\frac{\int^{m_2}_{m_1} dm
   f(m)L^\alpha_{\nu}(m)\tau(m)}{\int^{m_2}_{m_1} dm f(m) m},
\label{eq:lnu1}
\end{equation}
when the main sequence lifetime of stars is  shorter than 
the star formation timescale, $\tau(m)< t_{\rm SF}(z)$.
In terms of $\langle \epsilon_\nu^\alpha\rangle$ we may also write 
Eq.~(\ref{eq:lnu1}) as $l^\alpha_\nu(z)=\langle
\epsilon_\nu^\alpha\rangle c^2/t_{\rm SF}(z)$.

On the other hand, when the star formation timescale is shorter
than the main sequence lifetime of stars, $t_{\rm SF}(z)<\tau(m)$, 
we find a different expression for $l^\alpha_\nu$ (see Eq.~(A6) of FK06):
\begin{equation}
l^\alpha_\nu = \frac{\int^{m_2}_{m_1} dm f(m)
   L^\alpha_{\nu}(m)}{\int^{m_2}_{m_1} dm f(m) m},
\label{eq:lnu2}
\end{equation}
and $l^\alpha_\nu$ no longer depends on $z$ as long as $f(m)$ does not
depend on $z$. From Eqs.~(\ref{eq:lnu1}) and (\ref{eq:lnu2}) we find
that the former is roughly $\tau/t_{\rm SF}$ times the latter. In other
words, if one misused the latter form when $\tau<t_{\rm SF}$, one would
over-estimate the signal by a factor of $\approx t_{\rm SF}/\tau$, which can
be as large as a factor of 10 for short-lived, massive stars with
$\sim
100~M_\sun$.\footnote{\citet{salvaterra/ferrara:2003,magliocchetti/salvaterra/ferrara:2003,kashlinsky/etal:2004}
used Eq.~(\ref{eq:lnu2}) for $\tau<t_{\rm SF}$, and thus their predicted
amplitudes of NIRB are likely over-estimated by a factor of $\approx
t_{\rm SF}/\tau$.} 

For the precise calculation one should use both expressions depending on
the situation; 
however, to simplify the analysis, we shall use either 
Eq.~(\ref{eq:lnu1}) or (\ref{eq:lnu2}), depending on the ratio
of the stellar lifetime averaged over the initial mass spectrum and
weighted by the luminosity (since more massive, shorter lived stars will
contribute more to the overall luminosity),
$\langle{\tau}\rangle\equiv 
\int_{m_1}^{m_2}dmf(m)\tau(m)L/\int_{m_1}^{m_2}dmf(m)L$, to the star
formation timescale (where $L$ is the bolometric luminosity). 
In the simulations of
\citet{iliev/etal:2006,iliev/etal:2007,iliev/etal:2008}, the star
formation timescale takes on a universal value, $t_{\rm SF}\approx
20$~Myr.  For all the stellar populations, the luminosity-weighted lifetime
is shorter than $20$~Myr; thus, Eq.~\ref{eq:lnu2} will be our fiducial formula.

To compute $l_\nu^\alpha$ for each radiative process, 
we use a black-body for the stellar component,
$l^*_\nu$  (see Eq.~(6) of FK06).  This emission is cutoff above 13.6 eV,
so all of the ionizing photons go into producing emission in the nebula or
the IGM.  The expressions given in
\S~2.3, 2.4, and 2.5 of FK06 are used for the nebular processes. 
We then integrate $l^\alpha_\nu$ over a band of observed frequencies $\nu_1$ to
$\nu_2$ to obtain the band-averaged luminosity per stellar mass,
$\bar{l}^\alpha$, as
\begin{equation}
\bar{l}^\alpha(z)\equiv \int^{\nu_2(1+z)}_{\nu_1(1+z)}d\nu~l^\alpha_\nu(z).
\end{equation}

Following \citet{iliev/etal:2006, iliev/etal:2007,iliev/etal:2008}, we
assume all halos have a constant mass-to-light ratio.  
With the luminosities per stellar mass, $\bar{l}^\alpha(z)$, computed,
we obtain the 
luminosities of the halo, $L_h(z)$, by multiplying $\bar{l}(z)$ by the
total stellar 
mass per halo, $f_*M_h(\Omega_b/\Omega_m)$, where $M_h$ is the total
halo mass (including dark matter and baryons), and $f_*$ is the star
formation efficiency, which is the fraction of baryons that can form into
stars over the star formation timescale $t_{\rm SF}$. 
We find
\begin{equation}
\label{eq:ltom}
 \frac{L_h(z)}{M_h} = f_* \frac{\Omega_b}{\Omega_m}
\left\{\bar{l}^*(z) +
 (1-f_{\rm esc})\left[\bar{l}^{ff}(z)+\bar{l}^{fb}(z)+\bar{l}^{2\gamma}(z)+\bar{l}^{\rm Ly\alpha}(z)\right]\right\}, 
\end{equation}
where $f_{\rm esc}$ is the escape fraction of ionizing photons from the
halo.  Only those photons that do not escape into the IGM produce
nebular emission within the halo.

From this result one may conclude immediately that the NIRB power spectrum
from halos, which is proportional to $(L_h/M_h)^2$, is proportional to
$f_*^2$. Also, $L_h/M_h$ goes down as $f_{\rm esc}$ approaches unity, for
which all the ionizing photons would escape halos, and thus no nebular
emission would be left in halos.
The stellar properties, such as metallicities and initial mass spectra,
affect only $\bar{l}^\alpha$.
%%%%%%%%%%%%%%%%%%%%%%%%%%%%%%%%%%%%%%%%%%%%%%%%%%%%%%%%%%%%%%%%%%%%%%
\subsection{Stellar Populations}
\label{sec:massspectrum}
The simulations from \citet{iliev/etal:2008} define a quantity,
$f_\gamma$, which is proportional to the number of ionizing photons that
escape into the IGM: 
\begin{equation}
f_{\gamma} = f_* f_{\rm esc}N_i,
\label{eq:fgamma}
\end{equation}
where $N_i$ is the number of ionizing photons 
emitted per stellar atom.  When  modeling stellar populations in our calculations, we shall
assure that each of our models agrees with $f_\gamma=250$, which was used
in the simulations. 

\citet{iliev/etal:2008} have shown that this choice
of $f_\gamma$, combined with the universal star formation timescale of
$t_{\rm SF}=20~{\rm Myr}$, can reionize the universe successfully with
the resulting 
electron-scattering optical depth consistent with the WMAP data.
Within this framework, since $f_\gamma/t_{\rm SF}$ is the only free
parameter, models with the same $f_\gamma/t_{\rm SF}$ would produce the same
reionization histories.  (For the simulation case f250C on which our calculations here are
     based, for example, the globally-averaged ionized fraction of
     the IGM was found to be 50\% at $z = 8.3$ and 99\% at $z = 6.6$.)
To keep $f_\gamma/t_{\rm SF}$ constant, the
star formation efficiency must decrease as the escape fraction
increases.  The various populations that were modeled are shown in Table
\ref{tab:populations}. 

%%%%%%%%%%%%%%%%%%%%%%%%%%%%%%%%%%%%%%%%%%%%%%%%%%%%%%%%%%%%%%%
\begin{table}[t]
\begin{center}
\begin{tabular}{|l|l|l|l|l|l|l|l|}
\hline
Population & Initial Mass Spectrum & $m_1, m_2$ &
 $\langle\tau\rangle$ (Myr)& $N_i$ & $ f_{\rm esc}$ & $f_*$ \\
\hline
Pop III  & Salpeter & $3M_\sun$, $150M_\sun$  & 8.08 & 5600 & 0.22 & 0.2 \\
Pop III  & Larson, $m_c=250M_\sun$ & $3M_\sun$, $500M_\sun$ & 2.45
		 & 25000 & 0.1 & 0.1\\
Pop III  & Salpeter & $3M_\sun$, $150M_\sun$  &  8.08 & 5600 & 0.9 & 0.05\\
Pop III  & Larson, $m_c=250M_\sun$  & $3M_\sun$, $500M_\sun$ &  2.45 &
		     25000 & 1 & 0.01 \\
Pop II  & Salpeter & $3M_\sun$, $150M_\sun$ &  9.04& 2600 & 0.95 & 0.1\\
Pop II  & Larson, $m_c=50M_\sun$  &  $3M_\sun$,
	     $150M_\sun$ &  4.87 & 12000 & 0.9 & 0.023\\
Pop II  & Salpeter &  $3M_\sun$, $150M_\sun$ &  9.04& 2600 & 0.19 & 0.5\\
Pop II  & Larson, $m_c=50M_\sun$  &  $3M_\sun$,
	     $150M_\sun$ &  4.87 & 12000 & 0.098 & 0.21\\
\hline
\end{tabular}
\caption{%
Stellar populations (``Pop III'' are metal-free, and ``Pop II'' are
 metal-poor with the metallicity of $Z=1/50~Z_\sun$), parameters for
 initial  mass spectra,   
the luminosity-weighted main sequence lifetime of stars ($\langle\tau\rangle$), 
the corresponding number of ionizing photons per stellar atom ($N_i$),
escape fractions of ionizing photons ($f_{\rm esc}$), and 
the star formation efficiency ($f_*$).
Note that $f_{\rm esc}$ and $f_*$ are tuned such that 
the value of $f_\gamma=f_{\rm esc}f_*N_i$
is held fixed at  $f_\gamma=250$ which, when combined with the star
 formation timescale of $t_{\rm SF}=20~{\rm Myr}$, can reionize the
 universe such 
 that the resulting electron-scattering optical depth is consistent with
 the WMAP data.
}%
\label{tab:populations}
\end{center}
\end{table}
%%%%%%%%%%%%%%%%%%%%%%%%%%%%%%%%%%%%%%%%%%%%%%%%%%%%%%%%%%%%%%%%%%%%%%

We modeled both zero metallicity stars (Population III; $Z=0$) and low metallicity stars
(Population II; $Z=1/50~Z_\sun$) with either a heavy or a light initial mass spectrum, accompanied with
either a low escape fraction ($f_{\rm esc}\sim0.1$) or a high escape fraction
($f_{\rm esc}\sim1$).  
While we try to simulate a range of parameters, it is good to keep in mind that our choice of $f_*$ and
$f_{\rm esc}$  for a given $f_\gamma$ is basically arbitrary.  The level of
NIRB fluctuations can change 
 significantly when paired with different  assumptions for the
 metallicity, mass, and values for $f_{\rm esc}$ and $f_*$.

A lighter mass distribution of stars is represented by a Salpeter initial mass spectrum \citep{salpeter:1955}
\begin{equation}
f(m) \propto m^{-2.35}.
\label{eq:salpeter}
\end{equation}
We use mass limits of $m_1=3M_\sun$ and $m_2=150 M_\sun$ for this spectrum.
Heavier stars are represented by a Larson
initial mass 
spectrum \citep{larson:1998} 
\begin{equation}
f(m)\propto m^{-1}\left(1+\frac{m}{m_c}\right)^{-1.35},
\end{equation}
with $m_1 = 3M_\sun$, $m_2=500M_\sun$, and $m_c = 250M_\sun$ for Population III stars and $m_1 = 3M_\sun$,
 $m_2=150M_\sun$, and $m_c = 50M_\sun$ for Population II stars. 

\begin{figure}%[h!tb]
\centering \noindent
\plotone{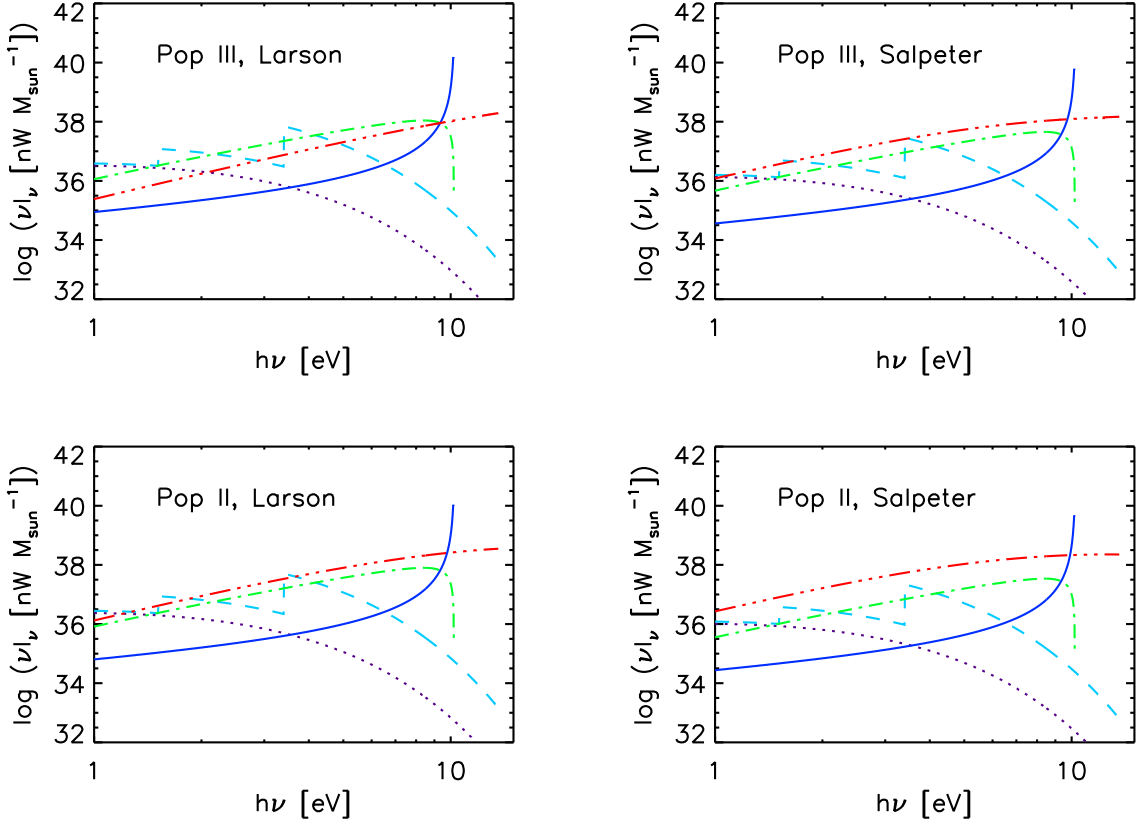}
\caption{%
 Luminosity spectrum per stellar mass. The stellar, $\nu l^*_{\nu}$ (triple-dot dashed red line), free-free,
 $\nu l_{\nu}^{ff}$ (dotted purple line), free-bound, $\nu l_{\nu}^{fb}$ (dashed light blue line), two-photon,
 $\nu l_{\nu}^{2\gamma}$ (dot dashed green line), and Lyman-$\alpha$
 emission, $\nu l_{\nu}^{\rm Ly\alpha}$ (solid dark blue line),
 are shown in units of nW~$M_\sun^{-1}$ as
 a function of the rest-frame energies.
 The stars are at $z=10$, but the redshift affects the profile of the
 Lyman-$\alpha$ line only, which was taken from Eq.~(15) of
 \citet{santos/bromm/kamionkowski:2002}. 
}%
\label{fig:lnu}
\end{figure}

\begin{figure}%[h!tb]
\centering \noindent
\plotone{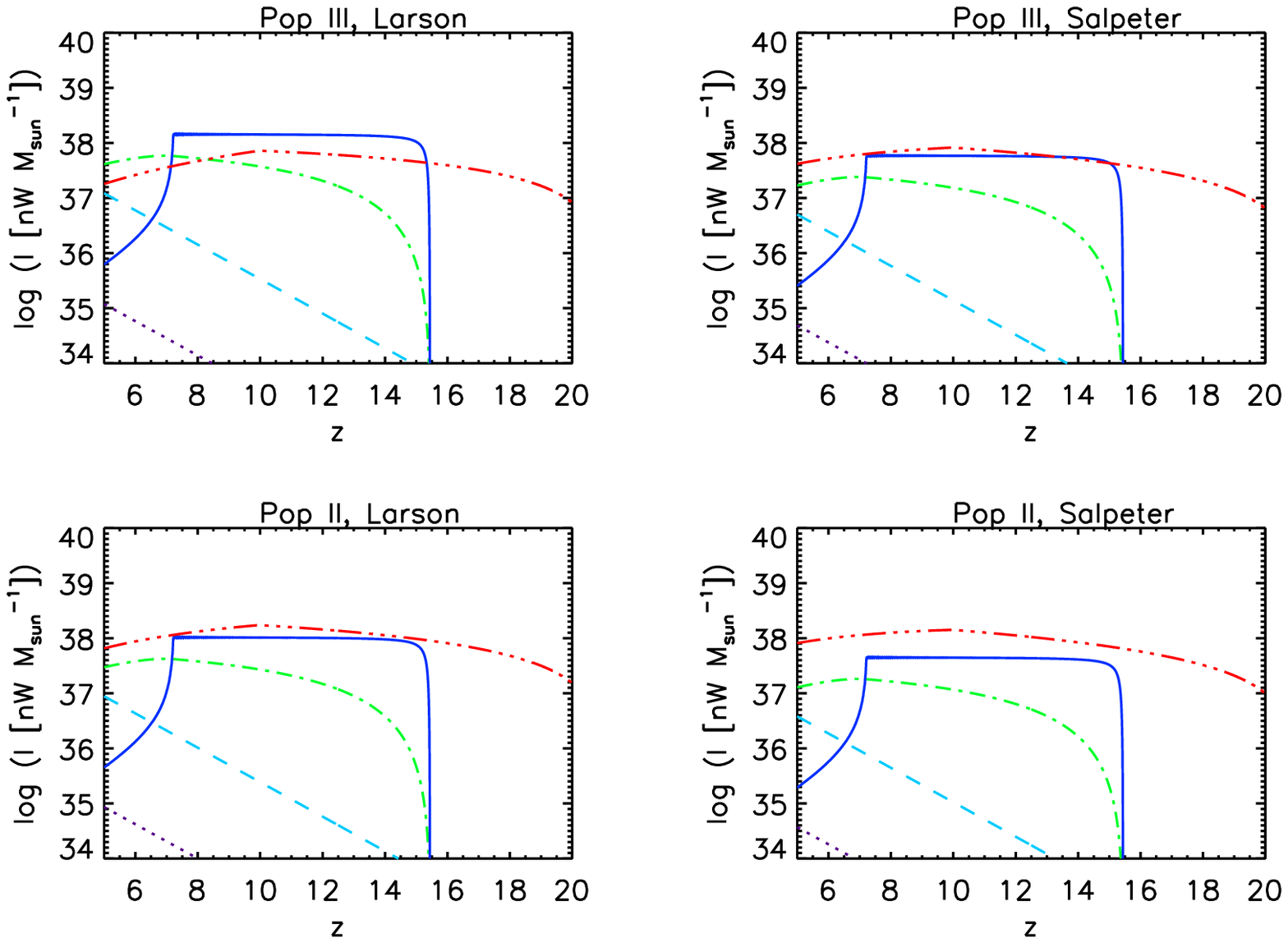}
\caption{%
 Luminosity per stellar mass averaged over a rectangular bandpass from
 $1-2~\mu{\rm m}$. The stellar, $\bar{l}^*$ (triple-dot dashed red line), free-free,
 $\bar{l}^{ff}$ (dotted purple line), free-bound, $\bar{l}^{fb}$ (dashed light blue line), two-photon,
 $\bar{l}^{2\gamma}$ (dot dashed green line), and Lyman-$\alpha$ emission,
 $\bar{l}^{\rm Ly\alpha}$ (solid dark blue line), are shown in units of nW~$M_\sun^{-1}$ as a
 function of redshifts.  Free-free and free-bound both decrease with
 redshift.  This is because both decrease with energy, and as redshift is
 increased, the bandwidth corresponds to higher rest-frame energies.  The
 initial rise in 
 Lyman-$\alpha$ is due to the wing of the line.  At $z\sim15.5$, the line
 hits the end of the band where there is no more Lyman-$\alpha$ emission.
 Stellar emission increases initially because there is more emission from
 the star as energy increases, and later decreases as the bandwidth begins
 to sample energies above 13.6 eV.  Two photon emission is cut off at
 $z\sim15.5$, which corresponds to the band sampling above 10.2 eV, above
 which there is no emission.  
% ({\it Top-left}) PopIII Salpeter, ({\it Top-right}) PopIII Larson,
% ({\it Bottom-left}) PopII Salpeter, and ({\it Bottom-right}) PopII Larson.
}%
\label{fig:lbar}
\end{figure}

In Figure~\ref{fig:lnu} we show $\nu l^\alpha_{\nu}$ (in units of
nW~$M_\sun^{-1}$), 
and in Figure~\ref{fig:lbar} we show $\bar{l}^\alpha(z)$ (in units of
nW~$M_\sun^{-1}$) averaged over a rectangular bandpass from
 $1-2~\mu{\rm m}$, for the stellar 
populations we consider in this paper.
In the relevant redshift range, $7\lesssim z\lesssim 15$, the stellar,
two-photon, and Ly$\alpha$ emission are the most dominant radiation
processes, and all of them are on the order of $\bar{l}\sim
10^{38}~{\rm nW}~M_\sun^{-1}  ~(20~{\rm Myr}/t_{\rm SF})$.

%%%%%%%%%%%%%%%%%%%%%%%%%%%%%%%%%%%%%%%%%%%%%%%%%%%%%%%%%%%%%%%%%%%%%%
%%%%%%%%%%%%%%%%%%%%%%%%%%%%%%%%%%%%%%%%%%%%%%%%%%%%%%%%%%%%%%%%%%%%%%
%%%%%%%%%%%%%%%%%%%%%%%%%%%%%%%%%%%%%%%%%%%%%%%%%%%%%%%%%%%%%%%%%%%%%%
\subsection{Luminosity Density from IGM}
\label{sec:lum}
Photons that do escape the halos go into producing emission in the HII
region surrounding the halo in the IGM (free-free, free-bound, two photon and Lyman-$\alpha$ emission).  The emission in the HII 
region can be found using the volume emissivity, $p({\nu})$, i.e.,
luminosity per comoving volume per frequency, or luminosity density per
frequency (see Eq.~(\ref{eq:def_emissivity}) for the precise definition).

Since all of the radiative processes  we discuss in this section are
proportional to the number density squared, we need to be careful about
the comoving versus proper quantities. 
The proper volume emissivity is proportional to the proper number
density squared, i.e., $p_{prop}\propto n_{prop}^2$. As the comoving
volume emissivity is $p_{com}=a^3 p_{prop}=p_{prop}/(1+z)^3$ and the comoving
number density is $n_{com}=a^3 n_{prop}=n_{prop}/(1+z)^3$, we obtain
$p_{com}\propto (1+z)^3n_{com}^2$. This factor of $(1+z)^3$ simply
reflects the fact that the IGM was denser at higher redshift, and thus
the IGM was brighter.  In the following derivations $n$
always refers to the comoving number density.

For free-free and free-bound emission, the volume emissivity is 
\begin{equation}
p_{\rm ff,fb}({\nu},z) = 4\pi (1+z)^3 n_e n_p \gamma_{\rm c} \frac{e^{-h\nu/kT_{\rm g}}}{T_{\rm g}^{1/2}},
\end{equation}
where $n_e$ and $n_p$ are the comoving number density of electrons
and protons respectively, 
$\gamma_{\rm c}$ is the continuum emission
coefficient including free-free and free-bound emission:
%%%%%%%%%%%%%%%%%%%%%%%%%%%%%%%%%%%%%%%%%%%%%%%%%%%%%%%%%%
\begin{equation}
  \gamma_{\rm c} \equiv f_k\left[\overline{g}_{ff}
  + \sum^{\infty}_{n=2}\frac{x_ne^{x_n}}{n}g_{fb}(n)\right],
\label{eq:gammac}
\end{equation}
%%%%%%%%%%%%%%%%%%%%%%%%%%%%%%%%%%%%%%%%%%%%%%%%%%%%%%%%%
where $x_n\equiv {Ry}/(kT_{\rm g}n^2)$, $\overline{g}_{ff}$ and $g_{bf}(n)$ 
are the Gaunt factors for free-free 
(which is thermally averaged) and free-bound emission, respectively,
$f_k$ is the collection of physical constants
which has a numerical value of $5.44\times 10^{-39}$ in cgs units, and $T_{\rm g}$
is the gas temperature, which we took to be $10^4$~K
(see \S~2.3 of FK06 for more details).

Using the charge neutrality, $n_e=n_p$, we write
\begin{equation}
 n_en_p=n_e^2=n_H^2X_e^2,
\end{equation}
where $n_H$ is the number density of hydrogen atoms and $X_e$ is the
ionization fraction, both of which are given in the simulation.
The volume emissivity is therefore given by
\begin{equation}
\frac{p_{\rm ff,fb}({\nu},z)}{n_H^2X_e^2} = 
4\pi (1+z)^3\gamma_c \frac{e^{-h\nu/kT_{\rm g}}}{T_{\rm g}^{1/2}}.
\end{equation}

The two-photon emissivity is
\begin{equation}
p_{2\gamma}({\nu},z) =
 (1+z)^3\frac{2h\nu}{\nu_{\rm Ly\alpha}}P(y)(1-f_{\rm Ly\alpha})\alpha_{\rm B}n_en_p, 
\end{equation}
A fraction of photons that make the $2-1$ transition, $(1-f_{\rm Ly\alpha})$,
go into two photon emission, while the remainder, $ f_{\rm Ly\alpha}$, produce
the Lyman-$\alpha$ line.  
The precise value of $f_{\rm Ly\alpha}$ depends slightly on the
temperature of gas, and for a gas at $10^4$~K the value of
$f_{\rm Ly\alpha}$  is 0.64 \citep{Spitzer:1978}.
Here, $\alpha_{\rm B}$
is the case B hydrogen recombination coefficient given by
\begin{equation}
\alpha_{\rm B}=\frac{2.06 \times 10^{-11}}{T_{\rm g}^{1/2}}\phi(T_{\rm g}) ~{\rm cm^3~s^{-1}},
\end{equation}
where $\phi(T_{\rm g})$ is given by \citet{Spitzer:1978}. Here, $P(y)$ is the
normalized probability per two photon decay that one photon 
is in the range $dy=d\nu / \nu_{\rm Ly\alpha}$, which can be fit as
(Eq.~(22) of FK06)
\begin{equation}
P(y)=1.307-2.627(y-0.5)^2+2.563(y-0.5)^4-51.69(y-0.5)^6,
\end{equation}
and $\nu_{\rm Ly\alpha}$ is the frequency of Lyman-$\alpha$ photons.  Using $n_H$ and $X_e$, we write the emissivity as
\begin{equation}
\frac{p_{2\gamma}({\nu},z)}{n_H^2X_e^2} = 
(1+z)^3\frac{2h\nu}{\nu_{\rm Ly\alpha}}P(y)(1-f_{\rm Ly\alpha})\alpha_{\rm B}.
\end{equation}

For Lyman-$\alpha$,
\begin{equation}
p_{\rm Ly\alpha}({\nu},z) = (1+z)^3f_{\rm Ly\alpha}h\nu_{\rm Ly\alpha}n_en_p \alpha_{\rm B}\phi(\nu-\nu_{\rm Ly\alpha}),
\end{equation}
where $\phi(\nu-\nu_{\rm Ly\alpha})$ is the line profile of the Lyman-$\alpha$ line, given in \citet{loeb/rybicki:1999,santos/bromm/kamionkowski:2002}.  Using $n_H$ and $X_e$, we get
\begin{equation}
\frac{p_{\rm Ly\alpha}({\nu},z)}{n_H^2X_e^2} = 
(1+z)^3f_{\rm Ly\alpha}h\nu_{\rm Ly\alpha} \alpha_{\rm B}\phi(\nu-\nu_{\rm Ly\alpha}).
\end{equation}

\begin{figure}%[h!tb]
\centering \noindent
\plotone{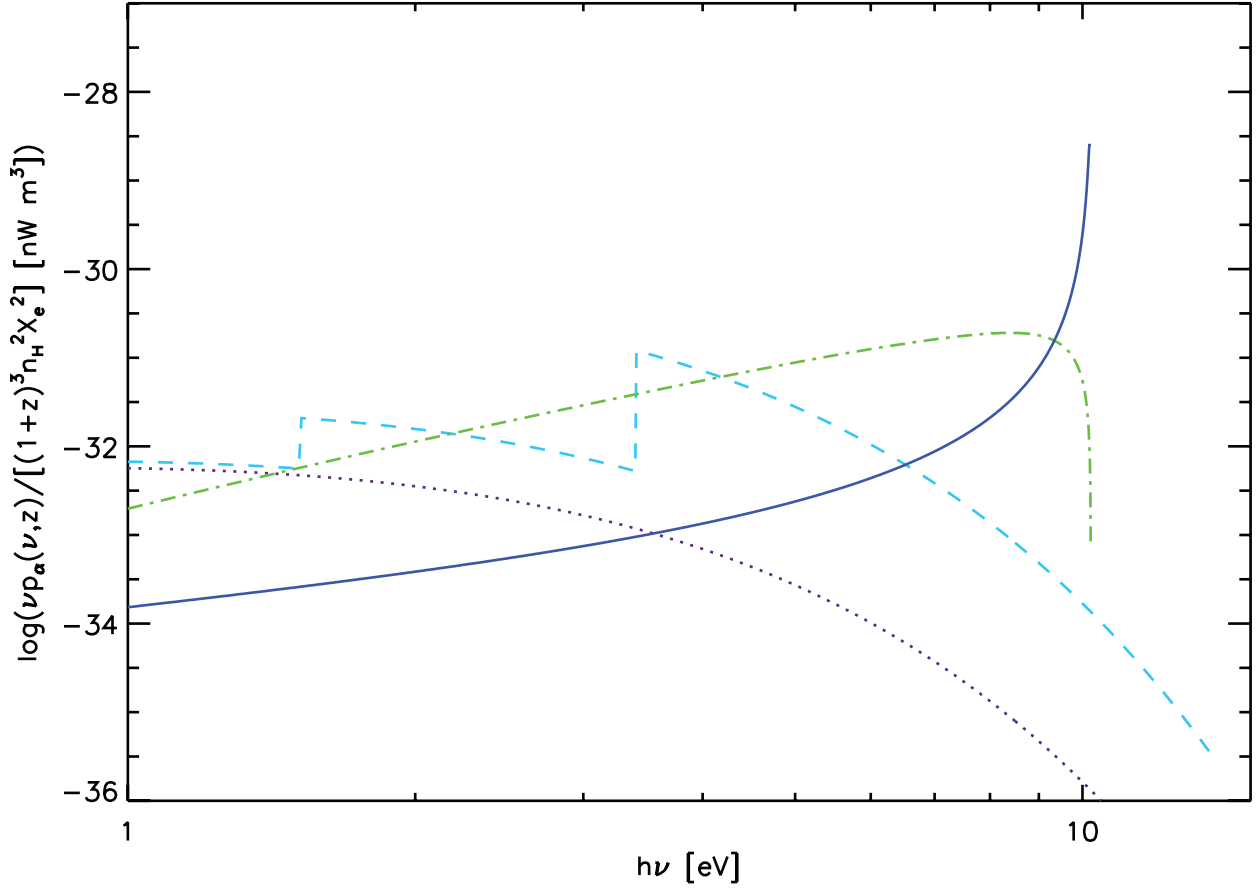}
\caption{%
 Volume emissivity spectrum of the IGM, $\nu p_\alpha(\nu,z)$, divided by $(1+z)^3n_H^2X_e^2$, for individual processes 
 in units of nW~m$^3$ as
 a function of the rest-frame energies. (Note that this quantity does
 not depend on $z$.) We use the ionized 
 gas temperature of $10^4$~K.  Free-free (dotted purple line), free-bound
 (dashed light blue line), two-photon (dot dashed green line), and
 Lyman-$\alpha$ emission (solid dark blue line) are shown.
}%
\label{fig:emissivityIGM}
\end{figure}

Collecting all the processes we obtain the volume emissivity of the IGM as
\begin{equation}
\label{eq:emissivityIGM}
\frac{p^{\rm IGM}({\nu},z)}{n_H^2X_e^2}
=
(1+z)^3\left\{
4\pi \gamma_c \frac{e^{-h\nu/kT_{\rm g}}}{T_{\rm g}^{1/2}}
+
\alpha_{\rm B}h\nu_{\rm Ly\alpha}\left[
(1-f_{\rm Ly\alpha})\frac{2\nu P(\nu/\nu_{\rm Ly\alpha})}{\nu^2_{\rm Ly\alpha}}
+
f_{\rm Ly\alpha}\phi(\nu-\nu_{\rm Ly\alpha})
\right]
\right\}.
\end{equation}
We are now in a position to find the emission of the IGM by
pairing these formulas with the hydrogen number densities ($n_H$) and
the ionization fractions ($X_e$) from the simulations.
In Figure~\ref{fig:emissivityIGM} we show 
$\nu p_\alpha(\nu,z)/[(1+z)^3n_H^2X_e^2]$ (in units of nW~m$^3$) for
individual processes 
as a function of the rest-frame energies.

%%%%%%%%%%%%%%%%%%%%%%%%%%%%%%%%%%%%%%%%%%%%%%%%%%%%%%%%%%%%%%%%%%%%%%
%%%%%%%%%%%%%%%%%%%%%%%%%%%%%%%%%%%%%%%%%%%%%%%%%%%%%%%%%%%%%%%%%%%%%%
\section{LUMINOSITY-DENSITY POWER SPECTRUM}
\label{sec:lpk} 
The three-dimensional power spectrum of over-luminosity
density, $\delta\rho_L({\mathbf x})$, is given by
\begin{equation}
\langle \widetilde{\delta\rho}_L({\mathbf k}) \widetilde{\delta\rho}^*_L({\mathbf
 k}') \rangle 
 = (2 \pi)^3 P_L(k) \delta^3 ({\mathbf k}-{\mathbf k}'),
\label{eq:PkL_def}
\end{equation}
where $P_L(k)$ is the luminosity-density power spectrum, and
$\widetilde{\delta\rho}_L({\mathbf k})$ is the Fourier transform of the
over-luminosity density field, $\delta\rho_L({\mathbf x})$.
The over-luminosity density field is related to the excess in the volume
emissivity over the mean, $\delta p(\nu,{\mathbf x})$, integrated over the
observed bandpass $\nu_1$ to $\nu_2$, as 
\begin{equation}
 \delta\rho_L({\mathbf x},z) = \int_{\nu_1(1+z)}^{\nu_2(1+z)}d\nu~\delta
  p(\nu,{\mathbf x},z). 
\end{equation}
In the following derivations we do not write $z$ explicitly for clarity.
%%%%%%%%%%%%%%%%%%%%%%%%%%%%%%%%%%%%%%%%%%%%%%%%%%%%%%%%%%%%%%%%%%%%%%
\subsection{Halo Contribution}
\label{sec:PLhalo}
How do we calculate the halo contribution from a given simulation box at
a given $z$?
For the halo contribution, $\delta\rho_L^{\rm halo}$, we have
\begin{equation}
\delta\rho^{\rm halo}_L({\mathbf x}) \equiv \left(\frac{L_h}{M_h}\right)
\frac{M_{\rm cell}({\mathbf x})-\overline{M}_{\rm cell}}{V_{\rm cell}},
\end{equation}
where $M_{\rm cell}({\mathbf x})$ is the total mass of halos within a given
cell, $V_{\rm cell}$ is the volume of each cell, and the bars
denote the volume average over the simulation box. 
Throughout this paper, we always include both the stellar contribution
as well as the nebular contribution when we refer to the ``halo
contribution.''

Since we assume that
halos have a constant mass-to-light ratio, $L_h/M_h$ does not depend on
${\mathbf x}$ or $M_h$ (but it depends on $z$), and is given by 
Eq.~(\ref{eq:ltom}). 
Since we assume a constant mass-to-light ratio, the luminosity density $\delta\rho_L^{\rm halo}$
is linearly proportional to the halo mass density,
$\delta\rho_M^{\rm halo}$, such that
$\delta\rho_L^{\rm halo}({\mathbf x})=(L_h/M_h)\delta\rho_M^{\rm halo}({\mathbf
x})$, where $\delta\rho^{\rm halo}_M({\mathbf x})$ is the {\it mass}
over-density of halos given by
\begin{equation}
 \delta\rho^{\rm halo}_M({\mathbf x}) \equiv
\frac{M_{\rm cell}({\mathbf x})-\overline{M}_{\rm cell}}{V_{\rm cell}}
=
\int dM_h~M_h 
\left[\frac{dn_h({\mathbf x})}{dM_h}
-\frac{d\bar{n}_h}{dM_h}
\right].
\end{equation}
Here, $dn_h({\mathbf x})/dM_h$ is the number density of halos per mass
within a cell at a location ${\mathbf x}$, and
$d\bar{n}_h/dM_h$ is its average.
Therefore, the luminosity-density power spectrum of halos,
$P^{\rm halo}_L(k)$, is simply proportional to the mass-density power
spectrum of halos, $P^{\rm halo}_M(k)$, as 
\begin{equation}
 P^{\rm halo}_L(k) = \left(\frac{L_h}{M_h}\right)^2 P^{\rm halo}_M(k).
\end{equation}
The shape of $P^{\rm halo}_L(k)$ is determined by that of the
halo mass-density power spectrum. In other words, one only needs to
compute $P^{\rm halo}_M(k)$ from simulations, and the analytical
calculations given in \S~\ref{sec:Lhalo} supply $L_h/M_h$ for a given stellar
population and 
observed bandpass.

Specifically, we compute $P^{\rm halo}_M(k)$  from the simulation as follows
\citep[e.g.,][]{jeong:2008}:
\begin{itemize}
 \item [(1)] Use the Cloud-In-Cell (CIC) 
mass distribution scheme to 
calculate the mass density field of halos on $256^3$ 
regular grid points, i.e., $M_{\rm cell}({\mathbf x})/V_{\rm cell}$, from the
       halo catalog.  
 \item [(2)] Fourier-transform the excess mass density, 
$[M_{\rm cell}({\mathbf x})-\overline{M}]/V_{\rm cell}$,
using {\sf
       FFTW}\footnote{{\sf http://www.fftw.org}}.
 \item [(3)] Deconvolve the effect of the CIC pixelization effect. 
We divide $P(\mathbf{k},z)\equiv|\delta(\mathbf{k},z)|^2$ 
at each cell by the Fourier transform of the CIC kernel squared:
\begin{equation}
W(\mathbf{k})=\prod_{i=1}^3 
\left[\frac{\sin\left(\frac{\pi k_i}{2 k_{N}}\right)}{\frac{\pi k_i}{2 k_{N}}}\right]^4,
\end{equation}
where  $\mathbf{k}=(k_1,k_2,k_3)$, and $k_N\equiv\pi/H$
is the Nyquist frequency 
($H$ is the physical size of the grid). In terms of the number of grids along one axis, $N_{\rm mesh}$, one may write $H=L_{\rm box}/N_{\rm mesh}$, and
$2k_N=N_{\rm mesh}(2\pi/L_{\rm box})=N_{\rm mesh}\Delta k$, where
       $\Delta k=2\pi/L_{\rm box}$ is the fundamental frequency of the
       box. (For our simulation, $L_{\rm box}=100~h^{-1}~{\rm Mpc}$.)
We also try a different deconvolution scheme that attempts to reduce the
       aliasing effect \citep{Jing:2005}:
\begin{equation}\label{eq:window}
W(\mathbf{k})=\prod_{i=1}^3 
\left[
1-\frac{2}{3}\sin^2 \left(\frac{\pi k_i}{2 k_{N}}\right)
\right].
\end{equation}
We then use $P_M(k)$ up to $k_{\rm max}$ below which both of the
       deconvolution schemes yield the same answer. We find
       $k_{\rm max}\sim 5~{\rm Mpc}^{-1}$.
 \item [(4)]  Compute $P_M(k,z)$ by taking the angular average of
CIC-corrected $P(\mathbf{k},z)$ within a 
spherical shell defined by
$k-\Delta k/2 < |\mathbf{k}| < k+\Delta k/2$.
\end{itemize}

In the previous work on NIRB fluctuations
\citep{kashlinsky/etal:2004,cooray/etal:2004} 
the linear bias model was used, i.e., $P^{\rm halo}_M(k)$ was assumed to be
linearly proportional to the underlying (linear) matter  power spectrum.
However, for such high redshifts halos are expected to be highly biased,
and thus {\it non-linear bias} cannot be ignored. In other words, it is
no longer correct to assume that $P^{\rm halo}_M(k)$ is linearly
proportional to the underlying  matter  power spectrum.

\begin{figure*}%[t]
\centering \noindent
\includegraphics[width=8cm]{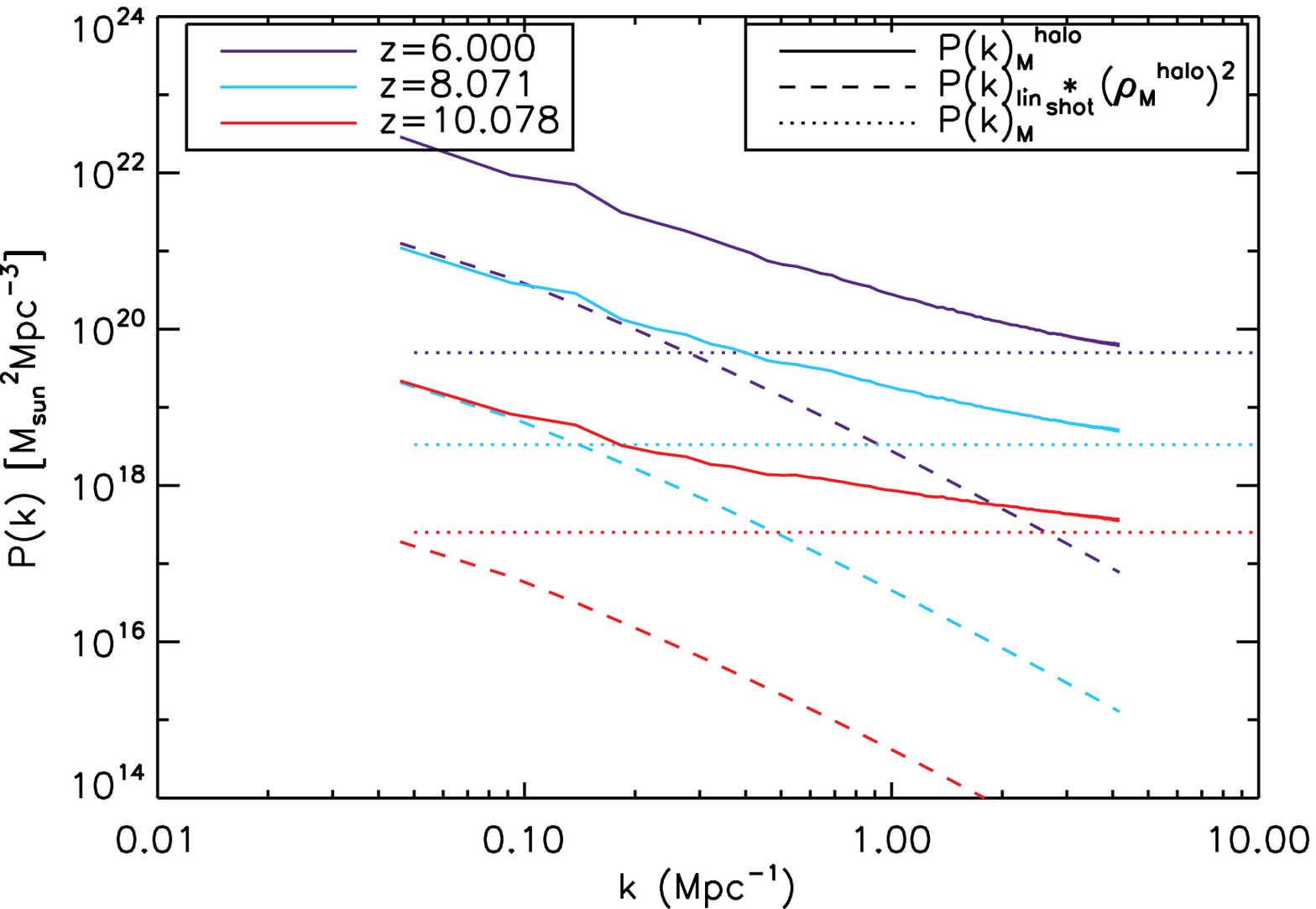}
\includegraphics[width=8cm]{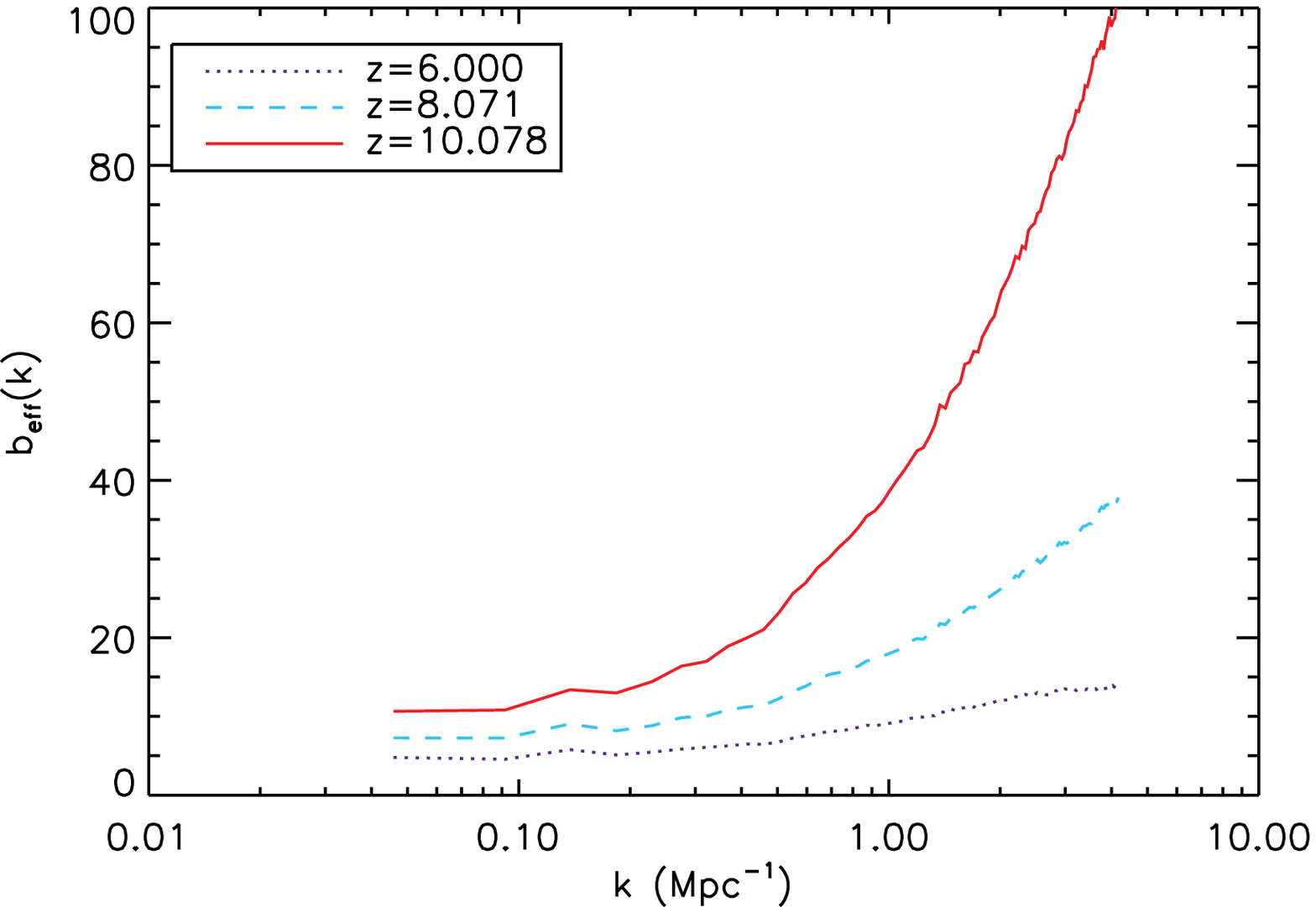}
%\vspace{5mm}
\includegraphics[width=8cm]{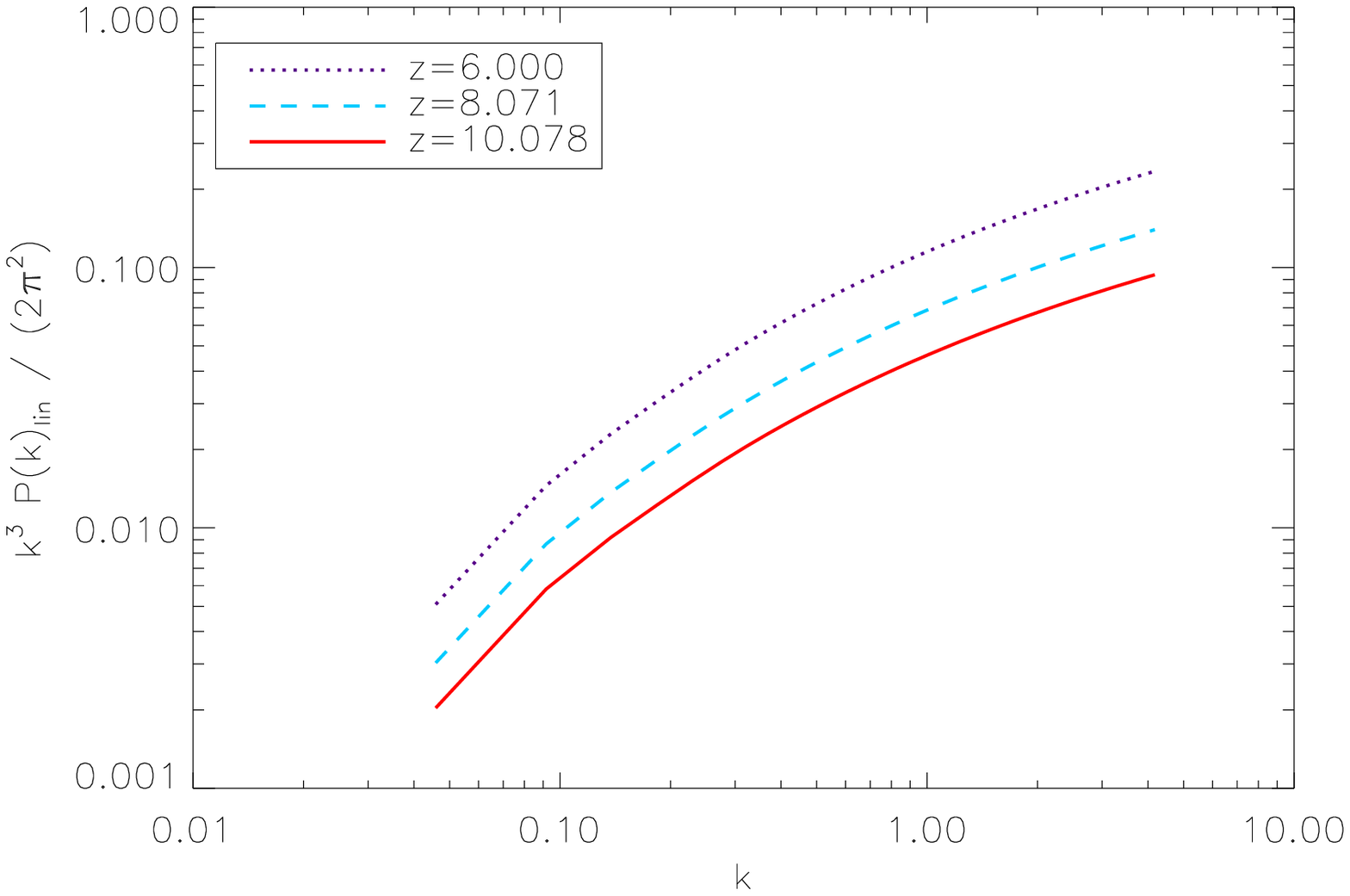}
\includegraphics[width=8cm]{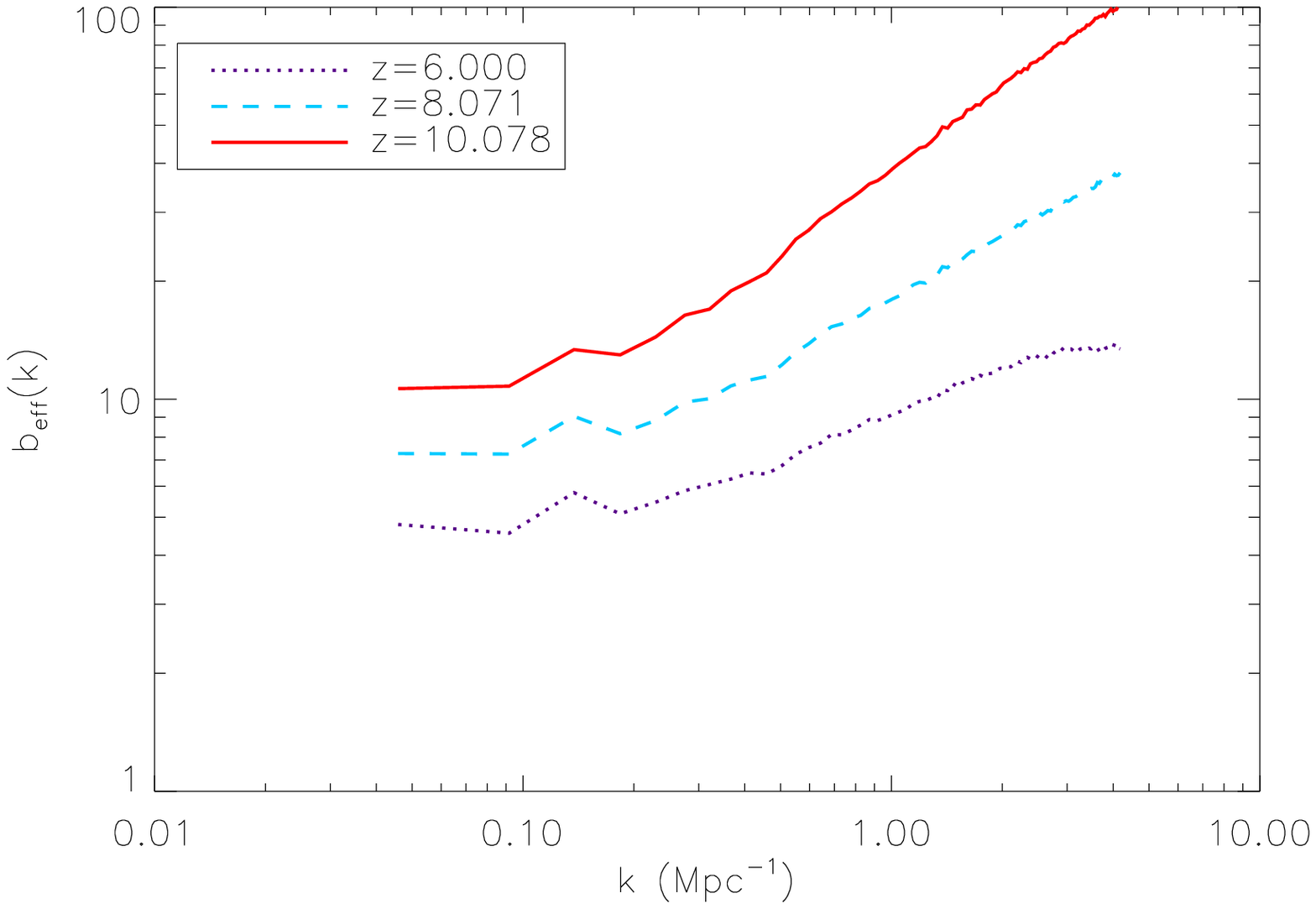}
\caption{%
Non-linear bias of the halo mass-density power spectrum. (This is not the
 luminosity-density power spectrum; see \S~\ref{sec:PLhalo} for the precise
 definition.)  
Top left panel: The power spectra of the halo mass density, $P_M^{\rm halo}(k)$
 are shown as 
the solid lines ($z=6$ to $10$ from top to bottom), the linear matter power
spectra times the mean halo mass density squared, $P_{\rm lin}(k)(\bar{\rho}_M^{\rm halo})^2$, are the
 dashed lines, and the shot noise power spectra, $P_M^{shot}$, are the
 dotted lines. Top right panel: We show the bias,
 $\sqrt{[P_M^{\rm halo}(k)-P_M^{shot}(k)]/[(\bar{\rho}_M^{\rm halo})^2P_{\rm lin}(k)]}$,
 ($z=10$ to $6$ from top to bottom). The bias
 increases significantly as we go to smaller 
 scales, and this effect has been ignored in the previous calculations
 of the power spectrum of NIRB fluctuations.  
 Note that the minimum halo mass resolved in the
 simulation is $2.2\times 10^9~M_\sun$. The degree of non-linear bias would be
 smaller for a smaller minimum mass \citep[see, e.g., Figure~6
   of][]{Trac/Cen:2007}.  
Bottom left panel: The linear power
  spectrum, $k^3P_{lin}(k)/(2\pi^2)$.  Bottom right panel: Same as top right
  panel, but on a log-log axis.
}%
\label{fig:bias}
\end{figure*}

To study this further, in Figure~\ref{fig:bias} we show 
$P_M^{\rm halo}(k)$ (in units of $M_\sun^2~{\rm Mpc}^{-3}$).  Also shown in
Figure~\ref{fig:bias} is the shot noise,
$P_M^{shot}$, where
$P_M^{shot}\equiv \int dM_h M_h^2d\bar{n}_h/dM_h$
($d\bar{n}_h/dM_h$ is the mean halo mass function), the linear
matter density fluctuations times the mean mass density squared 
($P_{\rm lin}(k)(\bar{\rho}_M^{\rm halo})^2$), where $\bar{\rho}_M^{\rm halo}$ is the mean mass density of halos within the
simulation box, and the bias, given by:
\begin{equation}
b_{\rm eff}(k) = \sqrt{\frac
 {P_M^{\rm halo}(k)-P_M^{shot}(k)}{(\bar{\rho}_M^{\rm halo})^2P_{\rm lin}(k)}}. 
\end{equation}

%%%%%%%%%%%%%%%%%%%%%%%%%%%%%%%%%%%%%%%%%%%%%%%%%%%%%%%%%%%%%%%%%%%%%%
%%%%%%%%%%%%%%%%%%%%%%%%%%%%%%%%%%%%%%%%%%%%%%%%%%%%%%%%%%%%%%%%%%%%%%
%%%%%%%%%%%%%%%%%%%%%%%%%%%%%%%%%%%%%%%%%%%%%%%%%%%%%%%%%%%%%%%%%%%%%%
\begin{figure}[t]
\centering \noindent
\plotone{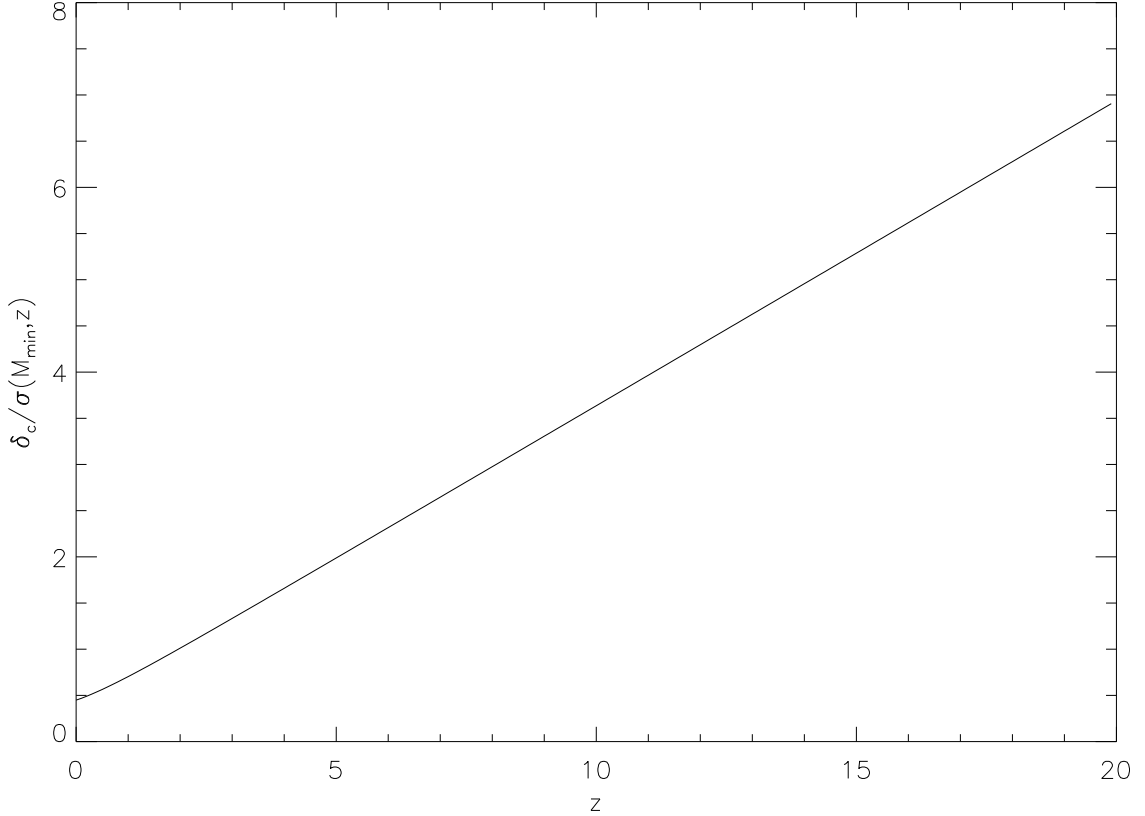}
\caption{%
$\delta_c/\sigma(M_{Min},z)$ versus redshift.  The halos resolved in our simulation, with
  $M>M_{\rm min}=2.2\times 10^9~M_\sun$, are located on rare peaks ($\delta_c/\sigma(M_{\rm min},z)\gtrsim 2.5$) at $z\gtrsim 7$.
}
\label{fig:sigmadel}
\end{figure}
%%%%%%%%%%%%%%%%%%%%%%%%%%%%%%%%%%%%%%%%%%%%%%%%%%%%%%%%%%%%%%%%%%%%%%
%%%%%%%%%%%%%%%%%%%%%%%%%%%%%%%%%%%%%%%%%%%%%%%%%%%%%%%%%%%%%%%%%%%%%%
\begin{figure}[t]
\centering \noindent
\plottwo{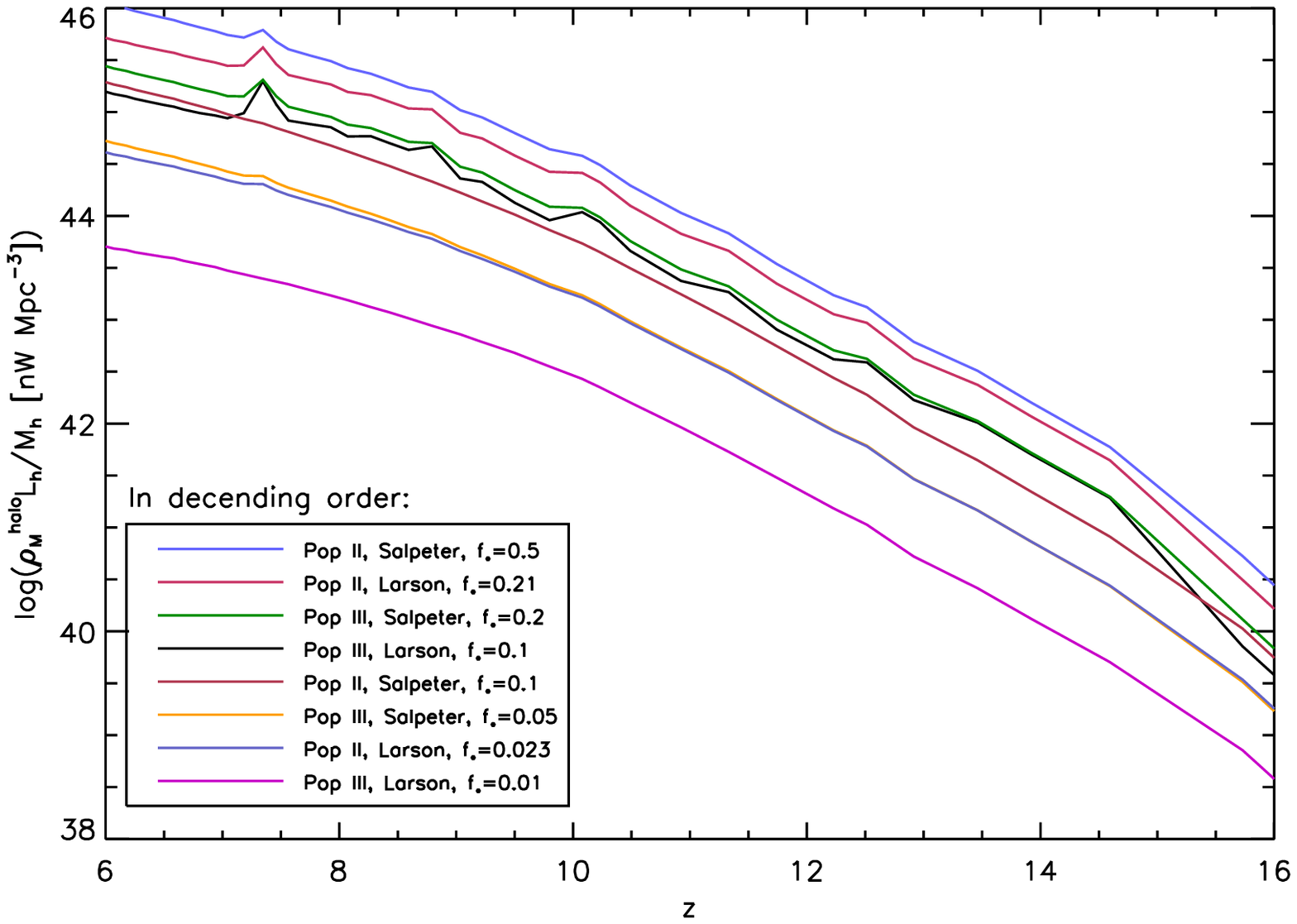}{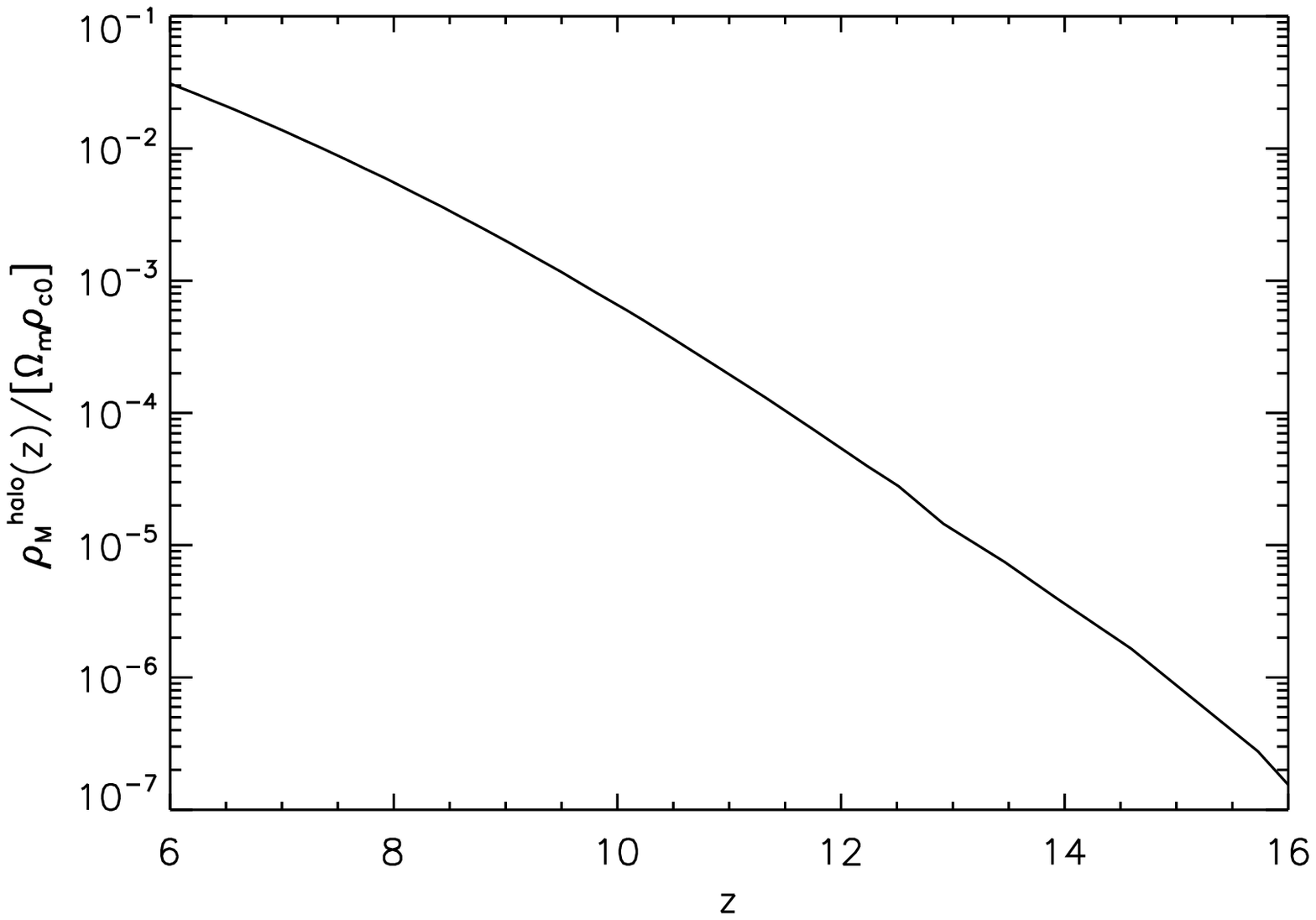}
\caption{%
({\it Left}) Mean halo luminosity density computed from our simulation,
 $\bar{\rho}_M^{\rm halo}(z)L_h(z)/M_h$, where $l^\alpha_\nu(z)$ is from
  equation (\ref{eq:lnu1}),  
in units of nW~Mpc$^{-3}$ as a function of redshifts, for various
 stellar populations given in Table~\ref{tab:populations}. The waves in the
 lines where $f_{\rm esc}$ are higher are from the discrete redshift sampling
 of the Lyman-$\alpha$ line.  We averaged
 the luminosity over a rectangular bandpass of $1-2~\mu{\rm m}$.
({\it Right}) Halo mass collapse fraction,
 $\bar{\rho}_M^{\rm halo}(z)/(\Omega_m\rho_{\rm c0})$,  
 as a function of
 redshifts. The redshift evolution of $\bar{\rho}_M^{\rm halo}L_h/M_h$ is
 essentially determined by that of $\bar{\rho}_M^{\rm halo}$.
}%
\label{fig:meanlumdensity}
\end{figure}
%%%%%%%%%%%%%%%%%%%%%%%%%%%%%%%%%%%%%%%%%%%%%%%%%%%%%%%%%%%%%%%%%%%%%%

By comparing $P_M^{\rm halo}(k)$ (with the shot
noise subtracted)  
with the power spectrum of linear matter density
fluctuations times $(\bar{\rho}_M^{\rm halo})^2$, we find that, on large
scales ($k\lesssim 0.1~{\rm Mpc}^{-1}$), they are 
related by $P_M^{\rm halo}(k)/(\bar{\rho}_M^{\rm halo})^2\approx b_1^2 P_{\rm lin}(k)$
with the linear bias factor being 
$b_1\simeq 5$ at $z=6$ to $b_1\simeq 10$ at $z=10$,  a
highly biased population. 
The bias increases monotonically as we go to smaller scales,
significantly boosting the power in the halo distribution relative to
the matter distribution. This changes the prediction for the shape of
the angular power spectrum qualitatively, compared with the previous
results given in the literature
\citep{cooray/etal:2004,kashlinsky/etal:2004}.  This behavior of
  non-linear bias with redshift is consistent with that expected from the halo
  model \citep{cooray/sheth:2002}.  These halos are very rare, located on
  high peaks with $\delta_c/\sigma(M_{min},z) \gtrsim 2.5$
  (see Figure \ref{fig:sigmadel}).

This motivates our writing $P^{\rm halo}_L(k)$ as
\begin{equation}
 P^{\rm halo}_L(k) =
  \left(\frac{\bar{\rho}_M^{\rm
   halo}L_h}{M_h}\right)^2b_{\rm eff}^2(k)P_{\rm lin}(k),
\label{eq:pkalt}
\end{equation}
where the pre-factor, $\bar{\rho}_M^{\rm halo}L_h/M_h$, is the mean
halo luminosity density. 
In the left panel of 
Figure~\ref{fig:meanlumdensity} we show $\bar{\rho}_M^{\rm halo}L_h/M_h$
(in units of nW~Mpc$^{-3}$) as a function of redshifts. We find that the
redshift evolution of  $\bar{\rho}_M^{\rm halo}L_h/M_h$ is very rapid; thus,
the redshift evolution of the halo luminosity density power spectrum,
$P^{\rm halo}_L(k)$, is dominated by that of the mean halo luminosity
density.

What determines the evolution of the mean halo luminosity density? The answer is
simple: it is determined by the rate at which the mass in the universe
collapses into halos. To show this, in the right panel of 
Figure~\ref{fig:meanlumdensity} we show the halo mass collapse fraction,
or the ratio of $\bar{\rho}_M^{\rm halo}$ 
to the mean comoving mass
density of the 
universe, $\Omega_m\rho_{\rm c0}$, where
$\rho_{\rm c0}=2.775\times 10^{11}~h^2~M_\sun~{\rm Mpc}^{-3}$ is the critical
density of the universe at the present epoch.
The evolution of the collapse fraction is fast, explaining the fast
evolution of the mean halo luminosity density.

As halos are discrete objects, and we do not expect to resolve
individual halos contributing to the diffuse NIRB, the observed NIRB
power spectrum is a sum of the clustering component and the shot noise
component.  If the shot noise dominates over the clustering component, 
it would be very difficult to ascertain information on the structure
from the signal of the NIRB.  The shot noise component can be estimated by
integrating the luminosity squared over the mass function:
\begin{equation}
P_L^{shot}
= 
\left(\frac{L_h}{M_h}\right)^2
P_M^{shot}
=
\left(\frac{L_h}{M_h}\right)^2
\int dM_h~M_h^2\frac{d\bar{n}_h}{dM_h},
\end{equation}
where we have again assumed that
each halo has a constant mass-to-light ratio, i.e., $L_h/M_h$
is independent of $M_h$. 
%%%%%%%%%%%%%%%%%%%%%%%%%%%%%%%%%%%%%%%%%%%%%%%%%%%%%%%%%%%%%%%%%%%%%%
\subsection{IGM Contribution} 
\label{sec:PLIGM}
For the IGM contribution, we have
 \begin{equation}
\delta\rho^{\rm IGM}_L({\mathbf x}) = \left(\frac{{p^{\rm IGM}}}{n_H^2X_e^2}\right)
\left[
C_{\rm cell}({\mathbf x})n_{\rm cell}^2({\mathbf x})X^2_{e,\rm cell}({\mathbf x})
-
\overline{(C_{\rm cell}n_{\rm cell}^2X_{e,\rm cell}^2)}\right],
\end{equation}
where $p^{\rm IGM}$ is the volume emissivity of the IGM, integrated over the
  observed frequencies, i.e., $p^{\rm IGM}\equiv
  \int_{\nu_1(1+z)}^{\nu_2(1+z)} d\nu p^{\rm IGM}(\nu)$,
$C_{\rm cell}$, $n_{\rm cell}$, $X_{e,\rm cell}$ are the clumping factor,
the comoving number density of hydrogen atoms, and the
  ionization fraction within a 
  cell, respectively. 
  We compute $n_{\rm cell}$ using
\begin{equation}
 n_{\rm cell} = \frac{\Omega_b}{\Omega_m}\frac{\rho_{M,\rm cell}}{\mu m_p},
\end{equation}
where $\mu=0.59$ and $m_p$ are the mean molecular weight of ionized gas and the
  proton mass, respectively. We have used the mass density of $N$-body
  particles per cell, $\rho_{M,\rm cell}$, multiplied by the baryon
  fraction, $\Omega_b/\Omega_m$, for computing the mass density of
  baryons per cell, as we have assumed that gas
  traces dark matter particles, i.e., $N$-body particles.
  The clumping factor, $C_{\rm cell}\equiv n^2_{actual}/n_{\rm cell}^2$, relates
  the actual density squared to the square of the density averaged
  within a cell. In other words, $C_{\rm cell}$ captures the sub-grid
  clumping that is not resolved by the simulation. 

Following \citet{iliev/etal:2007}, we make a simplifying assumption 
that $C_{\rm cell}$ takes on the same value everywhere in the simulation,
and evolves with redshift $z$ as $C_{\rm cell}(z)=26.2917e^{-0.1822z+0.003505z^2}$; thus, we have
 \begin{equation}
\delta\rho^{\rm IGM}_L({\mathbf x}) = 
26.2917e^{-0.1822z+0.003505z^2}
\left(\frac{p^{\rm IGM}}{n_H^2X_e^2}\right)
\left[
n_{\rm cell}^2({\mathbf x})X^2_{e,\rm cell}({\mathbf x})
-
\overline{(n_{\rm cell}^2X_{e,\rm cell}^2)}\right].
\end{equation}
Note that $p^{\rm IGM}/(n_H^2X_e^2)$ does not
  depend on ${\mathbf x}$, and is given by Eq.~(\ref{eq:emissivityIGM})
  integrated over a rectangular bandpass of $1-2~\mu$m in the observer's frame.
%%%%%%%%%%%%%%%%%%%%%%%%%%%%%%%%%%%%%%%%%%%%%%%%%%%%%%%%%%%%%%%%%%%%%%
\section{RESULTS}
\label{sec:results}
%%%%%%%%%%%%%%%%%%%%%%%%%%%%%%%%%%%%%%%%%%%%%%%%%%%%%%%%%%%%%%%%%%%%%%
\subsection{Luminosity-density Power Spectrum}
\label{sec:results_pk}
\begin{figure}%[t]
\centering \noindent
\plottwo{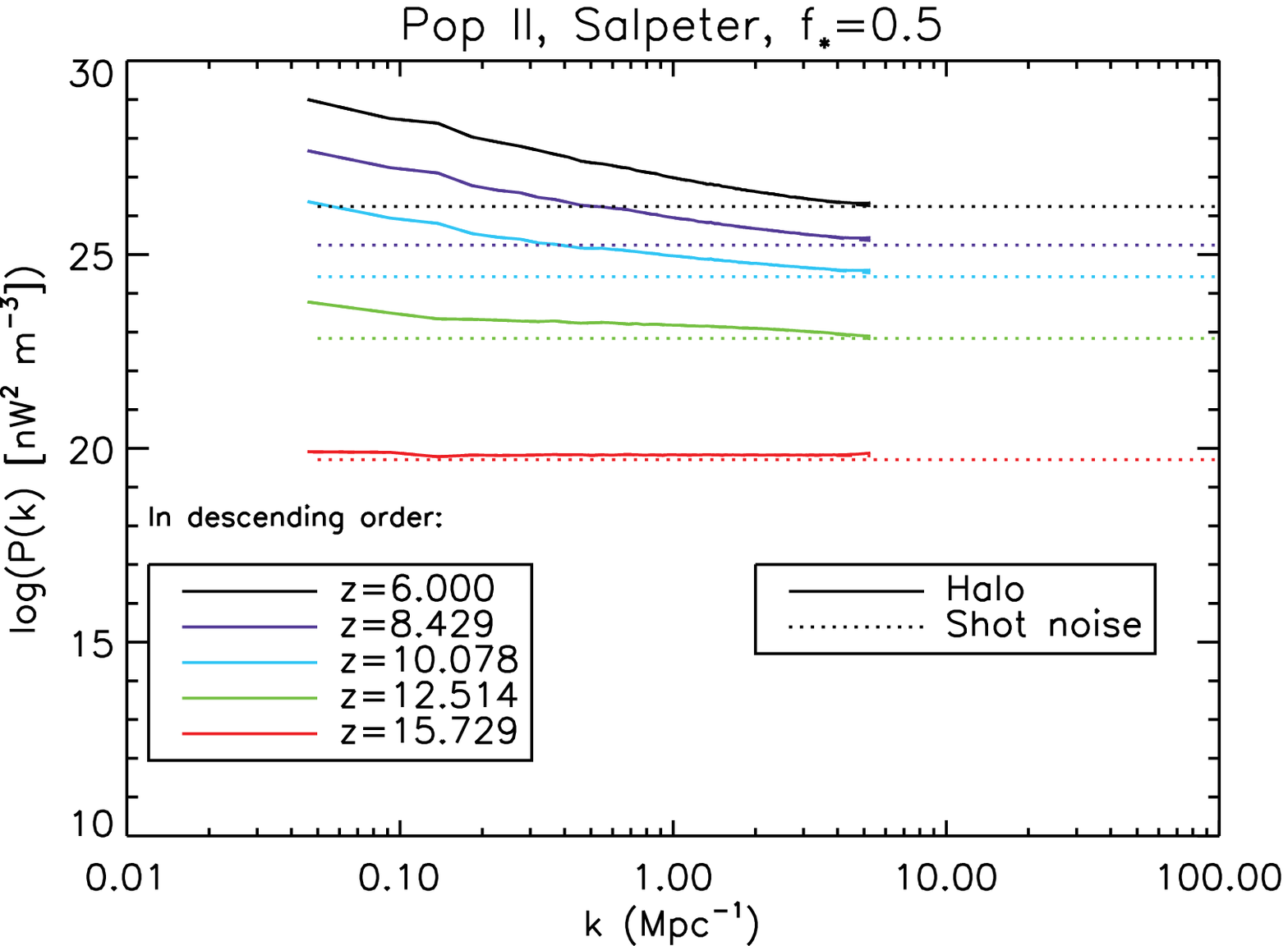}{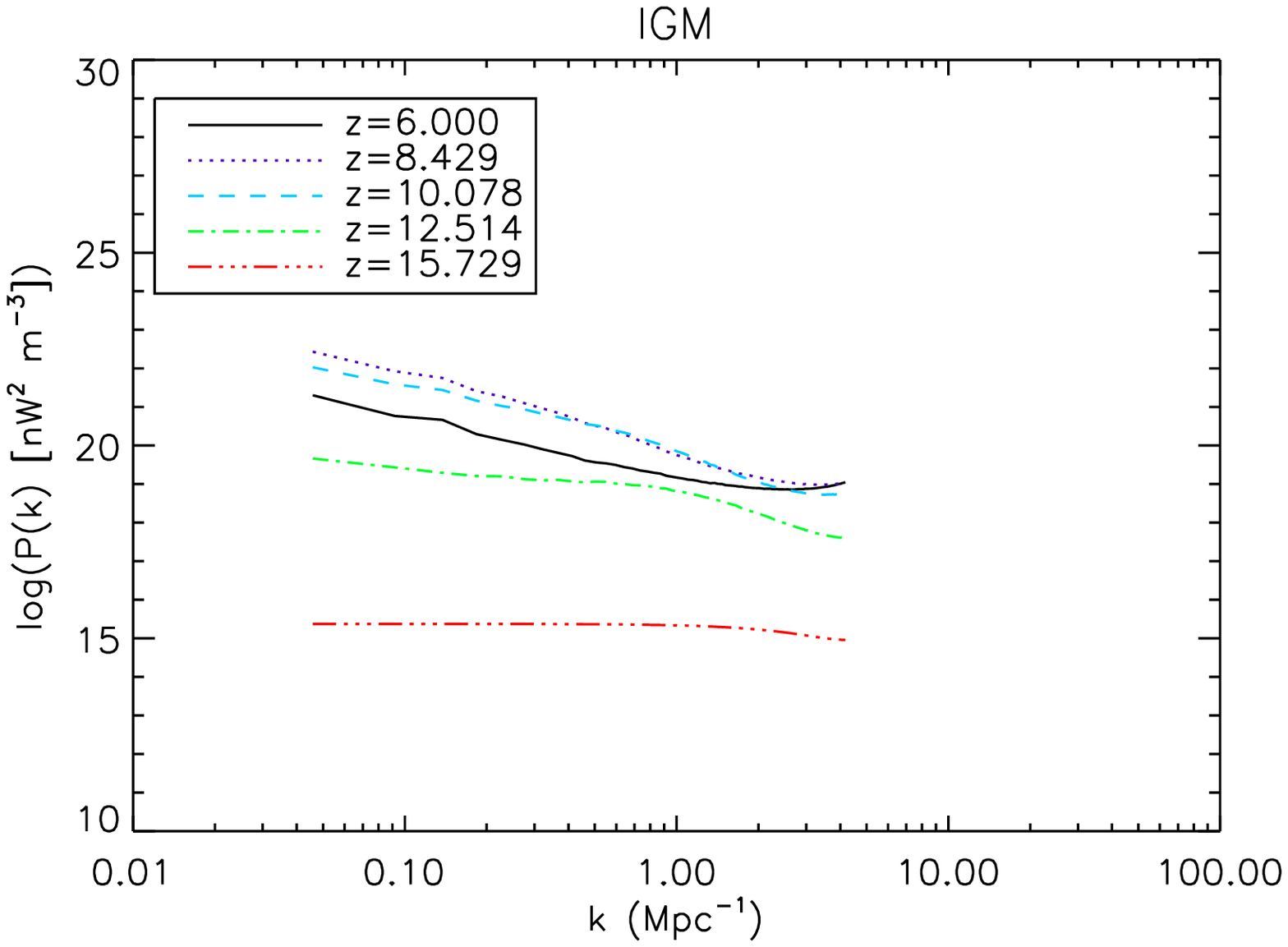}
\caption{%
({\it Left}) Luminosity-density power spectrum of halos with Pop II stars obeying a
 Salpeter initial mass spectrum, $f_{\rm esc}=0.19$, and $f_*=0.5$, assuming
 a rectangular bandpass from 
 $1-2$ $\mu{\rm m}$.    
The shot noise for the halo contribution is also shown as the dotted lines.
({\it Right})  Luminosity-density power spectrum of the IGM. 
The ionization
 fraction of the IGM reaches 0.5 at about $z\sim 8.3$.
 On large scales where the shot noise is sub-dominant, we find
 $P_L(k)\propto k^{-3/2}$, which yields $C_l\propto l^{-3/2}$ or
 $l^2C_l\propto l^{1/2}$ (see \S~\ref{sec:results_cl}).
}%
\label{fig:pk1}
\end{figure}

\begin{figure}%[t]
\centering \noindent
\plottwo{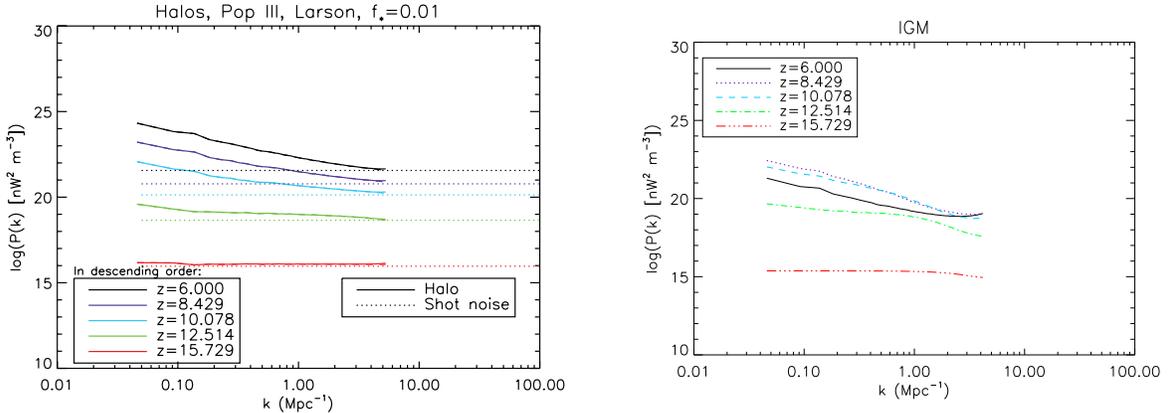}{powerspectrumIGM.eps}
\caption{%
({\it Left})
The same as the left panel of 
Figure \ref{fig:pk1} with Pop III stars with the Larson initial mass spectrum,
$f_{\rm esc}=1$, and $f_*=0.01$.
({\it Right})
The same as the right panel of Figure \ref{fig:pk1} for comparison.
}%
\label{fig:pk2}
\end{figure}

In Figures \ref{fig:pk1} and \ref{fig:pk2} we show the
luminosity-density power spectra, $P_L(k)$, for halos and their
associated HII regions in the IGM for two of our populations:
Population II stars with a Salpeter initial mass spectrum with
$f_{\rm esc}=0.19$ and $f_*=0.5$ (Figure~\ref{fig:pk1}) and Population III
stars with a 
Larson initial mass spectrum with $f_{\rm esc}=1$ and $f_*=0.01$
(Figure~\ref{fig:pk2}), assuming a rectangular bandpass from $1-2$
$\mu{\rm m}$.

%%%%%%%%%%%%%%%%%%%%%%%%%%%%%%%%%%%%%%%%%%%%%%%%%%%%%%%%%%%%%%%%%%%%%%
%%%%%%%%%%%%%%%%%%%%%%%%%%%%%%%%%%%%%%%%%%%%%%%%%%%%%%%%%%%%%%%%%%%%%%
%%%%%%%%%%%%%%%%%%%%%%%%%%%%%%%%%%%%%%%%%%%%%%%%%%%%%%%%%%%%%%%%%%%%%%

The luminosity-density power spectra of halos are approximately
power-laws over the entire range of wavenumbers that the simulation covers. 
At the highest redshift bin, $z\sim 16$, the power spectrum is entirely
dominated by the shot noise at all scales. The lower the redshifts are,
the more power in excess of the shot noise we observe on large scales
(because the shot noise is most important on small scales). The growth
of the power spectrum is partly driven by the growth of linear matter
fluctuations as well as that of halo bias, i.e., the clustering of halos
is biased relative to the underlying matter distribution. As we have
shown in the previous section, the bias of halos that we observe in the
simulation is highly 
non-linear, and thus has an important implication for the predicted
shape of the observed power spectrum of NIRB fluctuations.
However, as we have shown in \S~\ref{sec:PLhalo}, the evolution of 
$P_L(k)$ is almost entirely driven by the fast growth of the mean halo
luminosity density, $\bar{\rho}_M^{\rm halo}(z)L_h(z)/M_h$
(see Eq.~(\ref{eq:pkalt}) and the left panel of
Figure~\ref{fig:meanlumdensity}). As a result $P_L(k)$ grows by about six orders
of magnitude at $k=0.1~{\rm Mpc}^{-1}$ from $z\sim 13$ to $z\sim 6$,
which is much faster than the growth expected from the growth of bias
times the matter power spectrum.

The luminosity-density power spectrum of the IGM increases
quickly as the
mean ionization fraction, $\bar{X}_e$, 
approaches 0.5 (at about $z\sim 8.3$ for this
particular simulation), especially on larger
scales.  As the ionization fraction increases, the luminosity of the HII
region would also increase (because luminosity is proportional to
$X_e^2$).  Moreover, since we 
are looking at the over-luminosity-density power spectrum of the IGM, the greatest power results when there is the greatest difference between luminous 
regions and the average luminosity of the IGM; thus, the power spectrum
of $X^2_en^2$ grows rapidly as $\bar{X}_e$ approaches 1/2. 
However, this rapid growth of the power stops when the entire IGM is
ionized ($\bar{X}_e=1$), in which case the over-luminosity-density power
spectrum of 
the IGM is simply proportional to $n^2$. 

The most interesting feature of the luminosity-density power spectrum of
the IGM is  a ``knee'' feature, which is at $k\sim 2~{\rm Mpc}^{-1}$
at $z\sim 16$, and moves to $k\lesssim 1~{\rm Mpc}^{-1}$ at $z\lesssim
10$.  This ``knee'' is caused by the typical size of HII bubbles:
 the knee wavenumber 
is inversely proportional to the typical size of
the  bubbles.   
At the highest redshift bin, $z\sim 16$, the bubbles are  nearly
Poisson-distributed, and thus the power spectrum is flat up to
the knee scale, $k\sim 2~{\rm Mpc}^{-1}$, beyond which the power
decreases as one is looking at the scales inside the bubbles, which are smooth.
As the redshift decreases, the knee scale moves to
larger scales, 
 signifying a growth in the ionized bubbles with time until they
 merge. At the same time, the large-scale power also grows, and the
 shape of the HII region power spectrum is basically the same as that of
 the halo power spectrum, as the bubbles are created around the halos.    

Note that \citet{iliev/etal:2006,iliev/etal:2007} studied the power
spectrum of ionized gas density, and observed a similar trend. The power
spectrum of the luminosity density that we have presented here is the
four-point function of the ionized gas density (as the volume emissivity
is proportional to the ionized gas density squared), and thus it is different
from the power spectrum of the ionized gas density (which is quadratic
in density).

%%%%%%%%%%%%%%%%%%%%%%%%%%%%%%%%%%%%%%%%%%%%%%%%%%%%%%%%%%%%%%%%%%%%%%
\subsection{Angular Power Spectrum of NIRB Fluctuations}
\label{sec:results_cl}
What about the observable, the angular power spectrum of NIRB fluctuations,
$C_l$?  We compute the angular power spectrum of NIRB fluctuations, 
$C_l$, by projecting $P_L(k)$ on the sky. We do this using Limber's 
approximation, and obtain (see Appendix~\ref{sec:cl_derivation} for the derivation)
\begin{equation}
C_l =
\frac{c}{(4\pi)^2}
\int \frac{dz}{H(z)r^2(z)(1+z)^4}
P_L\left(k=\frac{l}{r(z)},z\right),
 \label{eq:cl}
\end{equation}
where $r(z)=c\int^z_0 dz'/H(z')$ is the comoving distance.
We integrate Eq.~(\ref{eq:cl}) over  the range of redshifts that the
 simulation covers for both halos and the IGM, $z=6.0- 15.7$.  

%%%%%%%%%%%%%%%%%%%%%%%%%%%%%%%%%%%%%%%%%%%%%%%%%%%%%%%%%%%%%%%%%%%%%%
\begin{figure}%[h!tb]
\centering \noindent
\plottwo{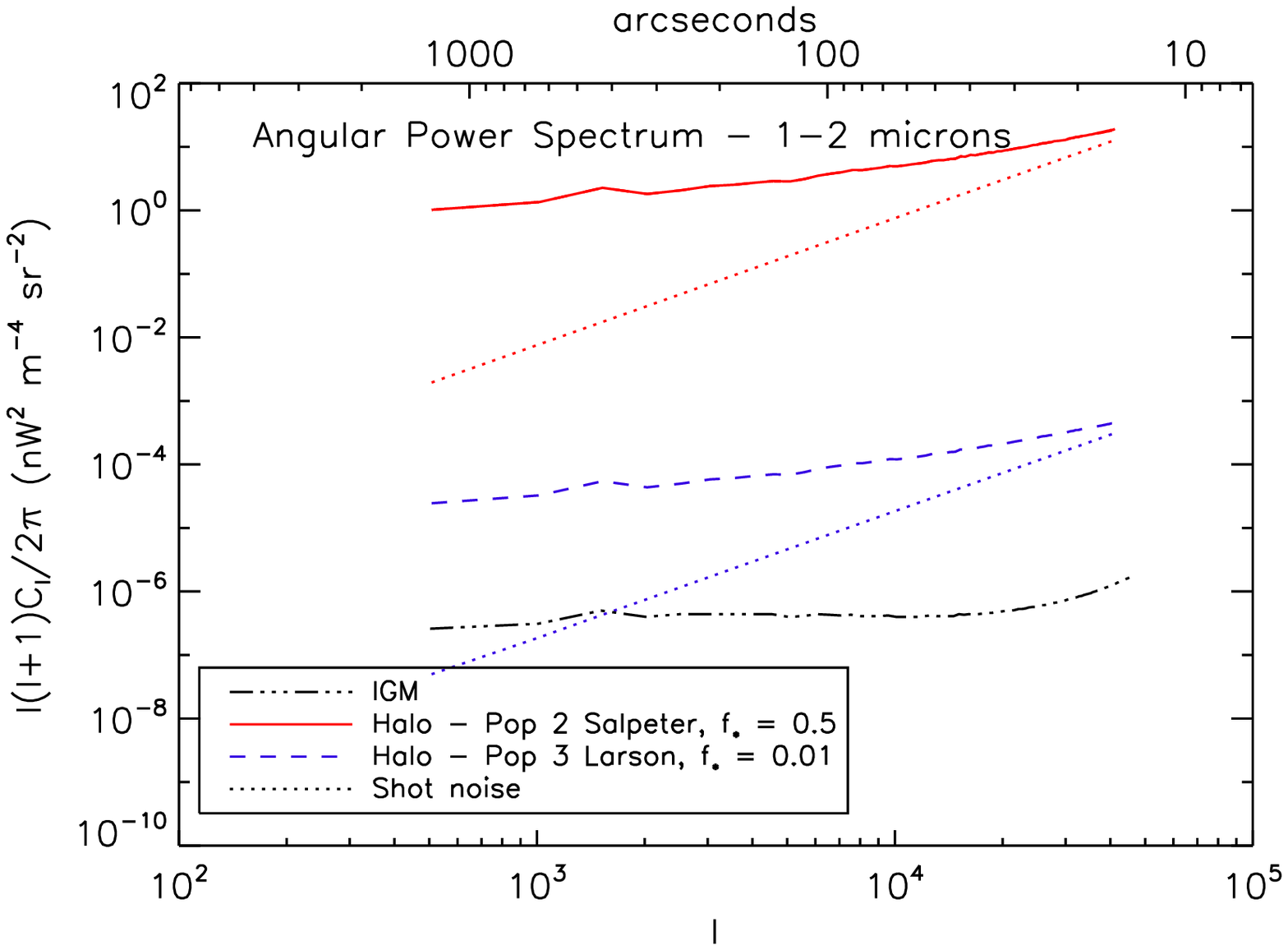}{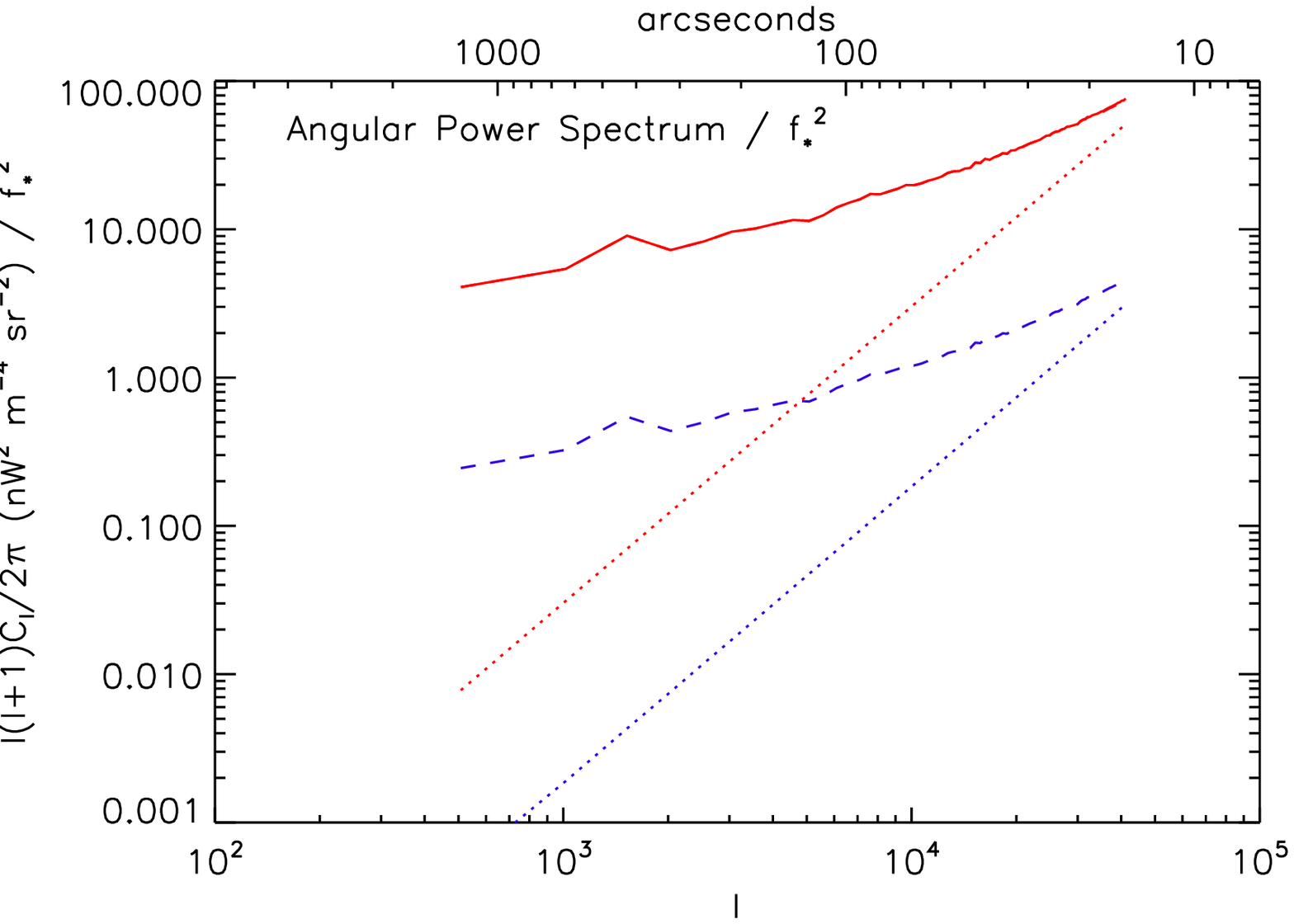}
\caption{%
Angular power spectra of NIRB fluctuations, $C_l$, from halos in comparison to
 the angular power spectrum of the IGM (the bottom line).  We show $C_l$
 from halos that 
 have Population 
II stars with a Salpeter mass spectrum and $f_*=
0.5$ (the angular power spectrum that has the highest
amplitude) and Population III
stars with a Larson mass function and $f_*=0.01$ (the angular
power spectrum with the lowest amplitude and which is closest to the
 angular power spectrum of the IGM).  
The dotted lines show the level of the shot noise.
 In
Figure \ref{fig:funcfstarfesc} we show how the amplitude of the power
spectrum changes between populations with various escape fractions of
 the ionizing 
photons into the IGM, $f_{\rm esc}$, and star formation efficiencies,
$f_*$.  (Right panel) Same as the left panel, except
  divided by $f_*^2$.  The IGM contribution is not shown.
}%
\label{fig:clchangepop}
\end{figure}
%%%%%%%%%%%%%%%%%%%%%%%%%%%%%%%%%%%%%%%%%%%%%%%%%%%%%%%%%%%%%%%%%%%%%%
%%%%%%%%%%%%%%%%%%%%%%%%%%%%%%%%%%%%%%%%%%%%%%%%%%%%%%%%%%%%%%%%%%%%%%
\begin{figure}%[h!tb]
\centering \noindent
\plotone{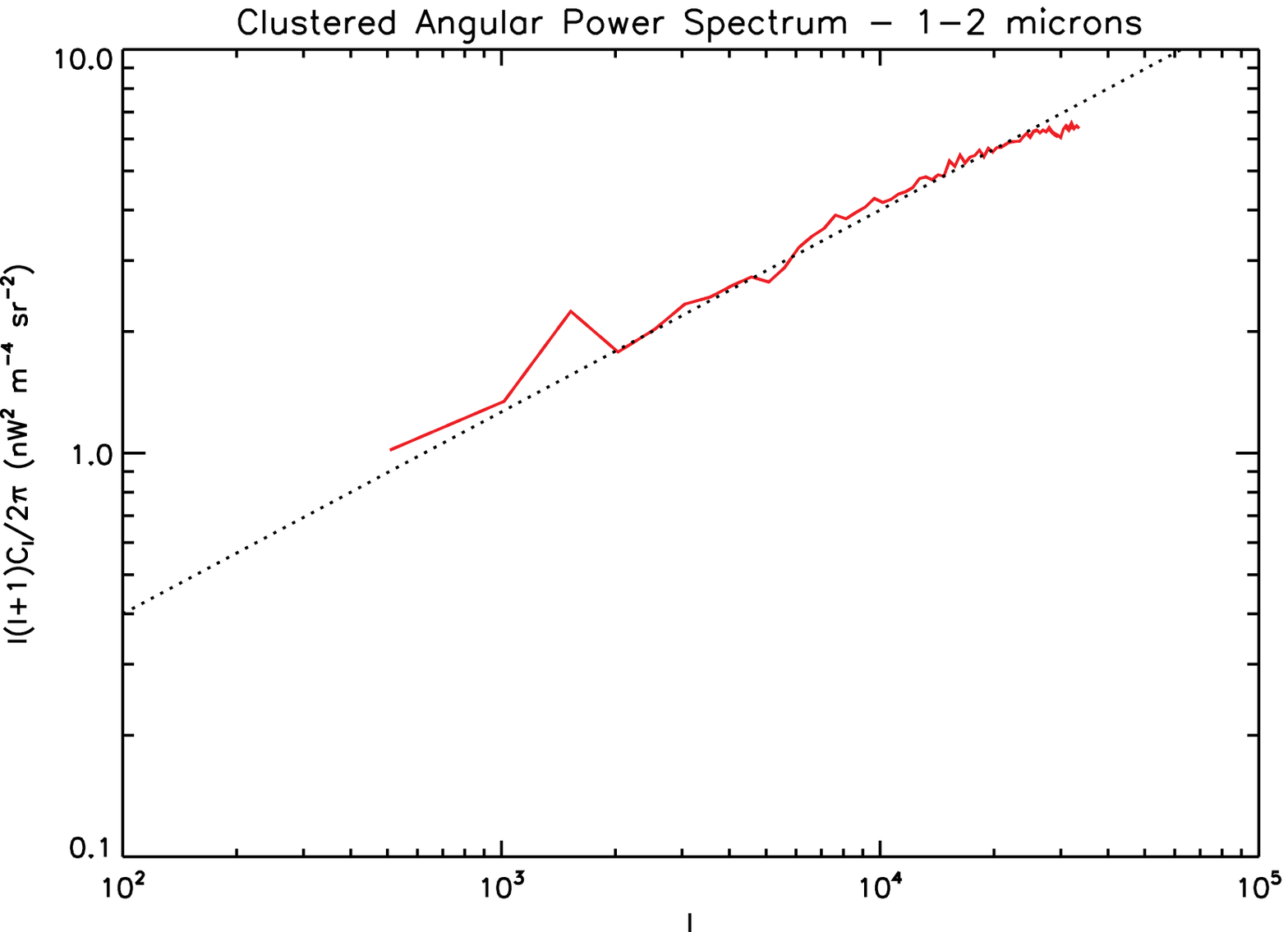}
\caption{%
The  angular power spectrum from the clustering of halos (solid line), 
i.e.,  the  angular power spectrum minus the shot noise contribution.
 The dotted  line has a slope of $l^{0.5}$. 
The clustered angular power spectrum  shows no
evidence of a turnover that was claimed to exist in the literature.
 This is because 
previous analytical models in the literature based their power spectrum
 on the linear 
bias model, which is not valid for this population, which has a high
level of non-linear bias. 
The minimum halo mass used in this calculation
 is $2.2\times 10^9~M_\sun$.  
}%
\label{fig:Clclustered}
\end{figure}
%%%%%%%%%%%%%%%%%%%%%%%%%%%%%%%%%%%%%%%%%%%%%%%%%%%%%%%%%%%%%%%%%%%%%%

In Figure 
\ref{fig:clchangepop} we show $l(l+1)C_l/(2\pi)$ for halos with Population
II stars with a Salpeter mass spectrum and $f_* = 0.5$ (the angular power
spectrum for halos with the highest amplitude) and for Population III stars
with a Larson mass spectrum and $f_* = 0.01$ (the angular power spectrum
for halos with the lowest amplitude), along with the angular
power spectrum of the IGM. 
The halo contribution at small scales, i.e., $l\gtrsim 10^4$, is
comparable to the shot noise contribution. 
When the shot noise is subtracted (see Figure~\ref{fig:Clclustered}), 
we find that 
$l(l+1)C_l/(2\pi)$ is nearly a power-law, $l(l+1)C_l/(2\pi)\propto
l^{0.5}$, {\it with no sign of a turn-over}, which
would be expected from the 
shape of the projected linear matter power spectrum. This is in a stark contrast
with the previous calculations
\citep{kashlinsky/etal:2004,cooray/etal:2004}, which predicted 
a turn-over at $l\sim 10^3$. They 
assumed that the
luminosity-density power spectrum was given by the linear bias model,
in which the halo power spectrum is a constant times the  matter
power spectrum. Our calculations, which are based on a realistic
simulation, indicate that the simple linear bias model is not valid for
these populations. This is expected, as these populations are very
highly biased, and therefore {\it non-linear bias} must also be large,
as demonstrated already in Figure~\ref{fig:bias}.

On the other hand, there is no freedom in changing the amplitude of 
the IGM power spectrum for a given simulation, i.e., a given $f_\gamma/t_{\rm SF}$;
thus, we show only one line for the IGM contribution in
Figure~\ref{fig:clchangepop} (the lowest line).
 For the parameter space explored here, the halo contribution
can be as low as being only slightly over an order of magnitude 
(for PopIII Larson with $f_{\rm esc}=1$ and $f_*=0.01$) 
to about $10^6$ times greater 
(for PopII Salpeter with $f_{\rm esc}=0.19$ and $f_*=0.5$) than 
the IGM contribution.  
If we were to increase $f_{\gamma}$ (which is possible using additional 
simulations in future work, although one has to make sure that the
resulting electron-scattering optical depth is consistent with the WMAP
data), a wider range of parameters $f_{\rm esc}$, $f_*$ 
and $N_i$ could result. This is a good news, as this gives us an
opportunity to study the physics of the reionization using the power
spectrum of NIRB fluctuations. Sensitive surveys may be able
to detect a change in the shape of the power spectra that would be a result
of the IGM power spectrum.  This may be one way of constraining $f_{\rm esc}$
observationally.

\begin{figure}[t]
\centering \noindent
\plottwo{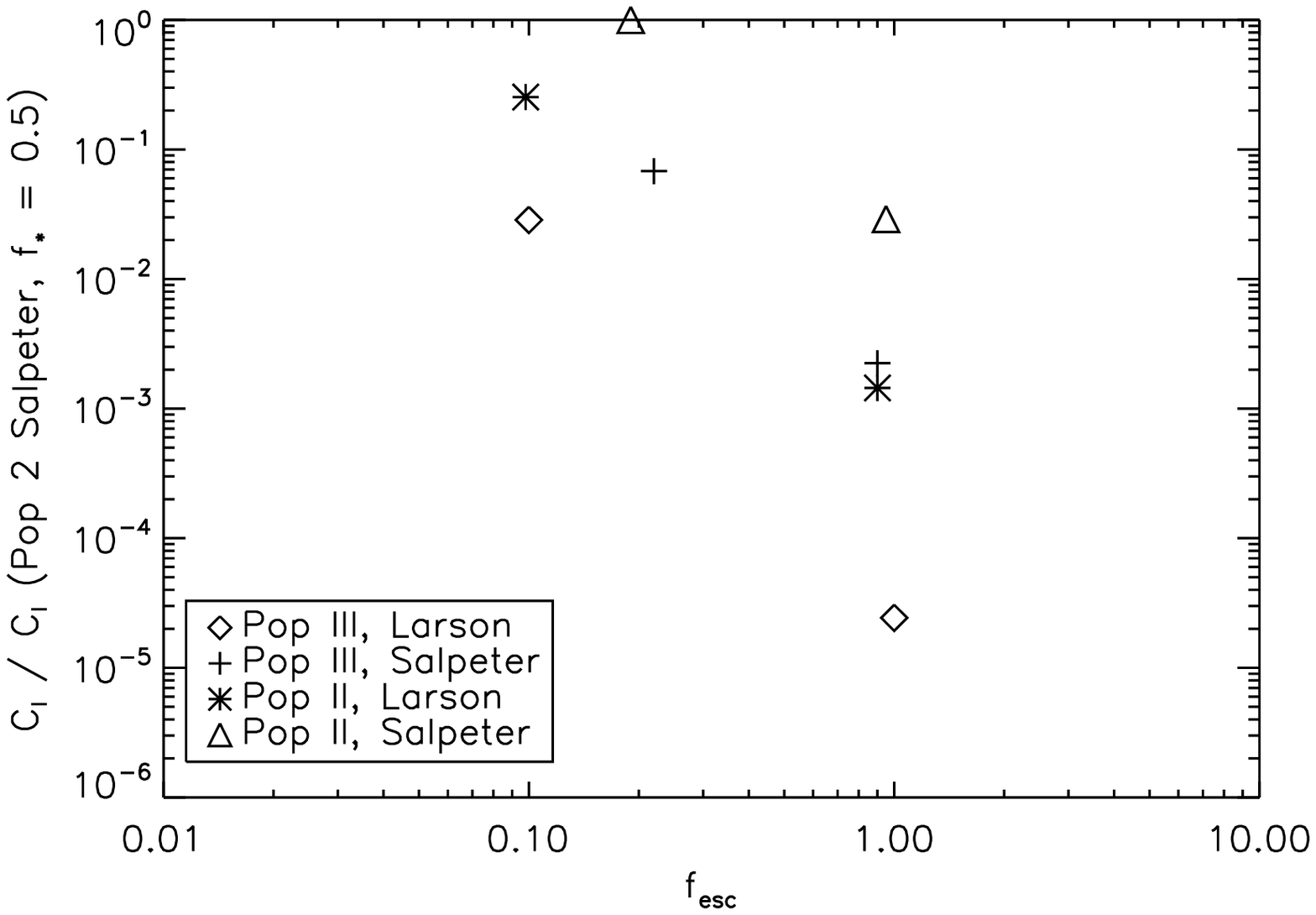}{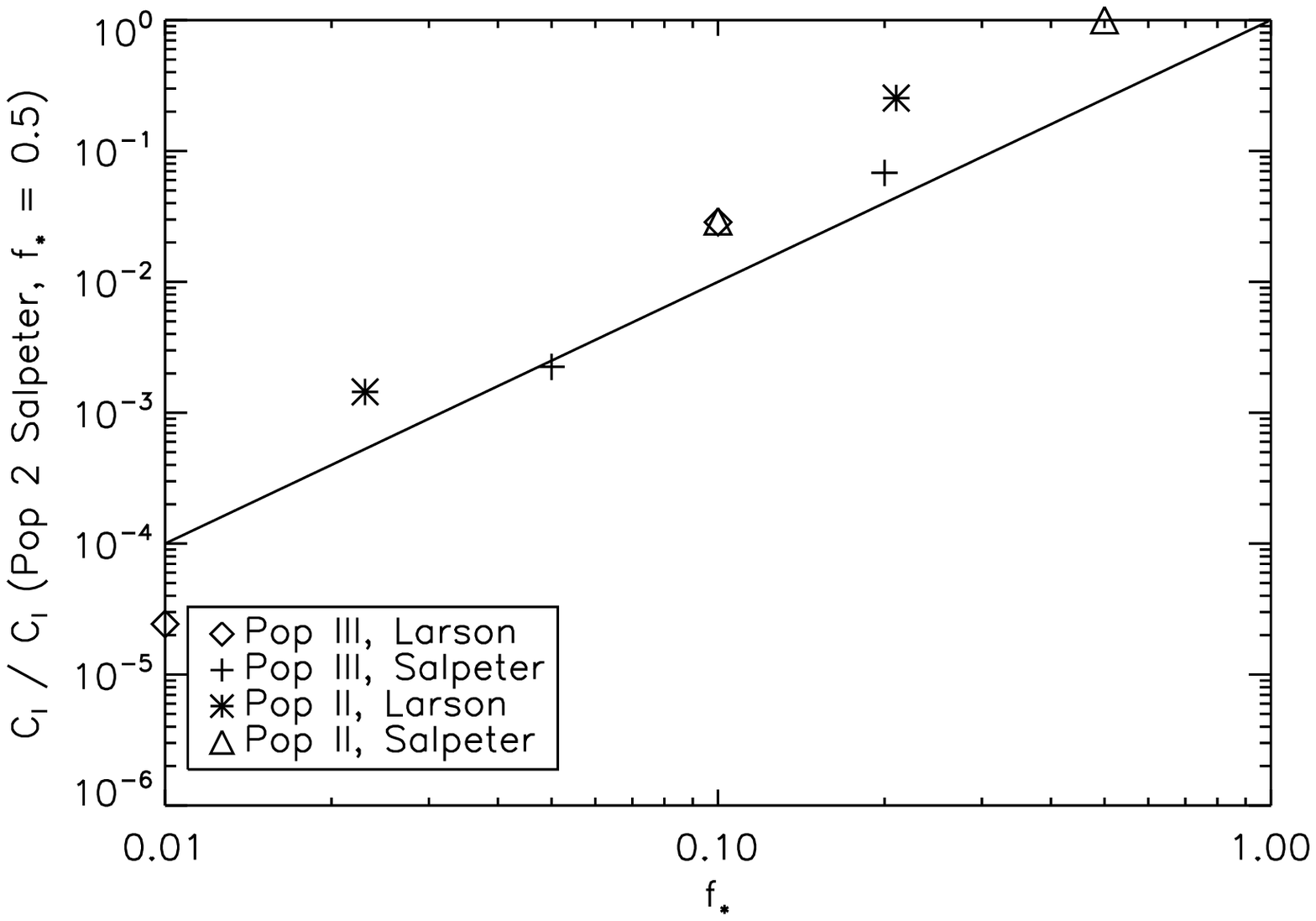}
\caption{%
(Left panel) The change of the angular power spectrum as
 a function of 
the escape fraction, $f_{\rm esc}$, for our selected samples of stellar
 populations. 
Each amplitude is scaled with relation to the angular power
spectrum of Population
II stars with a Salpeter mass spectrum and $f_* =0.5$.  (Since the
shape of the angular power spectra are the same for all stellar populations, this ratio is the same for
all wave numbers.)
(Right panel) The dependence of the angular power spectrum on $f_*$.  
The solid line shows $C_l\propto f_*^2$.
Note that each stellar population has a different set of $f_*$,
 $f_{\rm esc}$, and $N_i$, and thus both panels show a slice of the
 multi-parameter space.
}%
\label{fig:funcfstarfesc}
\end{figure}

In Figure \ref{fig:funcfstarfesc}, we show the amplitude of the angular
power spectra of other
stellar populations scaled to the angular power
spectrum of Population II stars with a Salpeter mass spectrum and $f_* =
0.5$.  As we find in Eq.~(\ref{eq:ltom}), the luminosity-density power spectrum
of halos is about proportional to $f_*^2$, and one of the terms in the power
spectrum (nebular contribution; the second term in Eq.~(\ref{eq:ltom}))
depends on  $(1-f_{\rm esc})$. Therefore, for a fixed
$f_\gamma=f_{\rm esc}f_*N_i$ and fixed $N_i$ (i.e., fixed stellar
population), the angular power spectrum of the halo contribution 
must always increase as we increase $f_*$, as increasing $f_*$ must be
accompanied by the corresponding reduction in $f_{\rm esc}$, both of which
will increase the power spectrum of the halo contribution.

%%%%%%%%%%%%%%%%%%%%%%%%%%%%%%%%%%%%%%%%%%%%%%%%%%%%%%%%%%%%%%%%%%%%%%
%%%%%%%%%%%%%%%%%%%%%%%%%%%%%%%%%%%%%%%%%%%%%%%%%%%%%%%%%%%%%%%%%%%%%%

As $C_l\propto f_*^2$, the parameter combinations that maximize $f_*$ tend
to give the largest $C_l$. For a fixed $f_\gamma=f_{\rm esc}f_*N_i$  this
means a lower $N_i$, i.e., lighter mass spectra with larger
metallicity (see the 5th column of Table~\ref{tab:populations}), and a
lower $f_{\rm esc}$. 
In reality, however, we should also take into account the fact that
heavier mass spectra produce more luminosity per stellar mass, i.e.,
more $\bar{l}$ in Eq.~(\ref{eq:ltom}). These factors explain 
the dependence of the predicted amplitudes of $l(l+1)C_l/(2\pi)$ 
(averaged over $\lambda=1-2~\mu{\rm m}$) on
parameters shown in Figure~\ref{fig:funcfstarfesc}. 
  Populations with higher $f_{\rm esc}$ have lower angular power spectrum.
  This is to be 
expected, because as $f_{\rm esc}$ increases, less photons are available to
create luminosity within the halo.  
%%%%%%%%%%%%%%%%%%%%%%%%%%%%%%%%%%%%%%%%%%%%%%%%%%%%%%%%%%%%%%%%%%%%%%
\section{VARYING THE MODEL: HALO CONTRIBUTION}
\label{sec:varysection}
In this section, we will focus on the halo contribution to the angular
power spectrum of NIRB fluctuations, and explore the effects of changing
various parameters. 
%%%%%%%%%%%%%%%%%%%%%%%%%%%%%%%%%%%%%%%%%%%%%%%%%%%%%%%%%%%%%%%%%%%%%%
\subsection{THE EFFECT OF THE STAR FORMATION TIMESCALE}
\label{sec:tsf}
As mentioned in section \ref{sec:Lhalo}, the star formation timescale will
affect the amplitude of the angular power spectrum.  We have
assumed in this work a constant star formation timescale of $t_{\rm SF}=20$~Myr
to make a consistent comparison between the halo and the IGM
contributions.
However, the amplitude of $C_l$ from halos depends sensitively
on this rather uncertain timescale, as the luminosity of halos is
proportional to $t_{\rm SF}^{-1}$, and thus $C_l\propto 1/t_{\rm SF}^2$. 
Motivated by this, in this section we
consider two other possibilities: 1) The star formation time scale is
shorter than the lifetime of the stars, in which case we will use equation
\ref{eq:lnu2} to compute the luminosity per mass, and 2) the star
formation is triggered by mergers, i.e., 
\begin{equation}
t^{-1}_{\rm SF}(z)=\frac{\int dM_h M_h(d^2n_h/dM_hdt)}{\int dM_h
M_h(dn_h/dM_h)}, 
\label{eq:mergertime}
\end{equation}
where $dn_h/dM_h$ is the mass function of dark matter
halos. 
For the Press-Schechter mass function, we can calculate $t_{\rm
SF}(z)$ analytically from
\begin{equation}
\label{eq:tsfanal}
 t_{\rm SF}^{-1}
=
H(z)\left|\frac{d\ln
     D}{d\ln(1+z)}\right|\left[\frac{\delta_c^2}{D^2(z)\sigma^2(M_{\rm
     min})}-1\right]
\approx 
H(z)\Omega_m^{0.55}(z)\left[\frac{\delta_c^2}{D^2(z)\sigma^2(M_{\rm
     min})}-1\right],
\end{equation}
where $\delta_c=1.68$, 
$D(z)$ is the growth factor of linear matter density fluctuations
normalized such that $D(0)=1$, 
$\sigma(M_{\rm min})$ is the present-day r.m.s. matter density
fluctuation smoothed over a top-hat filter that corresponds to the
minimum mass $M_{\rm min}$, and $\Omega_m(z)$ is the matter density
parameter at a given $z$. Note that interpreting this quantity as a
merger timescale makes sense only when we study the density peaks above
the r.m.s., i.e., $\delta_c/[D(z)\sigma(M_{\rm min})]>1$. (Otherwise
$t_{\rm SF}$ becomes negative.) This formula has a clear physical
interpretation: for a density peak of order the r.m.s. mass density
fluctuation, $\delta_c/[D(z)\sigma(M_{\rm min})]-1\approx 1$, the merger
timescale is of order the Hubble time, i.e., $t_{\rm SF}\approx
H^{-1}(z)$. The higher the peaks are, the shorter the merger timescale
becomes; thus, in this model, high-$z$ objects (for a given mass) have
shorter star formation timescales, and are brighter.

As the reionization history depends on $f_\gamma/t_{\rm SF}$, changing only
$t_{\rm SF}$ without the corresponding change in $f_\gamma$ results in a
different reionization history. For example, increasing $t_{\rm SF}$ by a
factor of 10 makes individual sources fainter by a factor of 10, and
thus it would result in a much slower reionization history.
To compensate this one would have to increase $f_\gamma$ by a factor of
10. Moreover, if we reduce $t_{\rm SF}$ by a large factor, it would make
individual sources brighter by a large factor, to the point where we
might start detecting these sources individually, e.g., as
Lyman-$\alpha$ emitters \citep{FK08}.

In this section, however, we explore the effects of $t_{\rm SF}$
for a given $f_\gamma$, to show how important this quantity is for
predicting the amplitude of NIRB fluctuations without any extra
information on reionization from WMAP or Lyman-$\alpha$ emitters.

\begin{figure}%[h!tb]
\centering \noindent
\plottwo{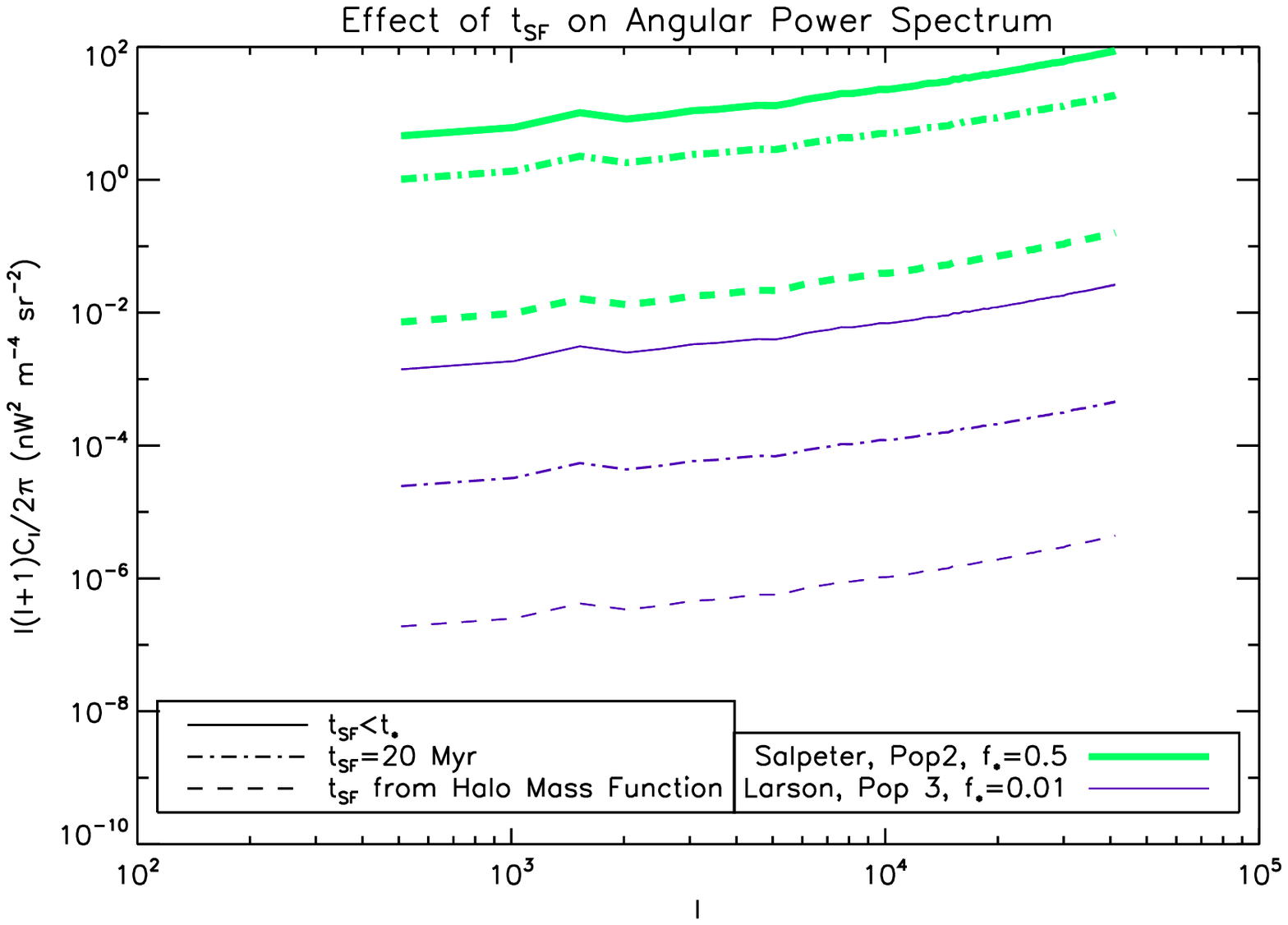}{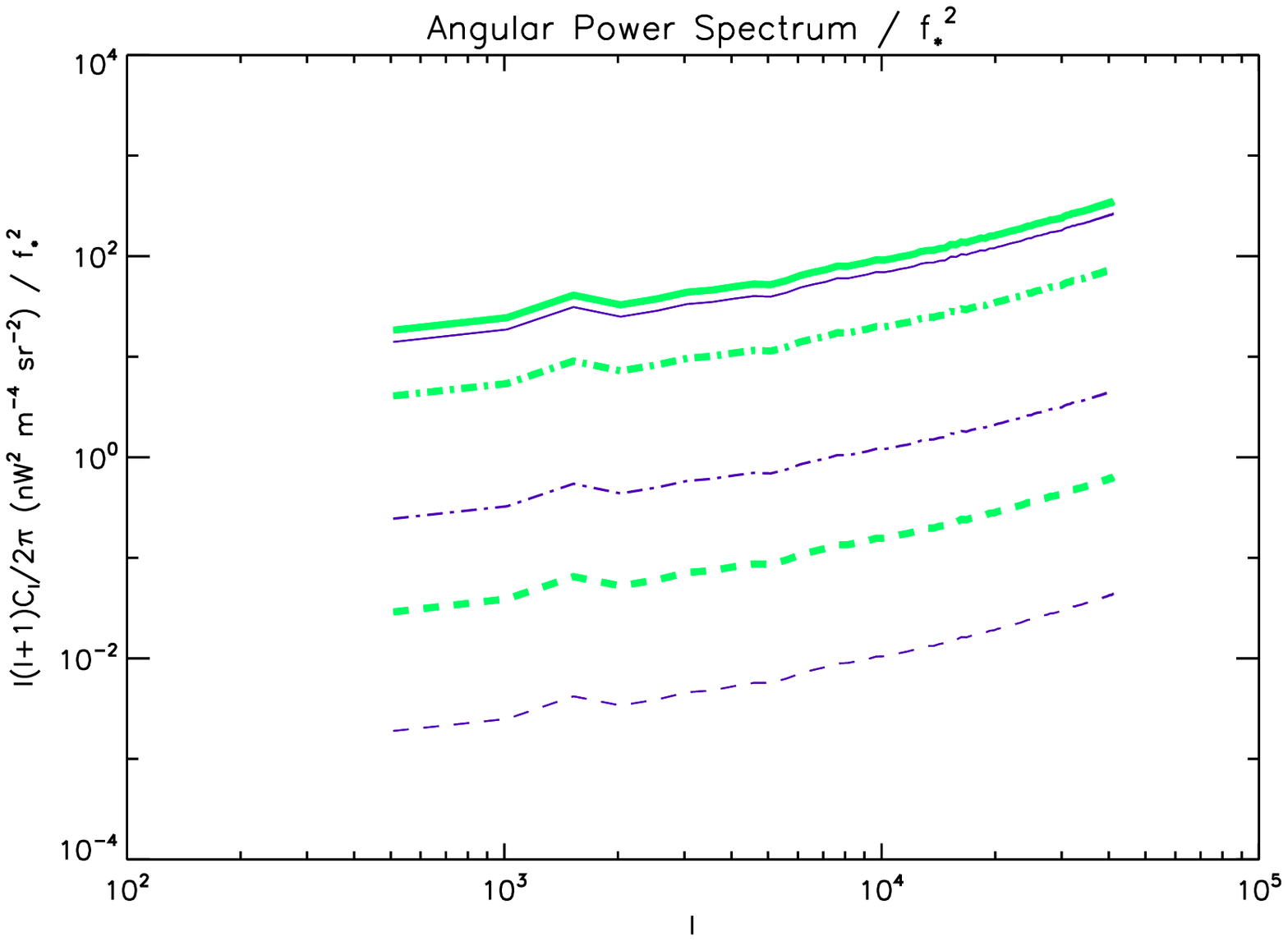}
\caption{%
The effect of the star formation time scale on the angular power
  spectrum.
We find the largest amplitudes when $t_{\rm SF}$ is shorter than the 
main sequence lifetime of stars, whereas we find the lowest amplitudes
  when  $t_{\rm SF}$ is given by the timescale of halo mergers
  (Eq.~\ref{eq:tsfanal}). 
The uncertainty due to
the star formation time scale is large and can lead to an uncertainty in
the angular power spectrum of a factor of $\approx 10^4$. This reflects
  our uncertainty in the mass to light ratio of galaxies
  that contribute to the NIRB.
 However, note that not all scenarios shown here yield the
  reionization 
  histories that are consistent with the WMAP data and the abundance of
  Lyman-$\alpha$ emitters.  (Right panel) Same as the left panel, except
  divided by $f_*^2$.
}%
\label{fig:TSFang}
\end{figure}

The angular power spectrum for various assumptions for the star formation
timescale is given in Figure \ref{fig:TSFang}.  The angular power spectrum
with the highest amplitude corresponds to when
the star formation time scale is shorter than the lifetime of the stars.
If the star formation timescale is given by the merger time of halos (Eq.~\ref{eq:tsfanal}), the
star formation timescale varies with redshift and we obtain the lowest
amplitude for the angular power spectrum, as the merger timescale at a
given redshift is usually comparable to the age of the Universe at the
same redshift.  Our assumption of
$t_{\rm SF}=20$ Myr lies between these two extremes.  

%%%%%%%%%%%%%%%%%%%%%%%%%%%%%%%%%%%%%%%%%%%%%%%%%%%%%%%%%%%%%%%%%%%%%%
%%%%%%%%%%%%%%%%%%%%%%%%%%%%%%%%%%%%%%%%%%%%%%%%%%%%%%%%%%%%%%%%%%%%%%

\begin{figure}%[h!tb]
\centering \noindent
\plottwo{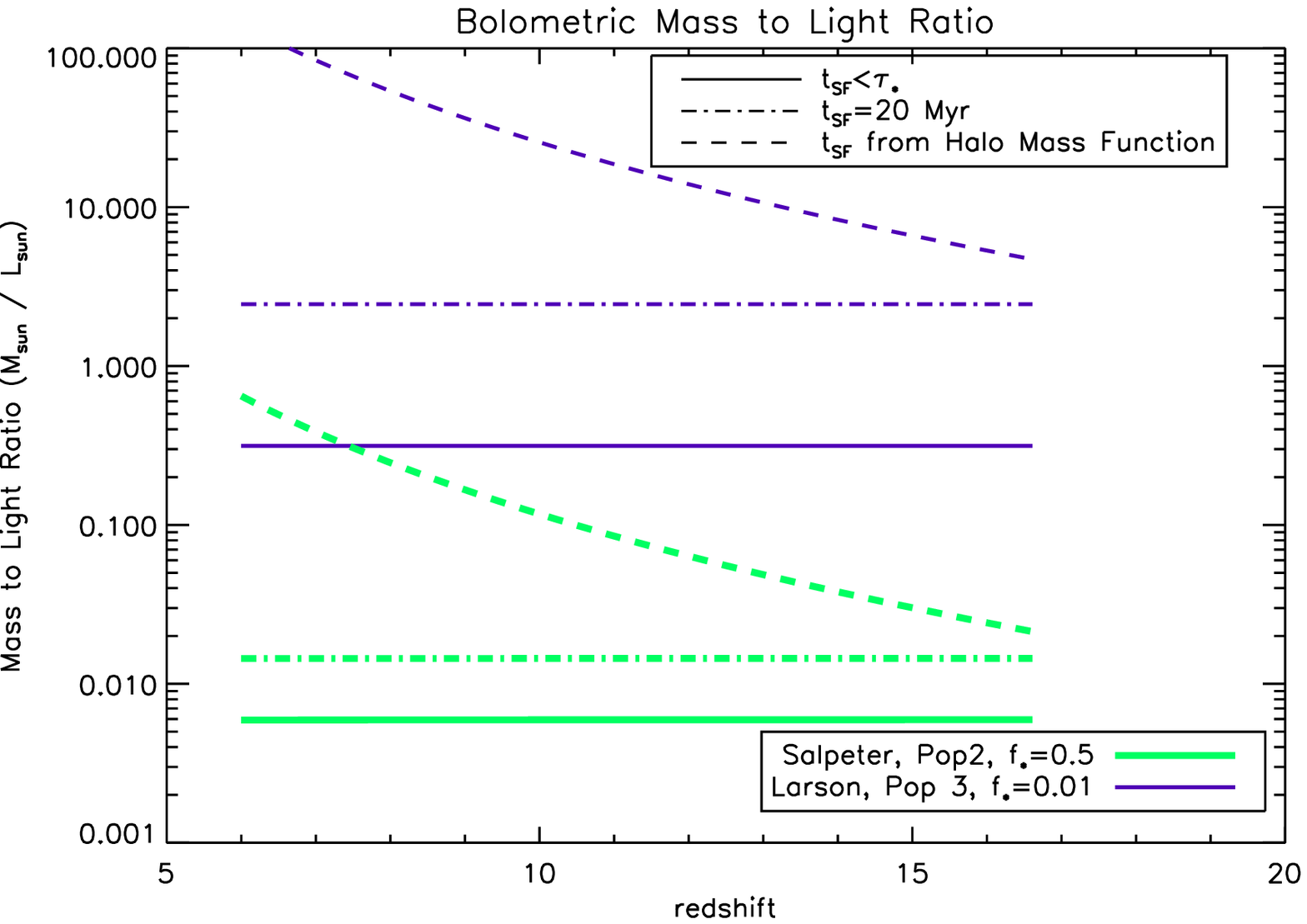}{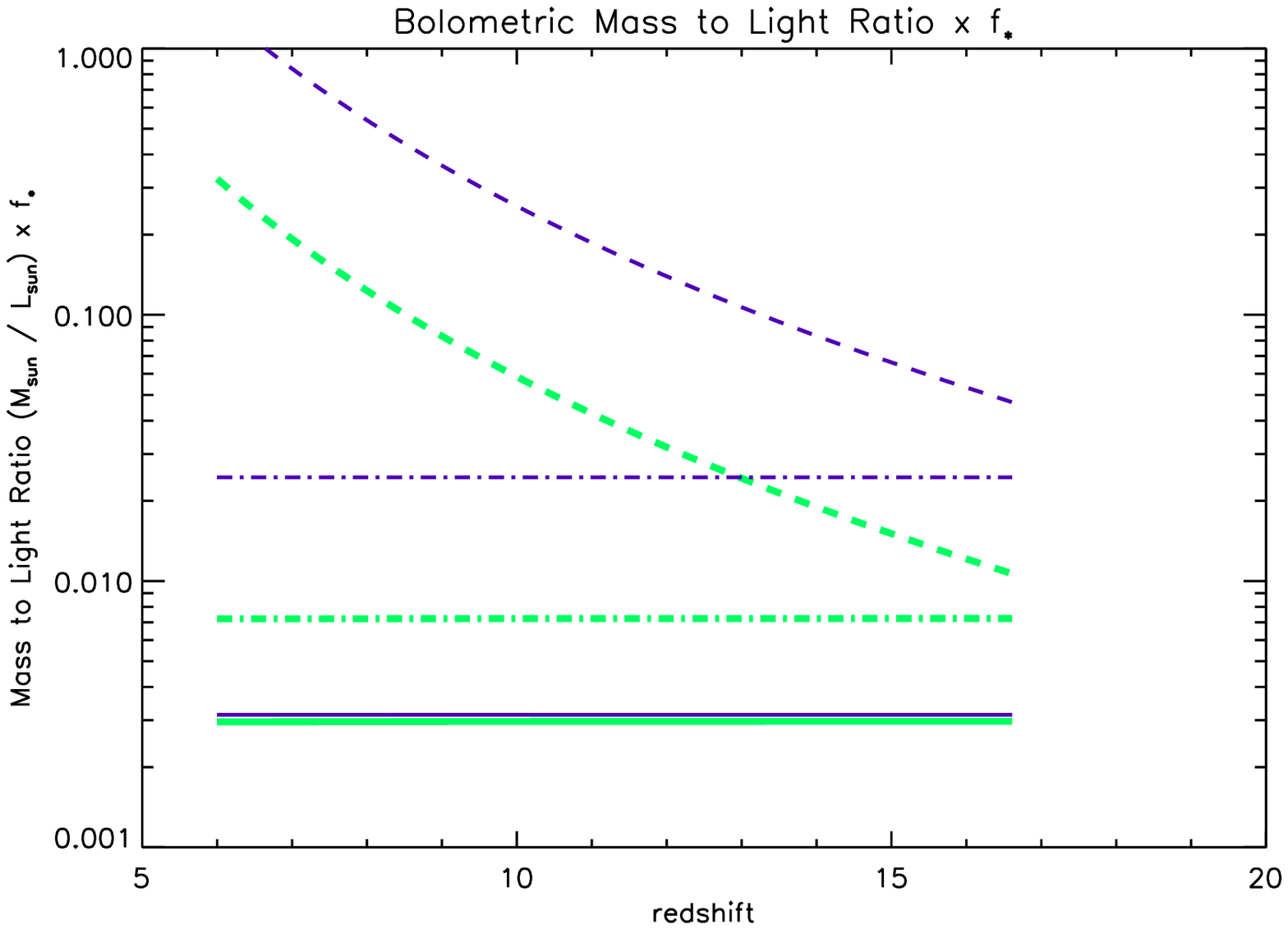}
\caption{%
The bolometric mass-to-light ratio for halos for various star formation
timescales. Uncertainty in the amplitude of the star formation time scale can
be equated to the uncertainty in the mass-to-light ratio, i.e., the
  nature of high-$z$ galaxies contributing to the NIRB.
The upper and lower sets of lines show the PopIII
  Larson and the PopII Salpeter, respectively.   (Right panel) Same as the left panel, except
  multiplied by $f_*$.
}%
\label{fig:MLratio}
\end{figure}

We can further quantify the uncertainty in $C_l$ from
$t_{\rm SF}$ by looking at the mass-to-light ratio of the galaxies
(see
Figure \ref{fig:MLratio}). 
We know very little about the
nature of high-$z$ galaxies contributing to NIRB. We don't know what the mass-to-light ratio is for these
populations.
An uncertainty of a factor of 100 in the star formation
timescale will correspond directly to an uncertainty in the mass-to-light ratio of
100, and an uncertainty of $10^4$ in the angular power spectrum.  Early
galaxies could be starbursts, with a mass-to-light ratio of less than 
0.1 to 1, or normal galaxies, with a mass-to-light ratio of $\gtrsim
10$. The amplitude of $C_l$ is, among other things, a sensitive probe of
the nature of high-$z$ galaxies.

%%%%%%%%%%%%%%%%%%%%%%%%%%%%%%%%%%%%%%%%%%%%%%%%%%%%%%%%%%%%%%%%%%%%%%
%%%%%%%%%%%%%%%%%%%%%%%%%%%%%%%%%%%%%%%%%%%%%%%%%%%%%%%%%%%%%%%%%%%%%%

%%%%%%%%%%%%%%%%%%%%%%%%%%%%%%%%%%%%%%%%%%%%%%%%%%%%%%%%%%%%%%%%%%%%%%
\subsection{THE EFFECT OF CHANGING $z_{\rm end}$}
\begin{figure}%[h!tb]
\centering \noindent
\centering \noindent
\plottwo{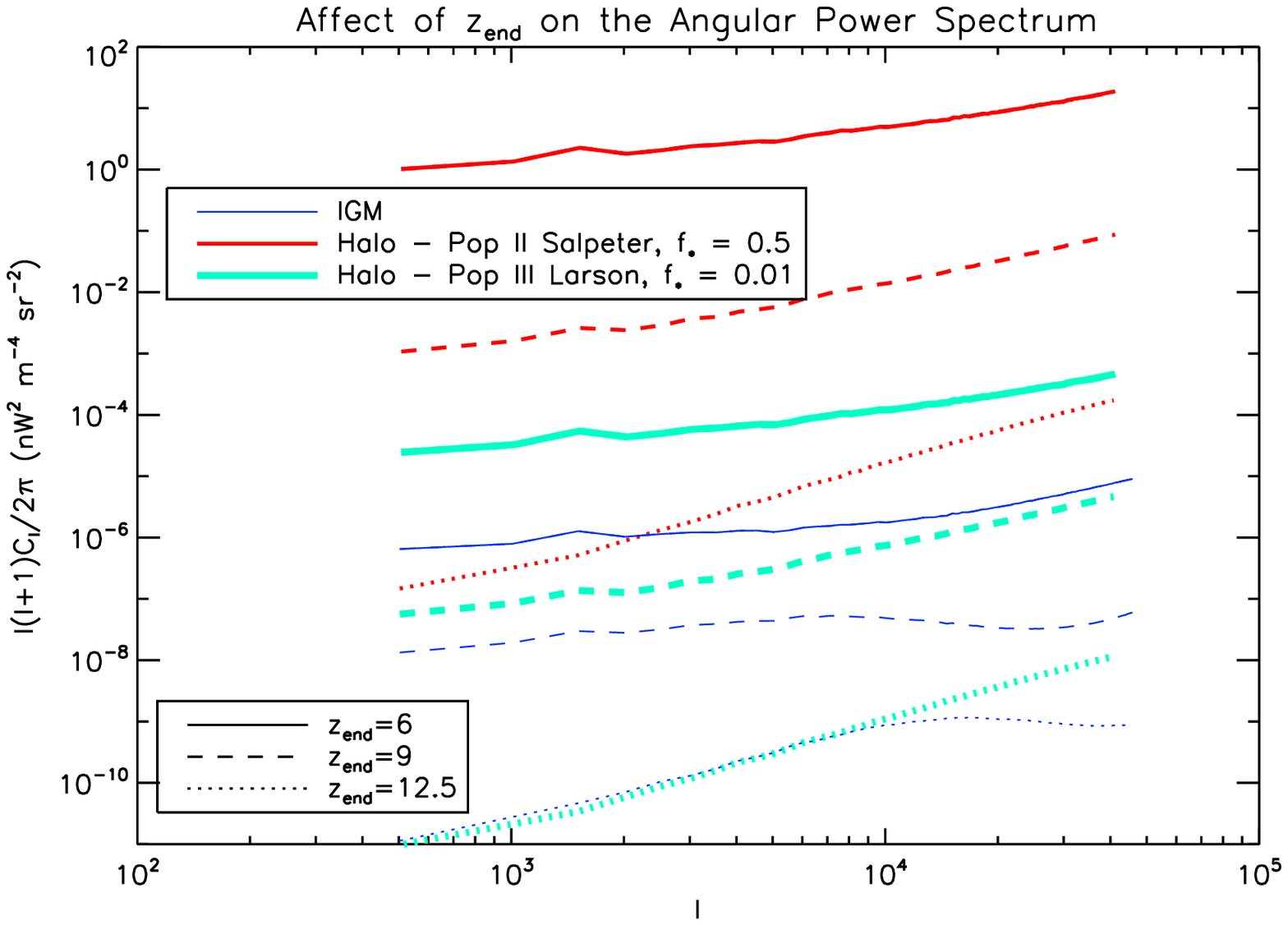}{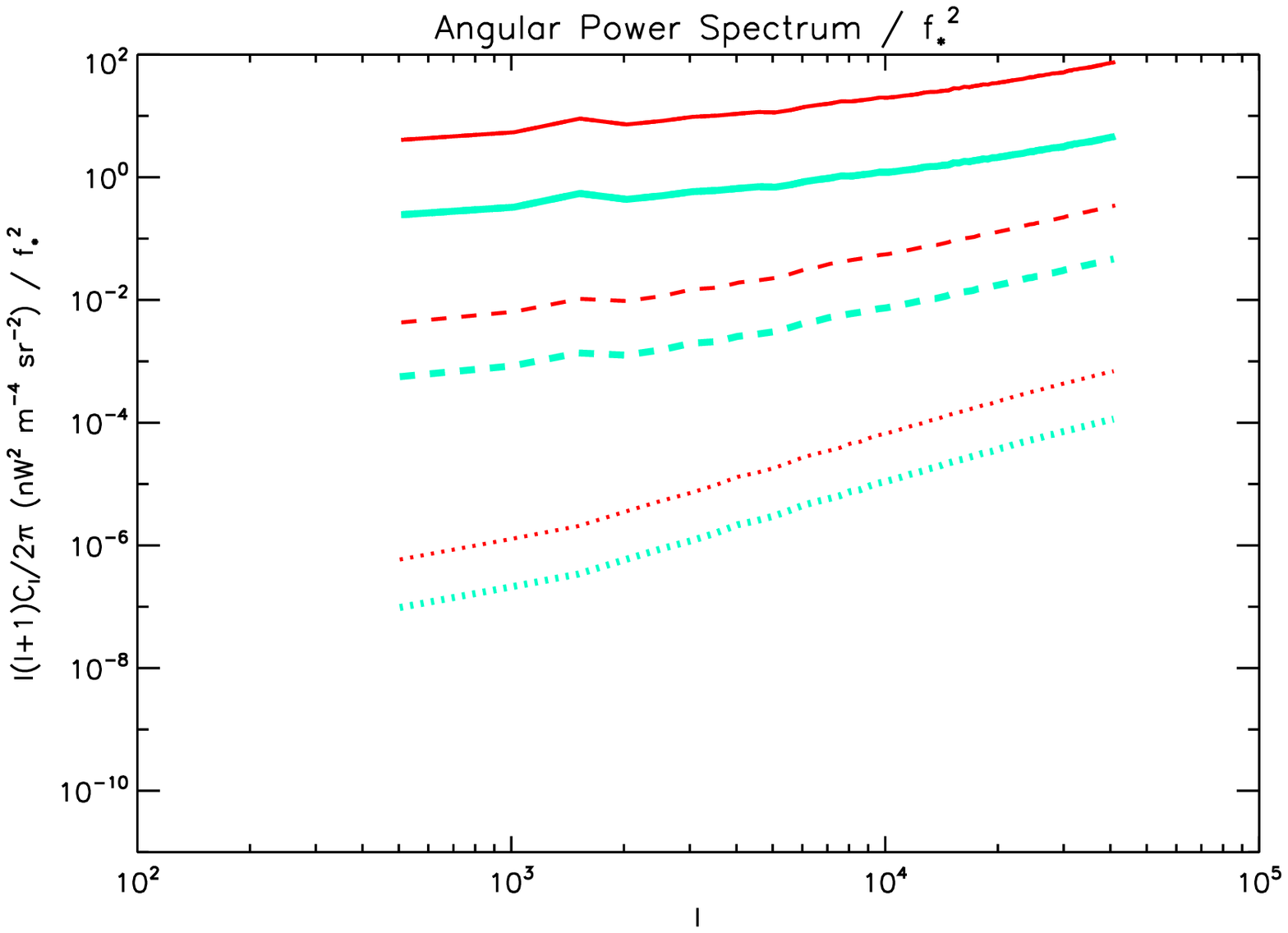}
\caption{%
The angular power spectrum for halos and IGM as $z_{\rm end}$ is
varied.  We show the angular power spectrum for the halos with the highest
and lowest
amplitude of the angular power spectrum (Population
II stars with a Salpeter mass spectrum and $f_* = 0.5$ and Population III stars
with a Larson mass spectrum and $f_* = 0.01$ respectively) and the IGM.  The angular power spectrum as shown throughout the rest of the
paper has $z_{\rm end} = 6$.  As $z_{\rm end}$ increases, the angular power
spectrum drops.  At very high redshifts, the angular power spectrum of the
IGM is higher than some of the angular power spectrum of the halos.  (Right panel) Same as the left panel, except
  divided by $f_*^2$.  The IGM contribution is not shown.
}
\label{fig:clzmin}
\end{figure}

The angular power spectra will also depend on what we choose for the end of
the star formation epoch, $z_{\rm end}$.  The effect of our choice of
$z_{\rm begin}$ is minimal, because at high redshift, the halos are smaller
and dimmer, contributing less to the angular power spectrum.  (See 
Figures \ref{fig:pk1} and \ref{fig:pk2}.) Since halos and IGM will contribute more to fluctuations at lower
redshifts, we find that the angular power spectrum dramatically drop as we stop
star formation at higher redshifts (see Figure \ref{fig:clzmin}). 

%%%%%%%%%%%%%%%%%%%%%%%%%%%%%%%%%%%%%%%%%%%%%%%%%%%%%%%%%%%%%%%%%%%%%% 
%%%%%%%%%%%%%%%%%%%%%%%%%%%%%%%%%%%%%%%%%%%%%%%%%%%%%%%%%%%%%%%%%%%%%%

The shape of the angular power spectrum also changes as we vary $z_{\rm
end}$.  As 
$z_{\rm end}$ increases, the angular power spectrum of the halos steepens.  The shape of the angular
power spectrum from the IGM can also affect the overall slope of the
observed angular power spectrum if the halo contribution is close to that
of the IGM contribution.  If the
escape fraction is small, this effect in the change of shape from the
IGM will be less than if the escape fraction is large.  When $z_{\rm end}$
is very large, the amplitude of the angular power spectrum of the IGM could
even be higher than that of the halos.  

%%%%%%%%%%%%%%%%%%%%%%%%%%%%%%%%%%%%%%%%%%%%%%%%%%%%%%%%%%%%%%%%%%%%%%
\subsection{LYMAN-$\alpha$ ATTENUATION}
The Lyman-$\alpha$ line can be attenuated by dust or neutral hydrogen.  
To understand this effect one would have to perform detailed
calculations of the radiation transport of Lyman-$\alpha$ photons,
including scattering of Lyman-$\alpha$ photons;
however, such calculations are usually quite complex and time-consuming.
Therefore, in this subsection we study the extreme limit of attenuation:
the case where all of the Lyman-$\alpha$
photons are absorbed or extinct.  How would this affect the angular
power spectrum? 
The effect of the {\it complete} Lyman-$\alpha$ attenuation is shown in Table
\ref{tab:lyaatten}.
%%%%%%%%%%%%%%%%%%%%%%%%%%%%%%%%%%%%%%%%%%%%%%%%%%%%%%%%%%%%%%%
\begin{table}[t]
\begin{center}
\begin{tabular}{|l|l|l|l|l|}
\hline
Population & Initial Mass Spectrum & $ f_{\rm esc}$ & $f_*$ & $C_{l,\: Ly\alpha
\:  atten}/C_{l,\: no\: atten}$\\
\hline
Pop III  & Salpeter & 0.22 & 0.2 &  0.848\\
Pop III  & Larson& 0.1 & 0.1& 0.632\\
Pop III  & Salpeter & 0.9 & 0.05& 0.975\\
Pop III  & Larson & 1 & 0.01 & 1\\
Pop II  & Salpeter & 0.95 & 0.1& 0.995\\
Pop II  & Larson  &  0.9 & 0.023& 0.974\\
Pop II  & Salpeter &  0.19 & 0.5& 0.926 \\
Pop II  & Larson  & 0.098 & 0.21& 0.825\\
IGM & & & & 0.448 \\
\hline
\end{tabular}
\caption{%
The effect of Lyman-$\alpha$ attenuation on the angular power spectrum.
Here, we assume complete attenuation (no production of Lyman-$\alpha$
photons).  The angular power spectrum is only slightly affected in most
cases, and is more affected in cases where the Lyman-$\alpha$ line was
strong to begin with (such as heavy Pop III stars).  The effect of
Lyman-$\alpha$ attenuation in the IGM is the highest, as the IGM does
 not have the stellar contribution, and is mainly dominated by the
 Lyman-$\alpha$ and two-photon emission.
}%
\label{tab:lyaatten}
\end{center}
\end{table}
%%%%%%%%%%%%%%%%%%%%%%%%%%%%%%%%%%%%%%%%%%%%%%%%%%%%%%%%%%%%%%%%%%%%%%
The effect of the Lyman-$\alpha$ attenuation is the greatest when the
Lyman-$\alpha$ line is 
the strongest (for heavy Pop III stars) and when the escape fraction is
smaller (so more photons stay within the halo to produce nebular
emission).  The effect of Lyman-$\alpha$ attenuation in the IGM is the
highest, because normally a higher fraction of emission is coming from the
Lyman-$\alpha$ line (in the halos, there is also stellar emission).
%%%%%%%%%%%%%%%%%%%%%%%%%%%%%%%%%%%%%%%%%%%%%%%%%%%%%%%%%%%%%%%%%%%%%%
\section{COMPARISON TO PREVIOUS WORK}
\label{sec:compare}

\begin{figure}%[h!tb]
\centering \noindent
\plotone{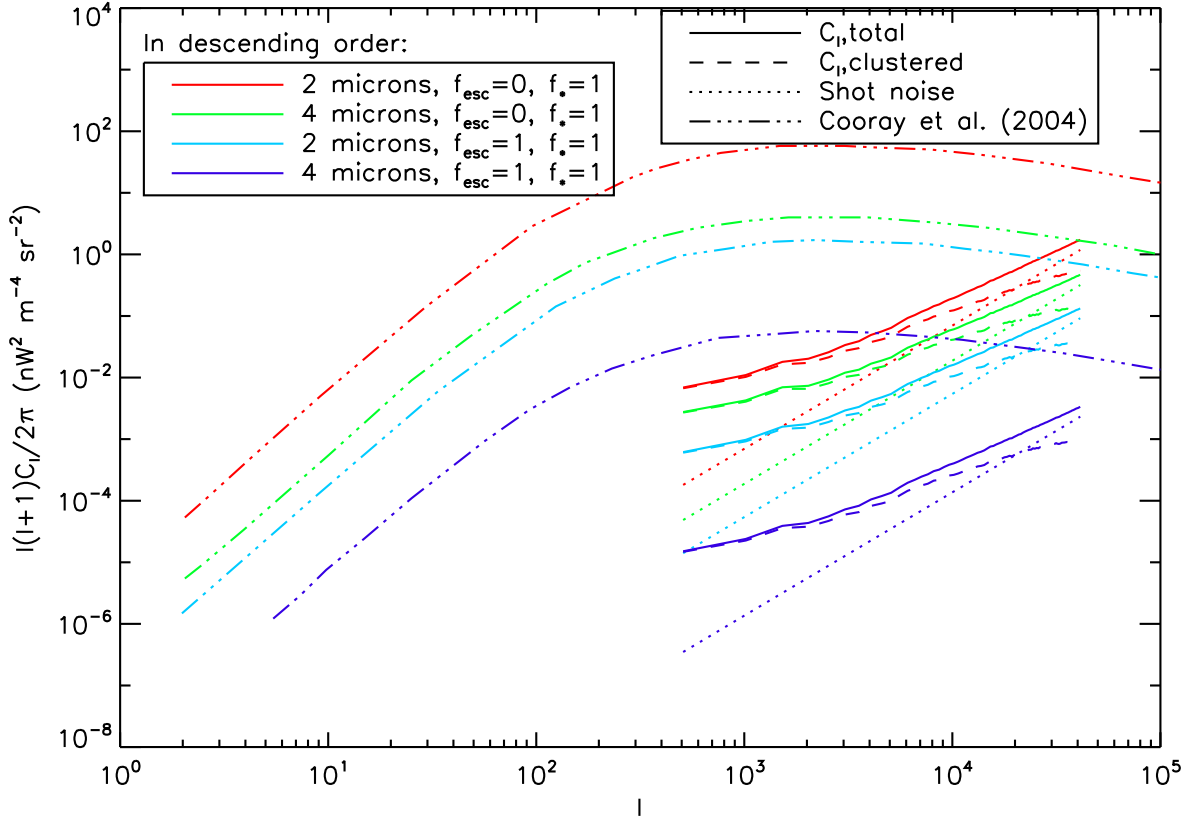}
\caption{%
Comparison to \citet{cooray/etal:2004} (shown as triple-dot dashed lines).
We show $l(l+1)C_l/(2\pi)$ where $C_l=\nu^2C_l^{\nu\nu}$ (see
 Eq.~\ref{eq:clonewave}),  from halos at  $z>10$ that host 
 only very massive  stars with $300~M_\sun$, at various wavelengths. 
The total angular power spectra from this work are shown as solid lines,
shot noise is shown as dotted lines, and the clustered angular power
spectra, which are the total power minus the shot noise components, are
 shown as dashed lines. Note that the amplitudes of $C_l$ shown here are much
 smaller than  those shown in the previous figures (despite a high star
 formation efficiency, $f_*=1$), as we have removed the most dominant,
 lower redshift contributions, $z<10$, in this figure, to be compatible
 with \citet{cooray/etal:2004}. See Figure~\ref{fig:clzmin} for the
 effects of changing the minimum redshift of star formation.  The mean
 intensity, $\nu I_\nu$, for this population of stars at $2 \mu{\rm m}$ and
 $4 \mu{\rm m}$ are $63$ and $16$ ~nW~m$^{-2}$ sr$^{-1}$  respectively,
 which is already ruled out by observations (see section \ref{sec:mean}).
}%
\label{fig:coor}
\end{figure}

\citet{cooray/etal:2004} made fully analytic predictions of the angular
power spectrum in the NIRB luminosity expected from the first stars in
halos. They ignored the IGM contribution, which we found to be small
relative to the halo contribution for
a range of parameters we have explored in this paper. 
They modeled halos with 300 solar mass stars 
for two cases: (1) an optimistic scenario - star formation in halos above 
$10^5$~K, halos forming stars from $z=10-30$, and a star formation efficiency
of 100\%; and  
(2) a pessimistic scenario - star formation 
beginning at 5000~K (so the bias is lower), halos forming stars from $z=15-30$, and a star
formation efficiency of 10\%.   Using the same stellar masses
($300~M_\sun$), we  
have compared our results from the simulation to the  
optimistic case from \citet{cooray/etal:2004} for two different escape
fractions, 0 and 1, and show the results in Figure \ref{fig:coor} for different
wavelengths.  As in \citet{cooray/etal:2004}, we use the star formation time scale given by the merger time
scale (see Eq.~\ref{eq:mergertime}).  The angular power spectrum here is 
\begin{equation}
 C_l^{\nu\nu'}
=
\frac{c}{(4\pi)^2}
\int \frac{dz}{H(z)r^2(z)(1+z)^2}
P_p\left(\nu(1+z),\nu'(1+z); k=\frac{l}{r(z)},z\right),
\label{eq:clonewave}
\end{equation}
which gives  the angular power spectrum at only one wavelength (rather
than that averaged over a certain bandpass).  The difference between this equation and Eq.~(\ref{eq:cl}) is a factor
of $(1+z)^2$ since we are no longer integrating over a range of
frequencies (see Appendix \ref{sec:cl_derivation} for the derivation).
Note that we do not show $C_l$ at $1~\mu{\rm m}$: at $1~\mu{\rm m}$,
the   emission comes from photons that are
  more energetic than $h\nu=13.6$~eV in the rest frame at $z>10$.
  Because of this, there should be no emission from the halos
  themselves, if one considers halos at $z>10$. (There would be
  contributions if one considered halos at lower redshifts, say, $z>6$.)

%%%%%%%%%%%%%%%%%%%%%%%%%%%%%%%%%%%%%%%%%%%%%%%%%%%%%%%%%%%%%%%%%%%%%%
%%%%%%%%%%%%%%%%%%%%%%%%%%%%%%%%%%%%%%%%%%%%%%%%%%%%%%%%%%%%%%%%%%%%%%

Since there were not enough halos in our simulation to create an accurate
power spectrum above $z=16.6$, our population of stars only goes from
$10<z<16.6$, while the model from \citet{cooray/etal:2004} included star
formation from $10<z<30$.  However, this should not make too much
of a difference, because halos at higher redshift do not contribute as much
to the angular power spectrum.  
In Figure \ref{fig:coor} we show the  angular power spectrum minus
the shot noise, which will give us the angular power spectrum of the
clustered component, which is directly comparable to the quantity from
\citet{cooray/etal:2004}.  We have
included all the nebular processes including the free-bound and
two-photon emission, which are 
important to the overall luminosity of the halo and which
\citet{cooray/etal:2004} have neglected.
The overall
amplitude of our angular power spectrum is lower than that which
\citet{cooray/etal:2004} predicted, by a large factor,
$10^3$. \footnote{This difference may be explained by the fact that 
\citet{cooray/etal:2004} actually rescaled the overall amplitude to fit
the mean intensity measured by the Infrared Telescope in Space (IRTS)
\citep{matsumoto/etal:2005} and 
the Diffuse Infrared Background Experiment (DIRBE) 
\citep{kashlinsky/odenwald:2000}. (A. Cooray, private communication.)}

In addition, the angular power
spectrum from \citet{cooray/etal:2004} peaks at about $l\sim1000$ and then
turns over. This is because \citet{cooray/etal:2004} did not take into
account the nonlinear bias 
in the halo power spectrum.  Nonlinear bias will increase the power at small
scales, especially at high redshifts, where galaxies were more highly biased.
We again refer to Figure \ref{fig:bias}, which shows the importance of
non-linear bias.  This greatly affects both the amplitude
and the shape of the angular power spectrum of the NIRB and should be
included.

%%%%%%%%%%%%%%%%%%%%%%%%%%%%%%%%%%%%%%%%%%%%%%%%%%%%%%%%%%%%%%%%%%%%%%
\section{OBSERVING THE FLUCTUATIONS IN THE NEAR INFRARED BACKGROUND}
\label{sec:obs} 
Interpretation of the NIRB data can be a challenging task.  Instrument emission, foregrounds and zodiacal light must all be taken into 
account.  Foreground stars and low-redshift galaxies, in addition to very
faint and the dim wings of galaxies, must be removed.  Much of the
differences in the existing measurements of the fluctuations from stars at
high redshift result from differences in how lower redshift galaxies are
accounted for.  Foreground galaxies are removed down to a limiting
magnitude (which is usually different between different 
observations).  Galaxies fainter than
this are taken into account using different methods.

There have been several observations of the NIRB.
\citet{kashlinsky/odenwald:2000} found fluctuations at the
wavelengths from ${\rm 1.25}$ to ${\rm 4.9} \: \mu
{\rm m}$ in the
images taken by the Diffuse Infrared Background Experiment (DIRBE) on
Cosmic Background Explorer (COBE), which were not consistent with the Galactic emission or instrument noise.
\citet{matsumoto/etal:2005} observed the NIRB using the Infrared Telescope in Space (IRTS).  They detected 
a clustering excess on scales of about $100'$ from 1.4 to 4 $\mu{\rm m}$, and an indication of a spectral jump from the high redshift Lyman 
cutoff.   This jump could indicate that Population III star formation
ended at about a redshift of $z\sim 9$.  Excess fluctuations were
detected, possibly from high redshift galaxies, at about 1/4 of the mean intensity. \citet{kashlinskyb/etal:2007b,kashlinsky/etal:2005} made observations of
the fluctuations of the NIRB using the Infrared Array Camera (IRAC) on the
Spitzer Space Telescope at 3.6, 4.5, 5.8 and 8 
$\mu{\rm m}$.  Sources were removed by clipping pixels
containing $\gtrsim 4\sigma$ peaks, as well as removing fainter sources
identified by SExtractor and convolved with the appropriate point spread
function of IRAC.  Since zodiacal light is not fixed in celestial
coordinates, it was removed by taking observations
six months apart in fields rotated by 180$^\circ$. They detected excess fluctuations ($0.1$~nW~m$^{-2}$ sr$^{-1}$ at 3.6 $\mu{\rm m}$) that were not consistent with instrument noise, dim wings of 
galaxies, zodiacal light, or galactic cirrus.  
They claim that it is possible that the excess fluctuations came from high redshift galaxies 
($z> 6.5$) or faint, low redshift galaxies.
However, since these fluctuations show little ($<10^{-3}$)
correlation 
  with the ACS source catalog maps, and the power spectrum of fluctuations is inconsistent with
the Hubble Space Telescope Advanced Camera for Surveys (ACS) catalog galaxies, they state it is unlikely that these fluctuations 
are from faint, low-$z$ galaxies \citep{kash/etal:2007}.  However, \citet{thompson/etal:2007b} claim that 
the color of the fluctuations detected by \citet{kashlinskyb/etal:2007b,kashlinsky/etal:2005} are
consistent with objects at $z<10$, and therefore not from a population
of high redshift stars.  

%%%%%%%%%%%%%%%%%%%%%%%%%%%%%%%%%%%%%%%%%%%%%%%%%%%%%%%%%%%%%%%%%%%%%%
\begin{figure}%[h!tb]
\centering \noindent
\plotone{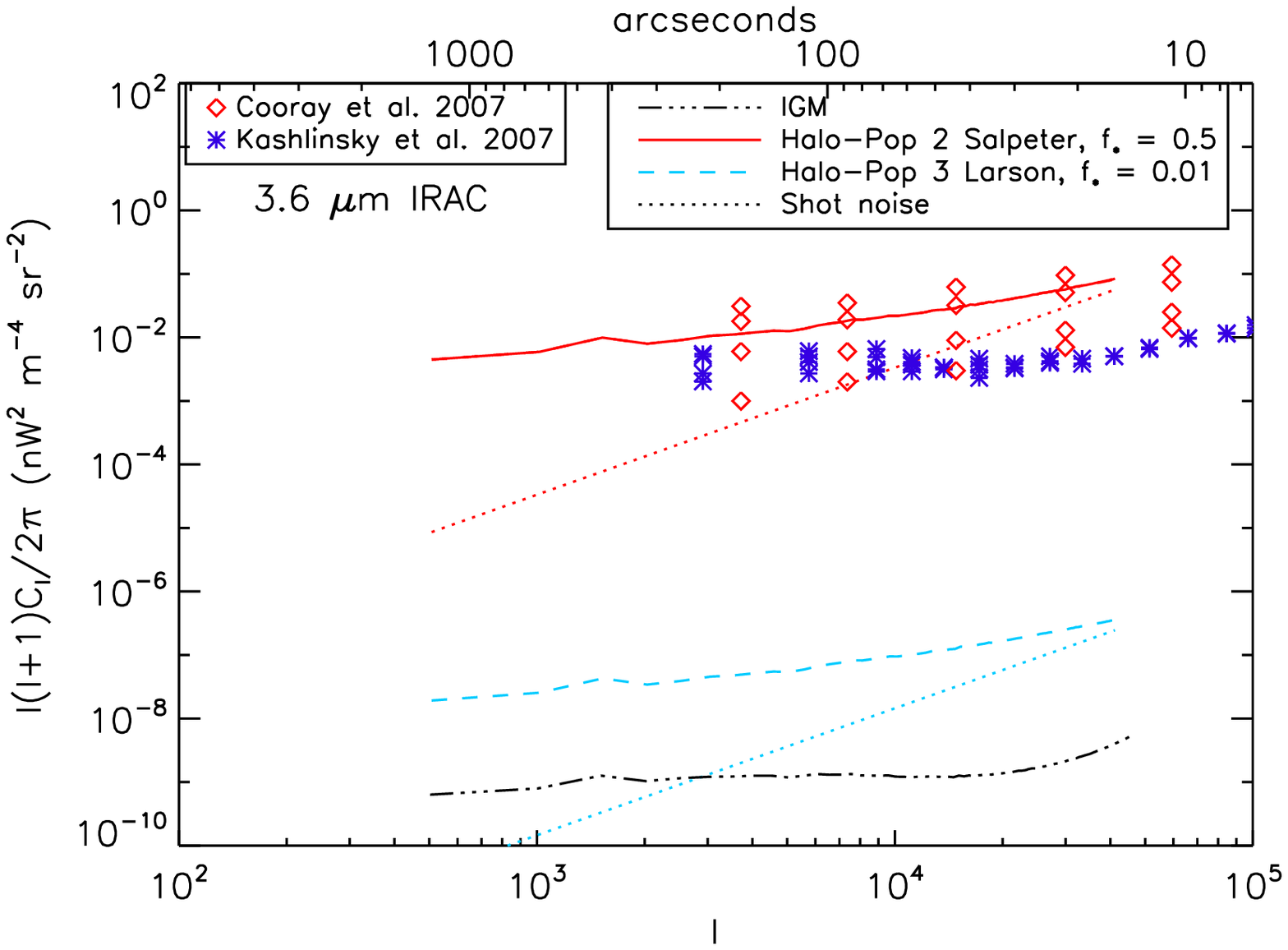}
\caption{%
Our models for the angular power spectra at $3.6$ $\mu{\rm m}$ (halo+IGM) are compared with observations
from \citet{kashlinskyb/etal:2007b} (their Figure 1, lower panel, shown as the
blue asterisks) and from \citet{cooray/etal:2007} (their Figure 1, images A,
B, C, and D, with
varying foreground galaxy cuts) shown as red diamonds. Most of our models
lie beneath current observations.  The mean intensity produced
  by Pop II stars with a Salpeter initial mass spectrum and $f_*=0.5$ is
  $\nu I_\nu = 15.1$ ~nW~m$^{-2}$ sr$^{-1}$, which is over current observations.  For Pop III
  stars with a Larson initial mass spectrum and $f_* = 0.01$, $\nu I_\nu =
  0.182$ ~nW~m$^{-2}$ sr$^{-1}$, which is allowed by observations (for more on the mean intensity,
  see section \ref{sec:mean}).
}%
\label{fig:3.6obs}
\end{figure}
%%%%%%%%%%%%%%%%%%%%%%%%%%%%%%%%%%%%%%%%%%%%%%%%%%%%%%%%%%%%%%%%%%%%%%
\begin{figure}%[h!tb]
\centering \noindent
\plotone{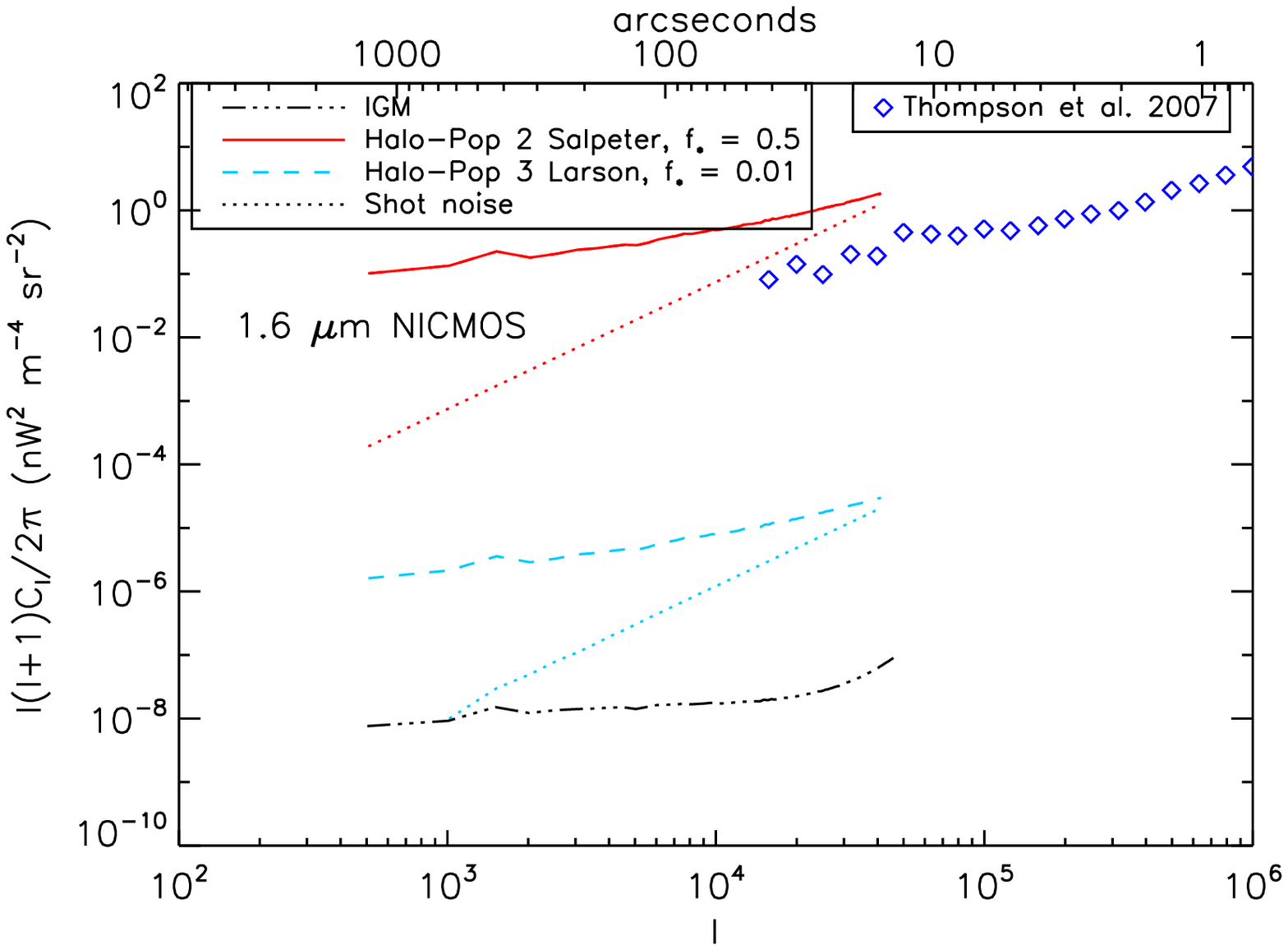}
\caption{%
Our models for the angular power spectra (halo+IGM) are compared with observations
from \citet{thompson/etal:2007a} (for all sources deleted) at $1.6$ $\mu{\rm m}$, which are shown by the
blue diamonds.  Again, most of our models lie beneath current
observations.  As in the case at $3.6  \mu{\rm m}$, the mean
  intensity from Pop II stars with a Salpeter initial mass spectrum and
  $f_*=0.5$ is high at $\nu I_\nu = 60.1$ ~nW~m$^{-2}$ sr$^{-1}$, while the mean intensity for Pop III
  stars with a Larson initial mass spectrum and $f_* = 0.01$ is $\nu I_\nu
  =0.802$ ~nW~m$^{-2}$ sr$^{-1}$.
}%
\label{fig:1.6obs}
\end{figure}
%%%%%%%%%%%%%%%%%%%%%%%%%%%%%%%%%%%%%%%%%%%%%%%%%%%%%%%%%%%%%%%%%%%%%%

\citet{cooray/etal:2007} observed the NIRB using IRAC at 3.6 $\mu{\rm m}$.  They masked the image to cut out faint, low redshift galaxies.  In
their most extensive masked image, they masked IRAC sources down to a
magnitude of 20.2 in addition to galaxies 
in ACS catalog.
They also discarded pixels that had a flux
$4\sigma$ above the mean.  

\citet{thompson/etal:2007a} also made observations of the fluctuations of
the NIRB using the Near Infrared Camera and Multi-Object Spectrometer
(NICMOS) camera on the Hubble Space Telescope at 1.1 and 1.6 $\mu{\rm m}$.  
The effects of zodiacal light were removed by dithering the camera.  After
removing galaxies down to the fainter ACS and NICMOS detection limit,
fluctuation power dropped two orders of magnitude in comparison to an
earlier paper by \citet{kashlinsky/etal:2002}.  Therefore,
\citet{thompson/etal:2007a} confirmed that the observed fluctuations reported by
\citet{kashlinsky/etal:2002}  in the 2MASS data are from low
  redshift galaxies ($z< 8$) (although  they are
unable to rule out contributions from galaxies in $8<z<13$).  
Yet, they concluded that an excess
fluctuation power in the NIRB of about $1-2$~nW~m$^{-2}$~sr$^{-1}$
could still be from the first stars. Their methodology would miss
fluctuations that are flat on scales above $100''$ or clumped on scales of a
few arc minutes. 

Our models are compared to the observations at $3.6$ $\mu{\rm m}$ by
\citet{kashlinskyb/etal:2007b,kashlinsky/etal:2005} and \citet{cooray/etal:2007} in Figures \ref{fig:3.6obs} and to  observations at $1.6$ $\mu{\rm m}$ from \citet{thompson/etal:2007a} in
Figure \ref{fig:1.6obs}.  For these observations, it is safe to treat them
as ``upper limits,'' as additional foreground contamination might still
exist.  At $3.6$ $\mu{\rm m}$, most of our predictions for the angular power
spectra are below the current observations, and are therefore still
viable candidates.  At $1.6$ $\mu{\rm m}$ we see similar results.  Therefore, it seems likely that early stars contribute at
very low levels to the fluctuations in NIRB.
Of course, other factors, such as the star formation time scale and the
minimum redshift that star formation occurs at, $z_{\rm end}$, can also 
affect which models can agree with observations.  

\begin{figure}[t]
\centering \noindent
\plottwo{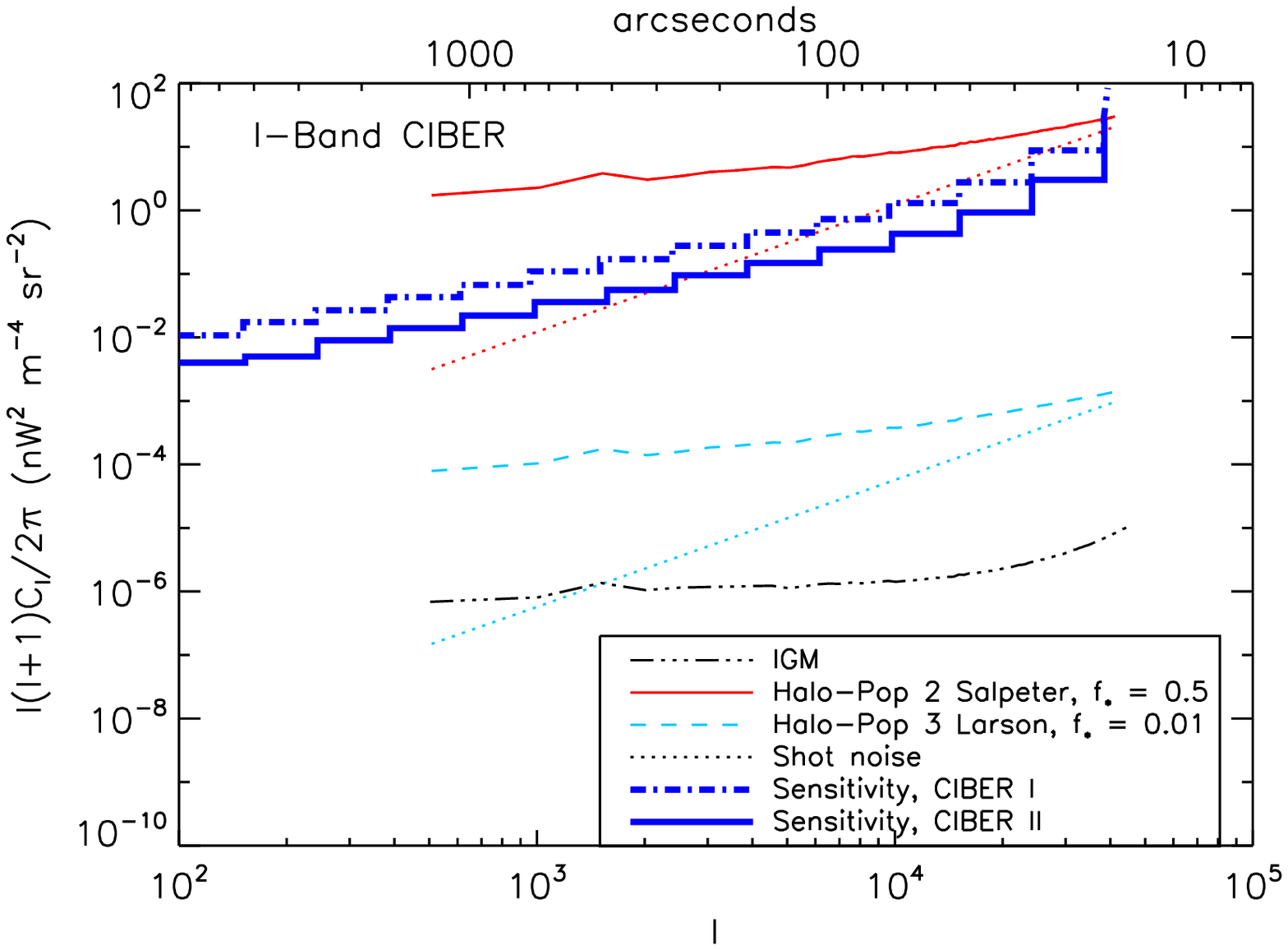}{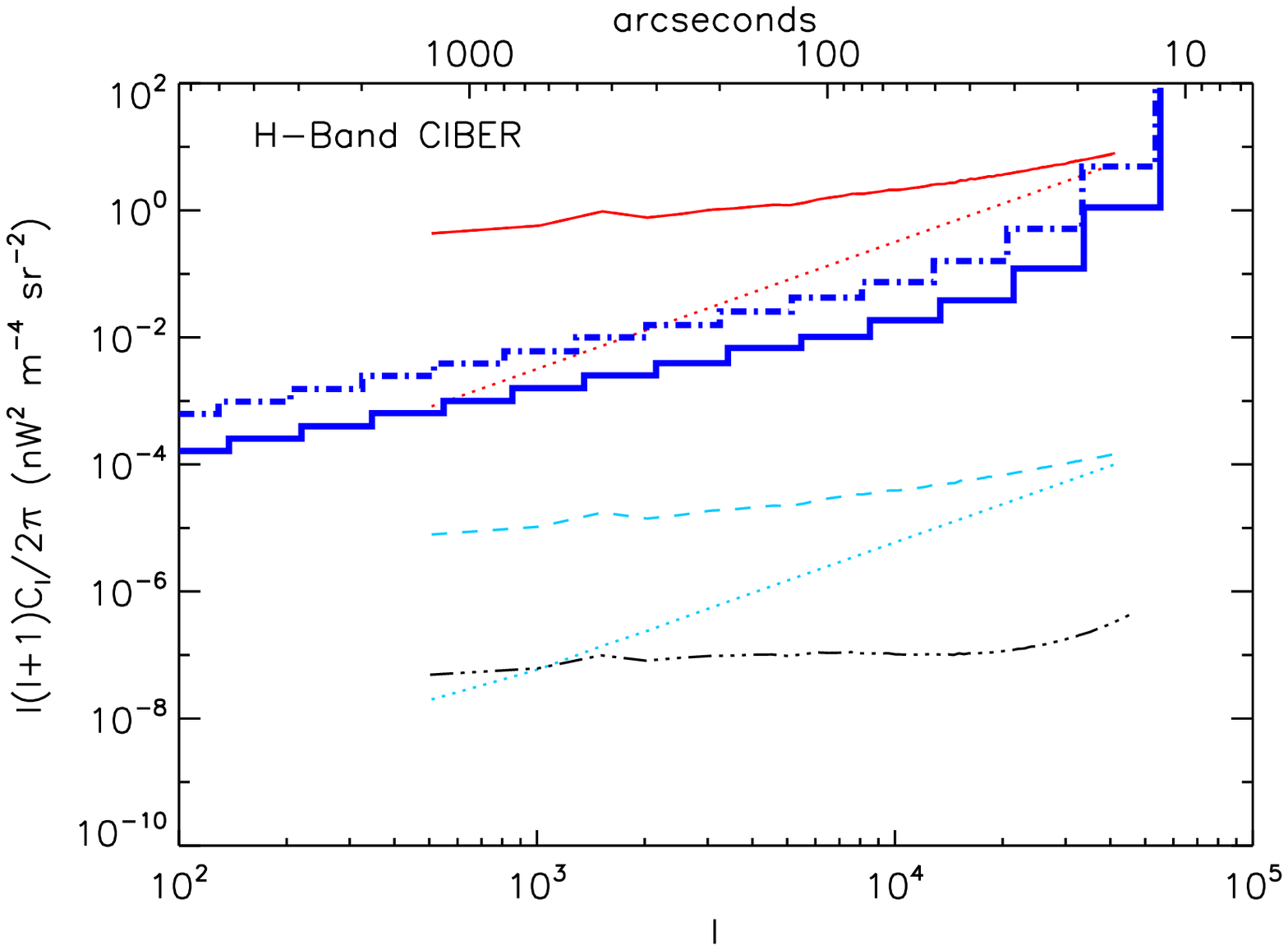}
\caption{%
Our models of the angular power spectrum (halos and the IGM) compared with
the sensitivities of upcoming CIBER missions (shown as the stepped blue
lines) from \citet{cooray/etal:2009}.  CIBER will increase sensitivity of measured fluctuations, but
still many of our models will lie beneath the detection limit.
}%
\label{fig:CIBER}
\end{figure}

Missions currently underway and future, more detailed experiments can make
better observations of the NIRB.  AKARI (previously known as ASTRO-F)
observed in 13 bands from 2-160 $\mu{\rm m}$ \citep{matsuhara/etal:2008}.  The
Cosmic Infrared Background Experiment (CIBER) will be able to obtain the
power spectrum from $7''$ to 2 degrees.  Combined with AKARI and Spitzer,
fluctuations 100 times fainter than IRTS/DIRBE will be able to be observed.
CIBER has two dual wide field imagers at 0.9 and 1.6 $\mu{\rm m}$.  An improved
CIBER II will also measure fluctuations in four bands from 0.5 to 2.1
$\mu{\rm m}$.  This
experiment will be pivotal to determine if the fluctuations observed are
from the first galaxies or have a more local origin
\citep{bock/etal:2006, cooray/etal:2009}.  Predictions for the
sensitivity of CIBER I  and II are shown in Figure
\ref{fig:CIBER} for both 0.9 and 1.6 $\mu{\rm m}$ (I and H-Band
respectively) \citep{cooray/etal:2009, bock/etal:2006}.  The sensitivity of CIBER
will be much better than any of the current observations, but still many
of our models lie beneath detection limits.

%%%%%%%%%%%%%%%%%%%%%%%%%%%%%%%%%%%%%%%%%%%%%%%%%%%%%%%%%%%%%%%%%%%%%%
%%%%%%%%%%%%%%%%%%%%%%%%%%%%%%%%%%%%%%%%%%%%%%%%%%%%%%%%%%%%%%%%%%%%%%

%%%%%%%%%%%%%%%%%%%%%%%%%%%%%%%%%%%%%%%%%%%%%%%%%%%%%%%%%%%%%%%%%%%%%%
\section{ADDITIONAL CONSTRAINTS FROM THE MEAN INTENSITY OF THE NEAR
  INFRARED BACKGROUND}
\label{sec:mean}

In addition to fluctuations, measurements have been taken of the mean
intensity of the NIRB.  Because these measurements rely
on an accurate subtraction of the zodiacal light, measurements of the mean
intensity of the NIRB are more difficult to perform.  Currently, the
interpretation of these
measurements is still highly controversial.  
Measurements of the excess in the NIRB (NIRBE) started out high ($70$~nW~m$^{-2}$~sr$^{-1}$) \citep{matsumoto/etal:2005} and have since declined.  The
most recent measurements are lower.  \citet{kashlinsky/etal:2007} report
that the mean intensity of the NIRBE must be greater than
1~nW~m$^{-2}$~sr$^{-1}$ to be consistent with fluctuations at 3.6 and 4.5 $\mu{\rm m}$.  \citet{thompson/etal:2007a}
report a residual NIRBE of $0.0^{+3}_{-0.3}$ at 1.1 and 1.6 $\mu{\rm m}$.
Fluctuations measured by \citet{cooray/etal:2007} imply that the mean NIRB
cannot be much more than 0.5~nW~m$^{-2}$~sr$^{-1}$ at 3.6~$\mu{\rm m}$.  Using
these limits, can we put additional constrains on the first stars?

We calculate the mean intensity of NIRB from \citep{peacock:1999}
%%%%%%%%%%%%%%%%%%%%%%%%%%%%%%%%%%%%%%%%%%%%%%%%%%%%%%%%%
\begin{equation}
 I_{\nu} =
 \frac{c}{4\pi} 
 \int 
 \frac{dz\, p([1+z]\nu, z)}{H(z) (1+z)},
\label{eq-inu-generic}
\end{equation}
%%%%%%%%%%%%%%%%%%%%%%%%%%%%%%%%%%%%%%%%%%%%%%%%%%%%%%%%%%
where $\nu$ is the observed frequency and $p(\nu, z)$ is 
given by Eq.~(\ref{eq:pnu}). 
The star formation rate contained in
$p(\nu, z)$,
is given by $\dot{\rho}_*(z)=\rho_*(z)/t_{\rm SF}(z)$, where
\begin{equation}
\rho_*(z) 
  = f_* \frac{\Omega_b}{\Omega_m} \bar{\rho}_M^{\rm halo}(z),
\end{equation}
where $\bar{\rho}_M^{\rm halo}(z)$ is the mean mass density collapsed into
halos taken from the simulation (which has the minimum halo mass of
$2.2\times 10^9~M_\sun$), and is shown in the right panel of
Figure~\ref{fig:meanlumdensity}.
For the star formation timescale, we use $t_{\rm SF}=20~{\rm Myr}$, so
that we can calculate the mean NIRB for models that are compatible with
the WMAP data. 
The star formation rates with various star formation
efficiencies are given 
in Table \ref{tab:vIvPop}.

%%%%%%%%%%%%%%%%%%%%%%%%%%%%%%%%%%%%%%%%%%%%%%%%%%%%%%%%%%%%%%%
\begin{table}[t]
\begin{center}
\begin{tabular}{|l|l|l|l|}
\hline
$f_*$ & $\dot{\rho}_*(z=6)$ &$\dot{\rho}_*(z=10)$ & $\dot{\rho}_*(z=15)$ \\
\hline
0.2 & $ 1.9 $ & $3.7\times10^{-2}$& $1.0\times10^{-4}$ \\
0.01 & $9.6\times10^{-2}$ & $1.9\times10^{-3}$ & $5.1\times10^{-6}$ \\
0.001 & $9.6\times10^{-3}$ & $ 1.9\times 10^{-4}$ & $ 5.1\times 10^{-7}$\\
\hline
\end{tabular}
\caption{%
   Values of the star formation rate computed from our simulation,
 $\dot{\rho}_*$, in units of $M_{\sun}$~yr$^{-1}$~Mpc$^{-3}$. We have
 used the star formation timescale of $t_{\rm SF}=20~{\rm Myr}$.
}%
\label{tab:vIvPop}
\end{center}
\end{table}
%%%%%%%%%%%%%%%%%%%%%%%%%%%%%%%%%%%%%%%%%%%%%%%%%%%%%%%%%%%%%%%%%%%%%%
\begin{figure}[t]%[h!tb]
\centering \noindent
\plottwo{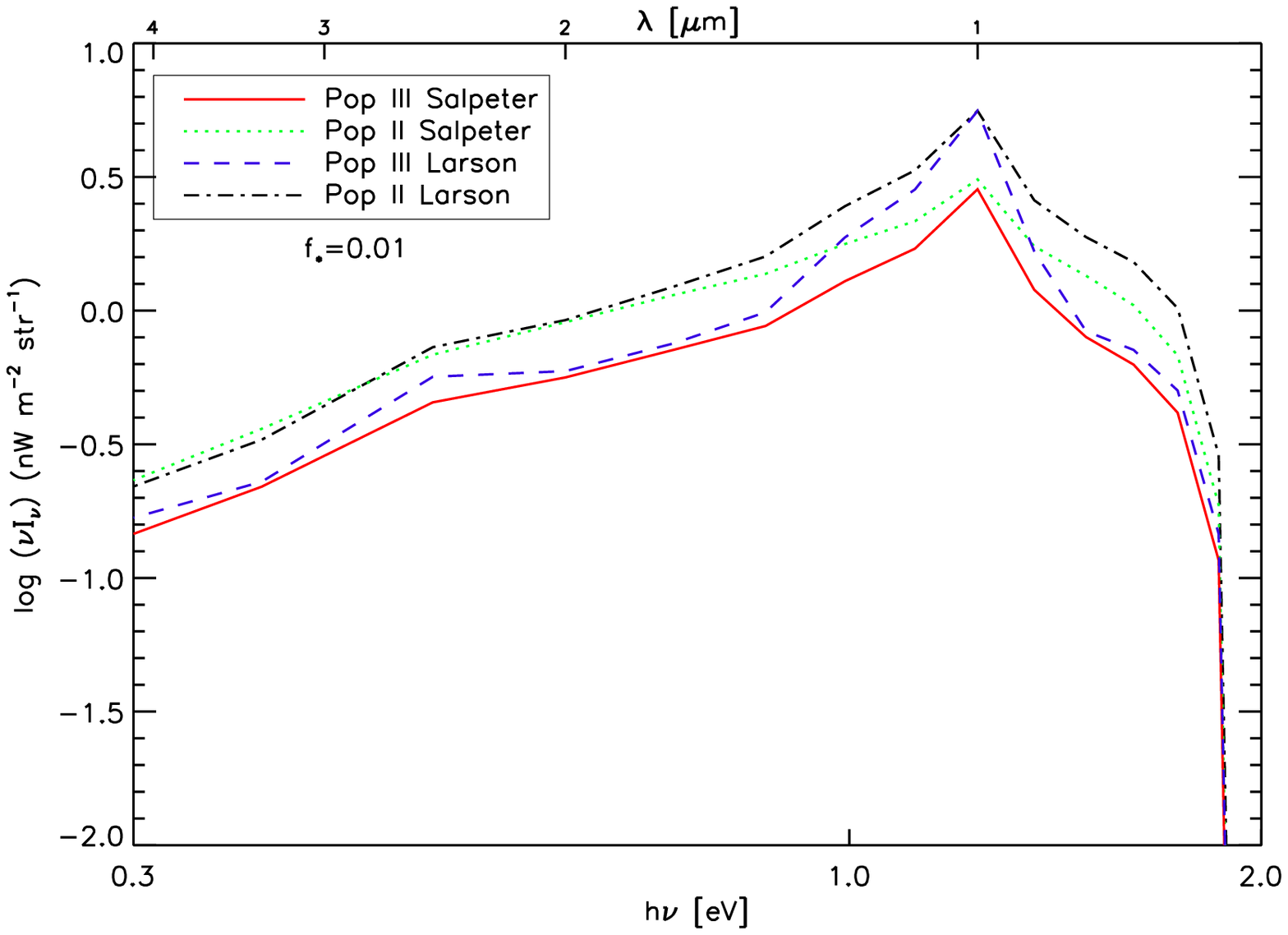}{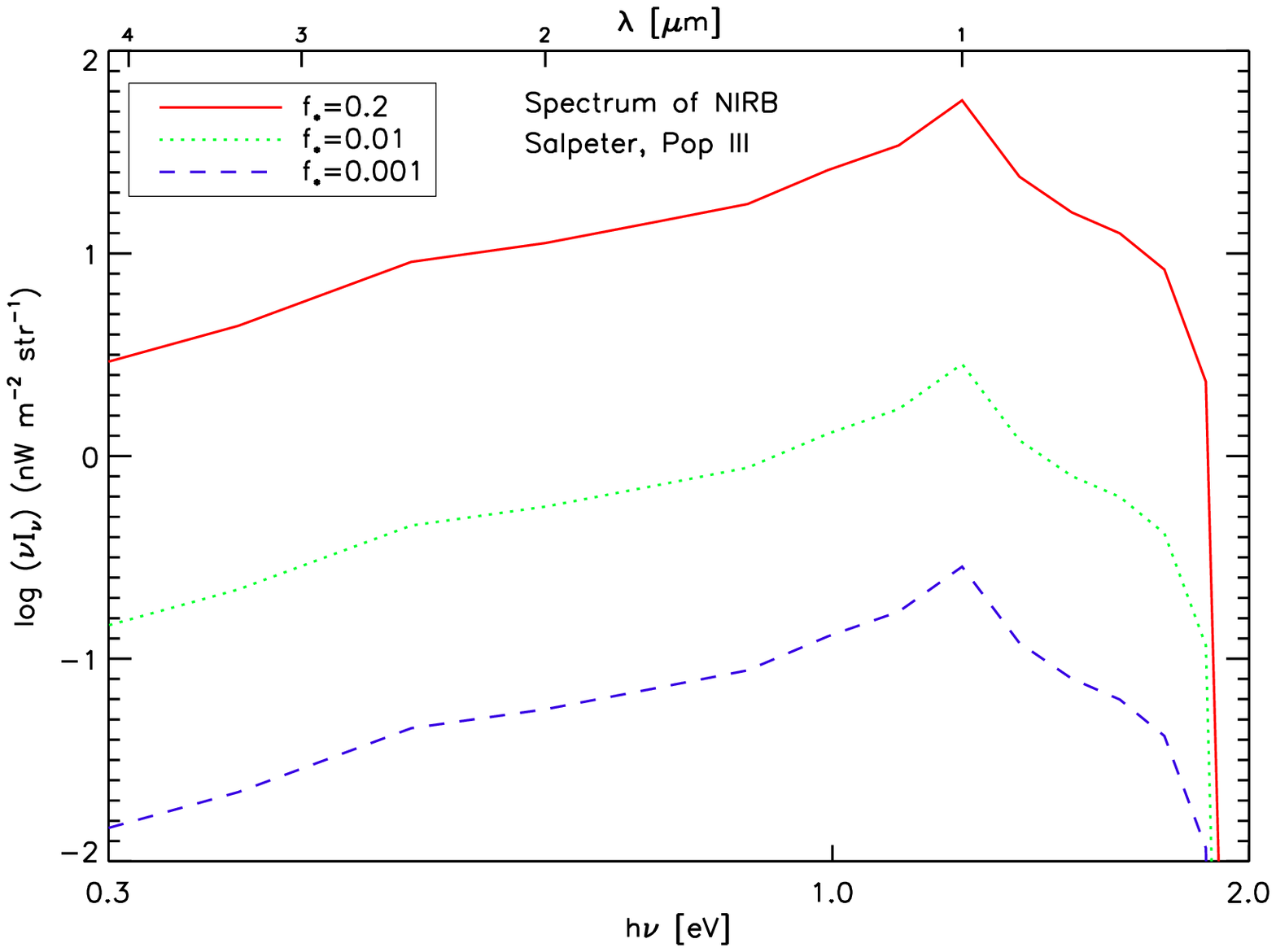}
\caption{%
Spectra of the NIRBE from various populations of stars over a redshift
range of 6 to 15.  (Left panel) Changing initial mass
 spectrum  and metallicity of the 
stars for a given star formation
 efficiency, $f_*=0.01$.  (Right panel) Changing the star formation
 efficiency for Population III stars with a Salpeter initial mass spectrum.  
Models with high star formation efficiency, $f_* = 0.2$, produce too
 high NIRB, and can be ruled out by the current upper limits from
 observations.   Note that if we divided these curves by $f_*$,
   they would become identical.  In other words, these curves differ solely
   due to the varying values of $f_*$.
}% 
\label{fig:NIRBpop}
\end{figure}
%%%%%%%%%%%%%%%%%%%%%%%%%%%%%%%%%%%%%%%%%%%%%%%%%%%%%%%%%%%%%%%%%%%%%%
\begin{figure}[t]%[h!tb]
\centering \noindent
\plotone{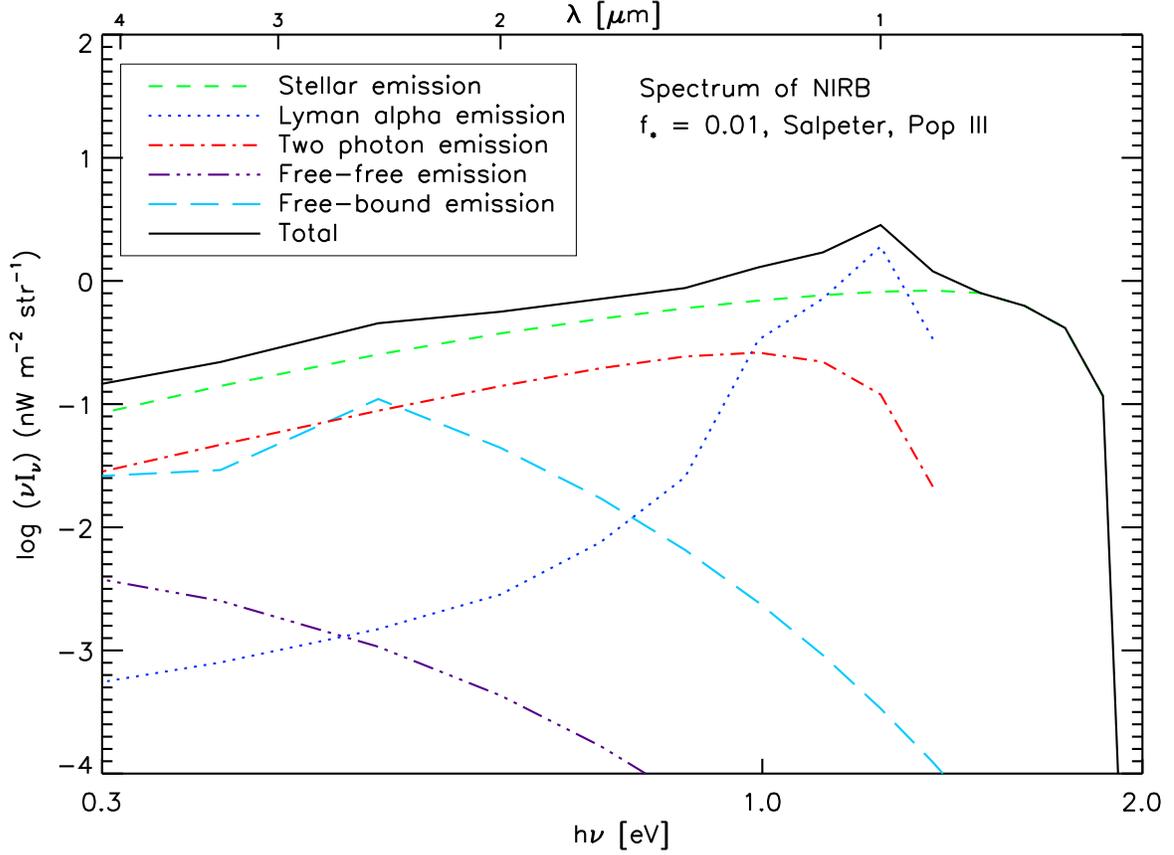}
\caption{%
Spectra of the NIRBE and how each component contributes to the overall
intensity,  over a redshift range of 6 to 15. 
}%
\label{fig:NIRB}
\end{figure}
%%%%%%%%%%%%%%%%%%%%%%%%%%%%%%%%%%%%%%%%%%%%%%%%%%%%%%%%%%%%%%%%%%%%%%

\begin{figure}[t]%[h!tb]
\centering \noindent
\plotone{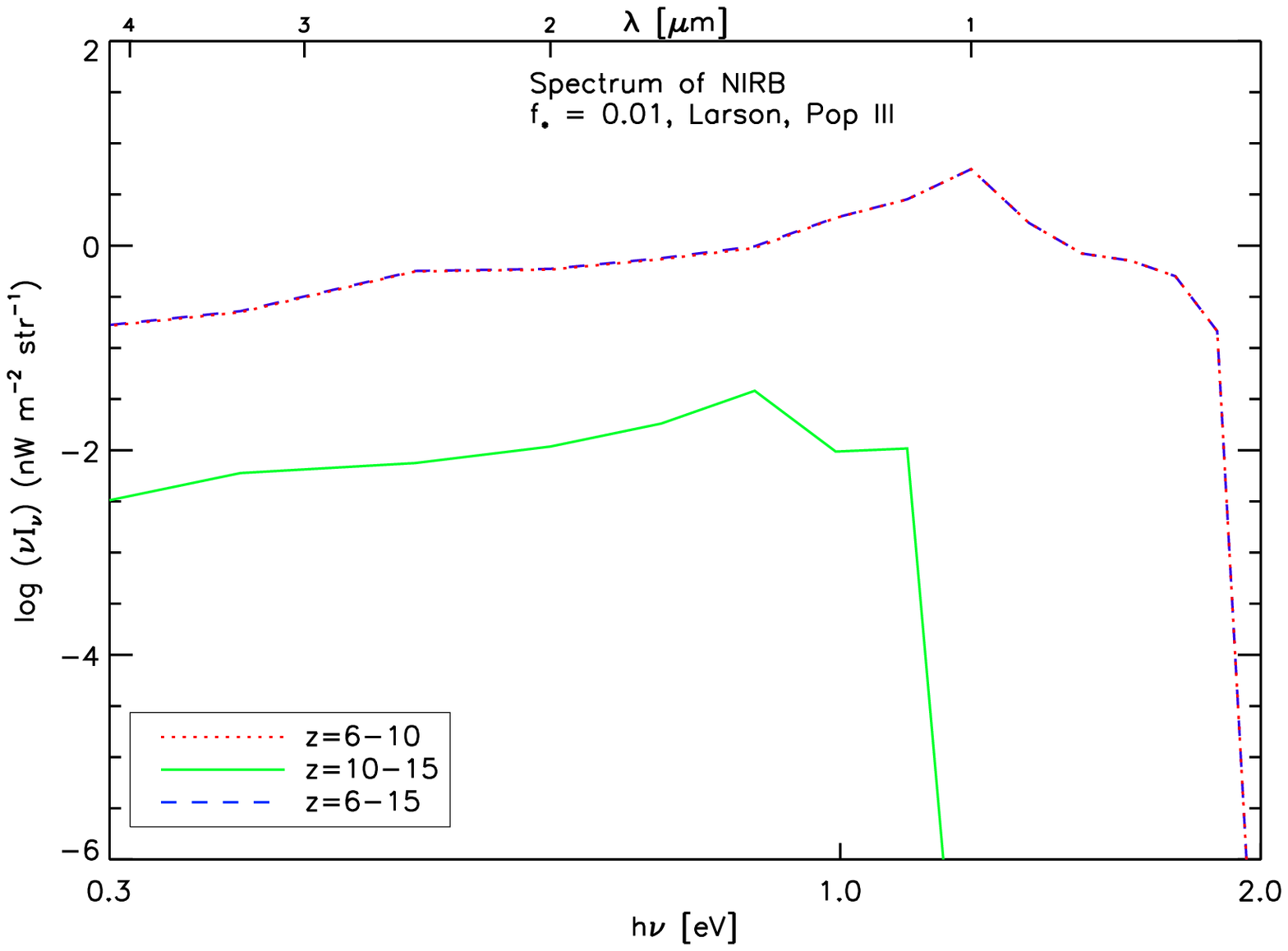}
\caption{%
Spectra of the NIRBE for populations of stars over two
 redshift bins, $z=6-10$ and $10-15$.
}
\label{fig:NIRBz}
\end{figure}

We can now calculate the mean intensity of NIRB for our models with
various stellar 
populations and values of $f_*$.  The value of $f_{\rm esc}$ does not matter
here because when calculating the mean intensity - it does not matter where
the photons are coming from - the halo itself or the IGM surrounding the
halo \citep{fernandez/komatsu:2006}. 
 The spectra of NIRB from various populations of stars
(with varying mass, metallicity, and $f_*$) over the
redshift range of $z=6-15$ (using simulation data up to the redshift
$14.6$) are shown in Figure \ref{fig:NIRBpop}, 
and their numerical values (integrated over $1-2~\mu{\rm
m}$) are tabulated in Table \ref{tab:vIvPopzvary}.
We also show the spectra of each radiation process in Figure \ref{fig:NIRB}.
Finally, we show the mean intensity from two redshift bins, $z=6-10$ and
$10-15$, in Figure \ref{fig:NIRBz}.  Lower redshift stars clearly 
dominate over stars  at higher redshifts.  This, combined with a sharp
break due to the Lyman limit as well as a bump due to the Lyman-$\alpha$
line, may be used to constrain $z_{\rm end}$.

%%%%%%%%%%%%%%%%%%%%%%%%%%%%%%%%%%%%%%%%%%%%%%%%%%%%%%%%%%%%%%%
\begin{table}[t]
\begin{center}
\begin{tabular}{|l|l|l|l|l|l|l|l|}
\hline
$f_*$ & Redshift Range & $\nu I_{\nu}$ &&&\\
&& Pop III Larson & Pop III Salpeter & Pop II Larson & Pop II Salpeter\\
\hline
0.2 & $6-15$ & $   32.2    $  &  $ 22.9 $      & $ 43.2 $  &  $ 32.2 $     \\
0.01 &  & $1.61   $   &  $ 1.15                $       &$ 2.16  $   &  $  1.61 $      \\ 
0.001 & & $ 0.161  $ &  $ 0.115 $   &  $0.216 $   &  $ 0.161 $       \\

\hline
\end{tabular}
\caption{%
   Values of the mean background intensity, $\nu I_{\nu}$, in
   units of nW~m$^{-2}$~sr$^{-1}$ for stars with different star formation efficiencies.  The mean is calculated as an average
 of $\nu I_{\nu}$ over 1 to 2 $\mu{\rm m}$.
}%
\label{tab:vIvPopzvary}
\end{center}
\end{table}
%%%%%%%%%%%%%%%%%%%%%%%%%%%%%%%%%%%%%%%%%%%%%%%%%%%%%%%%%%%%%%%%%%%%%%
%%%%%%%%%%%%%%%%%%%%%%%%%%%%%%%%%%%%%%%%%%%%%%%%%%%%%%%%%%%%%%%%%%%%%%
Assuming that our equation for the star
formation rate is accurate up to high redshifts, and
using the parameters of this simulation,
we can put constraints on the populations of first stars.  If we take
our upper limit for the mean 
intensity of the NIRBE to be $3$~nW~m$^{-2}$~sr$^{-1}$ (the upper limit
from \citet{thompson/etal:2007a}), we can rule out
most populations with high star formation efficiencies ($f_*=0.2$), unless star
formation is constrained to only high redshifts.  
This is consistent with
our fluctuation analysis - some of our models with high $f_*$ would produce
angular power spectra above the levels observed.  
If the star formation
efficiency is very low, say, $f_* = 0.001$, then the mean background would be
too small to detect.  Most of the change in the amplitude of the NIRBE is
from a change in the star formation efficiency $f_*$, while the metallicity
and initial mass spectra of the stars affect the shape of the spectra.
Therefore, an accurate measurement of the mean NIRBE can give information
on the star formation efficiency.  Further constraints on
the metallicity and mass may be possible with more precise observations
in the future.

\section{PREDICTIONS FOR FRACTIONAL ANISOTROPY} 
\label{sec:fractional}
As we have seen, the magnitude of the predicted angular power spectrum
depends on various parameters such as $f_*$, $t_{\rm SF}$, $f_{\rm
esc}$, and the initial mass spectrum. However, as the mean intensity also
depends on these quantities, one may hope that the ratio of the power spectrum
and the mean intensity squared would depend much less on these
astrophysical parameters. 

\begin{figure}[t]
\centering \noindent
\includegraphics[width=8cm]{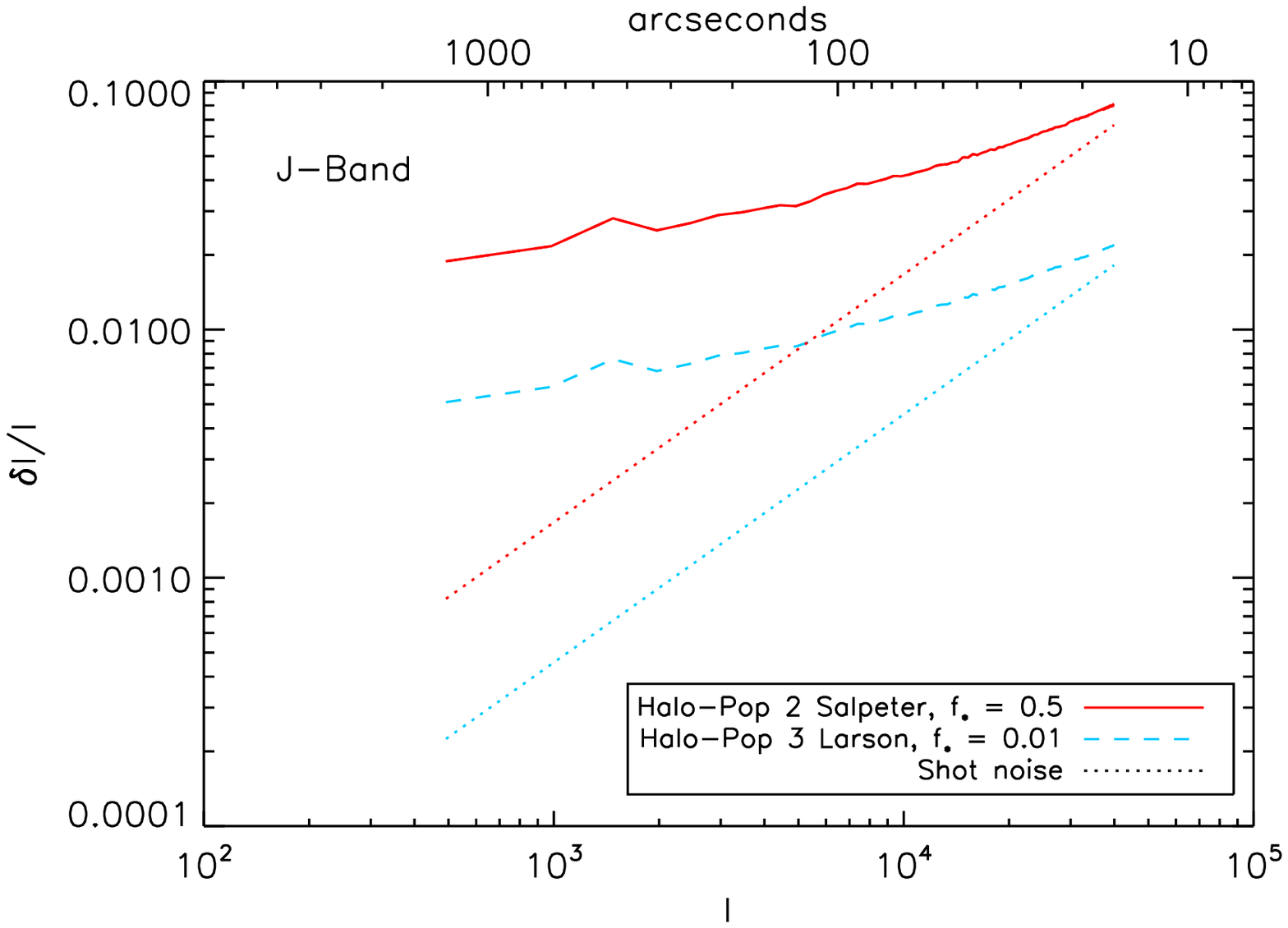}
\includegraphics[width=8cm]{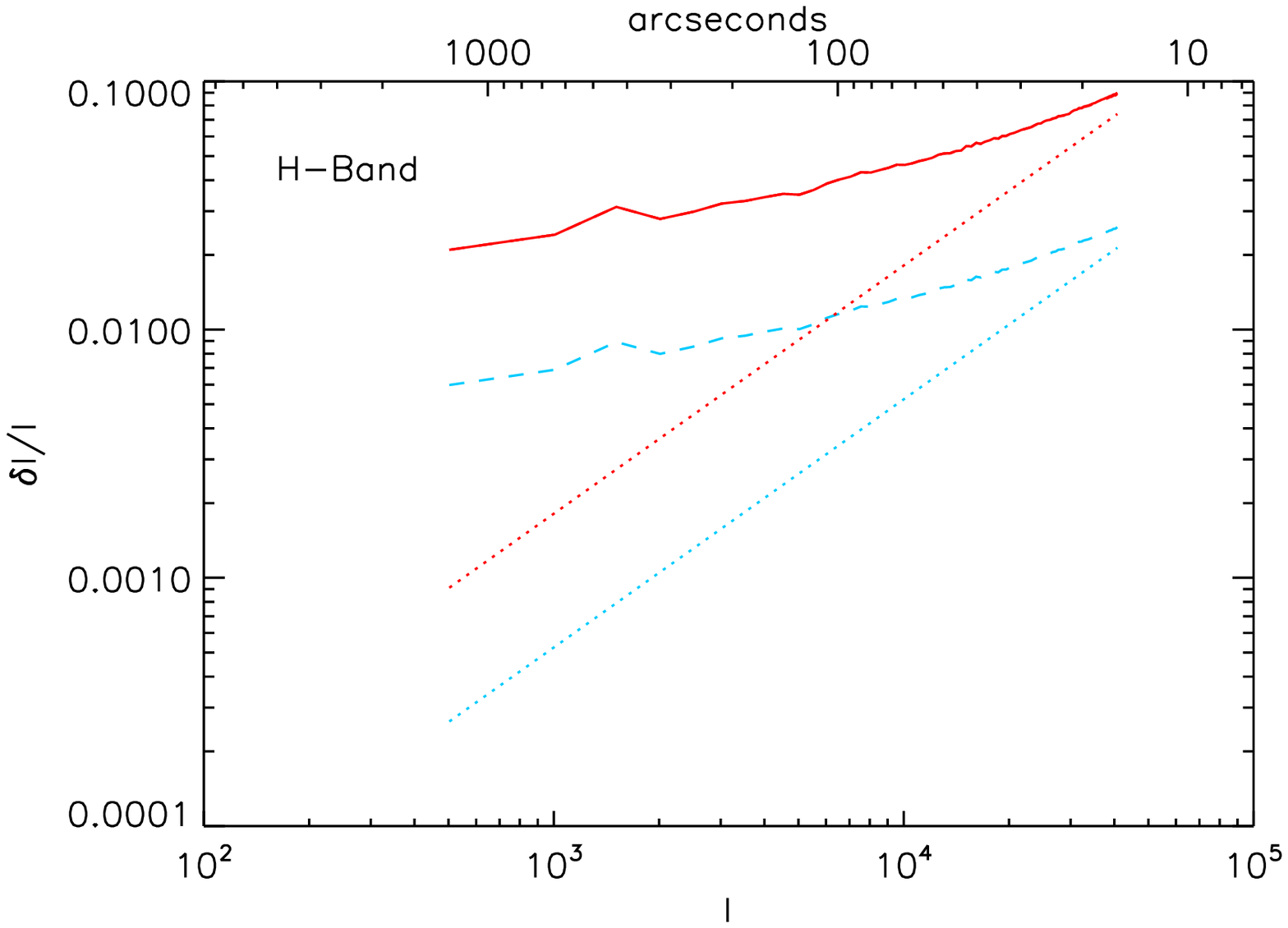}
\includegraphics[width=8cm]{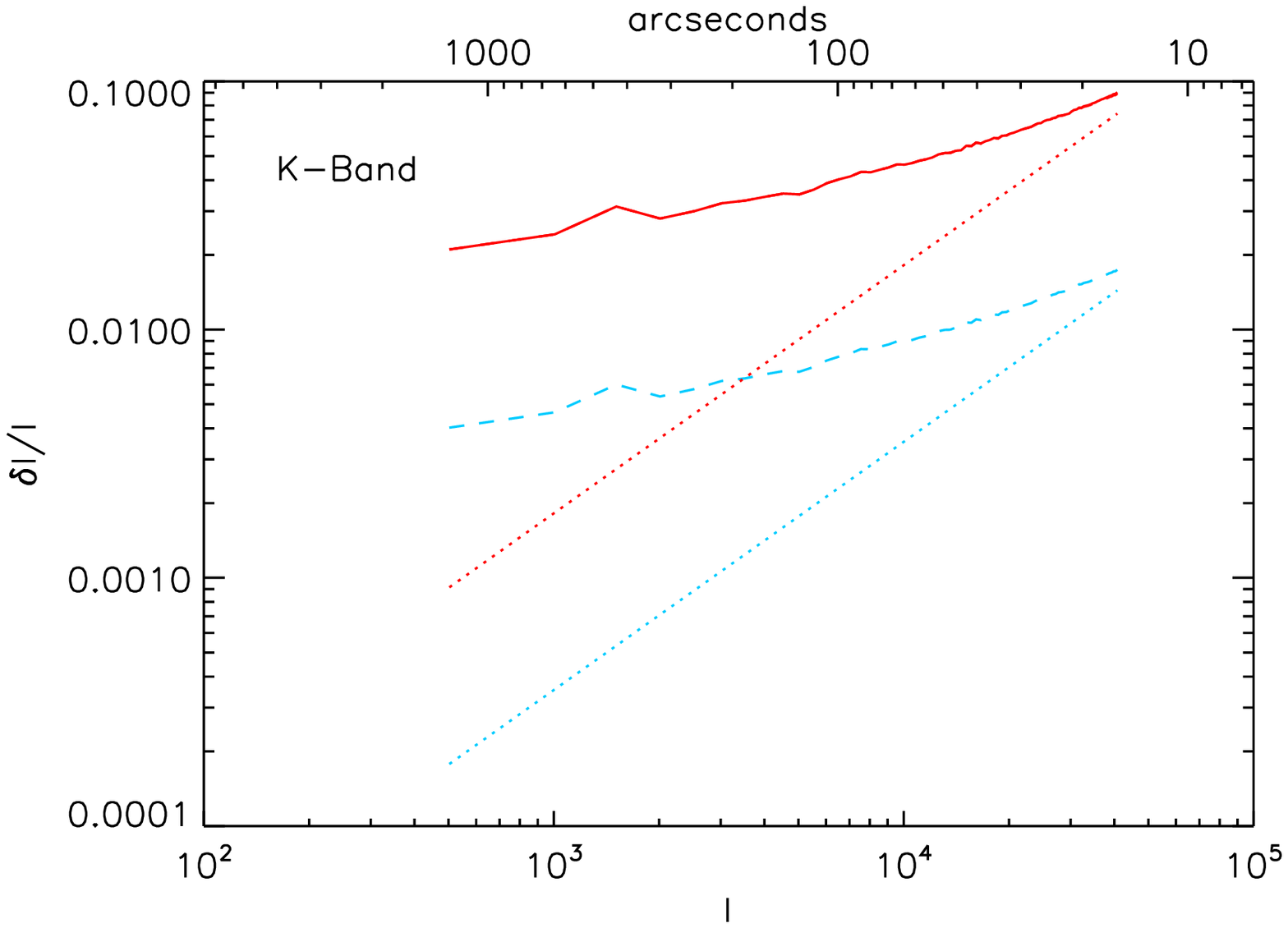}
\includegraphics[width=8cm]{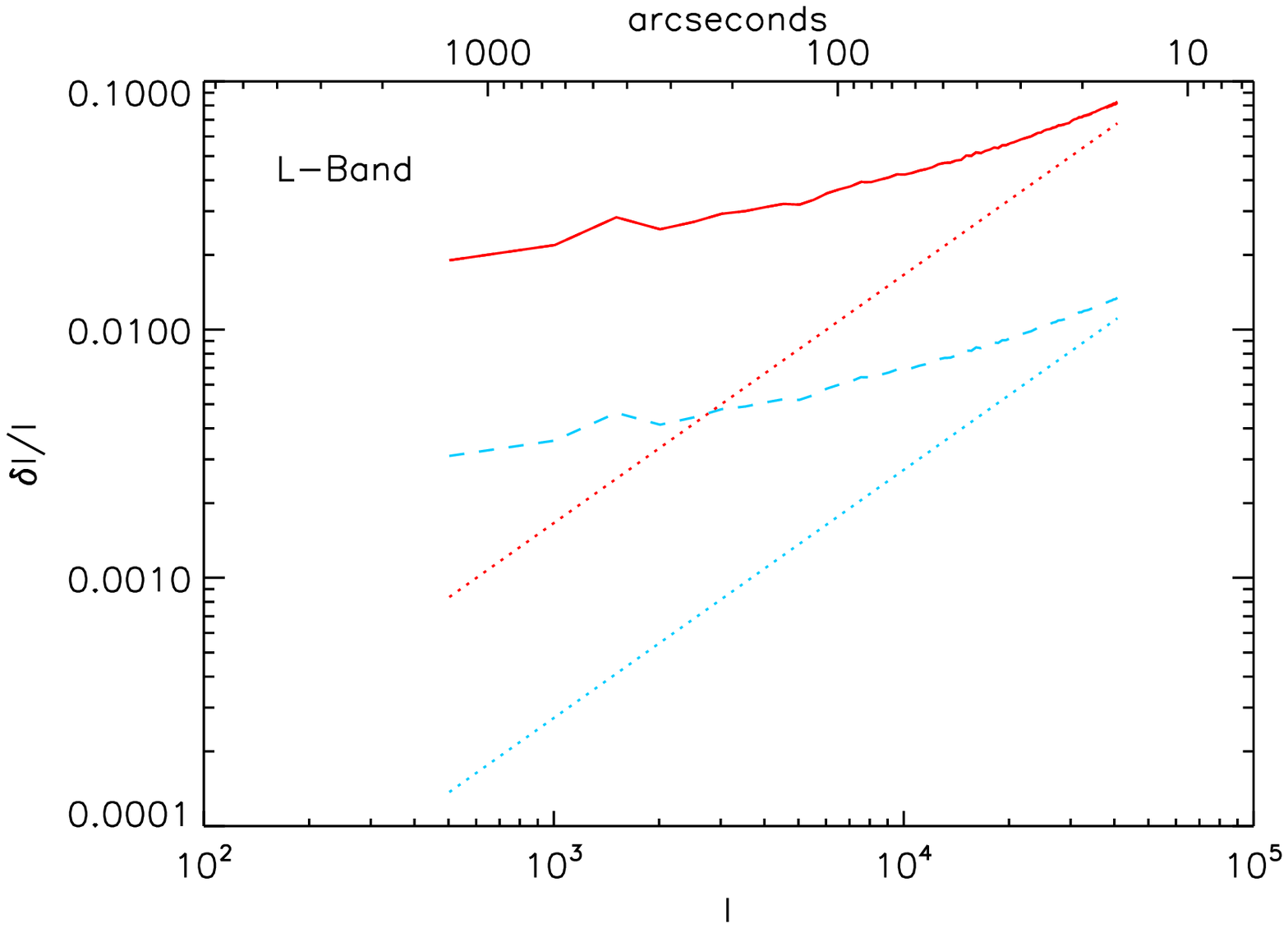}
\includegraphics[width=8cm]{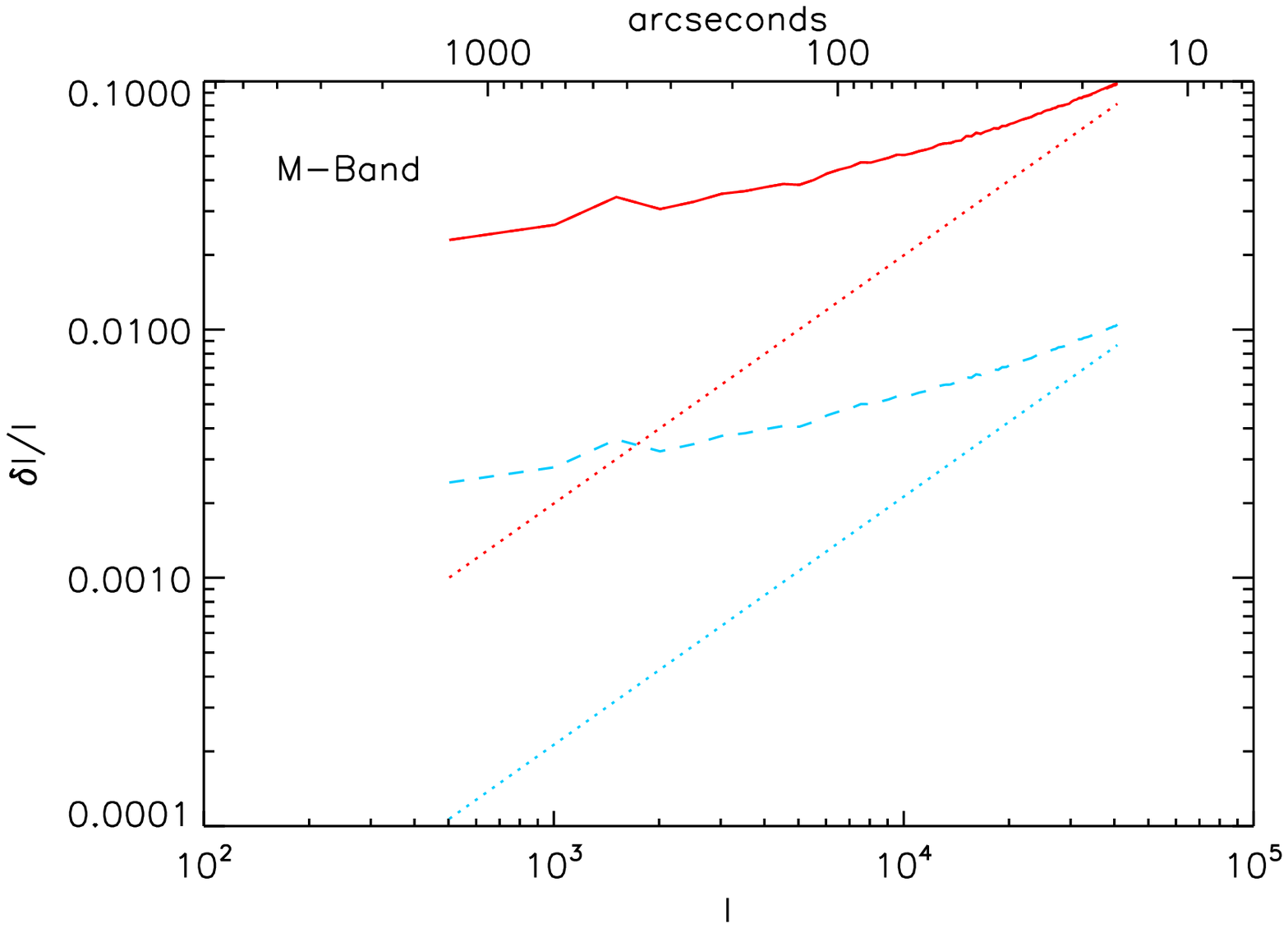}
\caption{%
Fractional anisotropy,
$\delta I/I = \sqrt{l(l+1)C_l/(2\pi I^2)}$, for different infrared bands,
  as labeled, versus the wavenumber $l$. This quantity depends primarily on the escape fraction, $f_{\rm
 esc}$.  The upper curves are for  $f_{\rm esc}=0.19$, while the lower
 curves are for $f_{\rm esc}=1$. 
 Therefore, the expected fractional anisotropy of the near infrared
 background is of order of a few percent for $f_{\rm esc}=0$, and can be lower by a
 factor of a few for $f_{\rm esc}=1$. Note that the dependence on $f_*$
 and $t_{\rm SF}$ cancels out exactly in $\delta I/I$.
}%
\label{fig:bands}
\end{figure}

Ignoring the IGM contribution and rewriting the halo contribution given
by equation~(\ref{eq:cl}), we get
\begin{eqnarray}
\nonumber
  C_l &=& \frac{c}{(4\pi)^2}\left(f_*\frac{\Omega_b}{\Omega_m}\right)^2\int
  \frac{dz}{H(z)r^2(z)(1+z)^4}\\
\nonumber
& &\times\left[\bar{\rho}_M^{halo}(z)
\left\{\bar{l}^*(z) +
 (1-f_{\rm
 esc})\left[\bar{l}^{ff}(z)+\bar{l}^{fb}(z)+\bar{l}^{2\gamma}(z)+\bar{l}^{\rm
       Ly\alpha}(z)\right]\right\}\right]^2\\
& &\times
b^2_{eff}\left(k=\frac{l}{r(z)},z\right)P_{\rm
  lin}\left(k=\frac{l}{r(z)},z\right).
\end{eqnarray}
By rewriting and averaging equation~(\ref{eq-inu-generic}) over a band, we get
\begin{eqnarray}
 I &=& \frac{c}{4\pi}\left(f_*\frac{\Omega_b}{\Omega_m}\right)\int
  \frac{dz}{H(z)(1+z)}\\
& &\times\bar{\rho}_M^{halo}(z)
\left[\bar{l}^*(z) +\bar{l}^{ff}(z)+\bar{l}^{fb}(z)+\bar{l}^{2\gamma}(z)+\bar{l}^{\rm Ly\alpha}(z)\right].
\end{eqnarray}
Therefore, in the ratio $C_l/I^2$, $f_*$ and $t_{\rm SF}$ (which is
related to $\bar{l}^\alpha$ as $\bar{l}^\alpha\propto 1/t_{\rm SF}$)
cancel out exactly.  
The dependence on the initial mass spectrum, $f(m)$, which determines 
$\bar{l}^\alpha$ via integral, nearly cancels out.
However, the dependence on $f_{\rm esc}$ does not cancel out: 
the power spectrum depends on $f_{\rm esc}$, whereas the mean intensity
does not. Therefore, we conclude that the ratio depends primarily on
$f_{\rm esc}$.  In Figure \ref{fig:bands} we show the
fractional   anisotropy,
  $\delta I/I\equiv \sqrt{l(l+1)C_l/(2 \pi I^2)}$, for various infrared
  bands. Here, $I$ is the mean intensity averaged over the bands defined
  in Table~\ref{table:bands}, which are taken from
  \citet{sterkin/manfroid:1992}.  We assume a rectangular bandpass.  
  The upper curves are for  $f_{\rm esc}=0.19$, while the lower curves are
  for $f_{\rm esc}=1$, which is consistent with the expectation: the
  ratio of the angular power spectrum to the mean intensity is lower for a higher $f_{\rm
  esc}$. We have checked that the ratio is nearly the same for different
  mass spectra for $f_{\rm esc}=0$, in which case the dependence on
  $\bar{l}(z)$ nearly cancels out. 

Note that for $f_{\rm esc}=0$ the ratio, $C_l/I^2$, may be regarded as
an weighted average of $b^2_{eff}(l/r)P_{\rm lin}(l/r)$.
We find $\delta I/I= \sqrt{l(l+1)C_l/(2 \pi I^2)}\approx 10^{-2}$ with a
weak dependence on $l$, i.e., $\delta I/I\propto l^{0.25}$. In other words,
the expected fractional anisotropy of the near infrared background is of
order a few percent for $f_{\rm esc}=0$, and can be lower by a factor of a few for
$f_{\rm esc}=1$. 
%%%%%%%%%%%%%%%%%%%%%%%%%%%%%%%%%%%%%%%%%%%%%%%%%%%%%%%%%%%%%%%
\begin{table}[t]
\begin{center}
\begin{tabular}{|l|l|l|}
\hline
Band & Center (microns) & Waveband (microns)\\
\hline
J & 1.25 & 1.1-1.4 \\
H & 1.65 & 1.5-1.8 \\
K & 2.2 & 2.0-2.4 \\
L & 3.5 & 3.0-4.0 \\
M & 4.8 & 4.6-5.0 \\
\hline
\end{tabular}
\caption{%
   Band definitions used for infrared bands.  These are
   given in Table 16.2 in \citet{sterkin/manfroid:1992}.
}%
\label{table:bands}
\end{center}
\end{table}
%%%%%%%%%%%%%%%%%%%%%%%%%%%%%%%%%%%%%%%%%%%%%%%%%%%%%%%%%%%%%%%%%%%%%% 

%%%%%%%%%%%%%%%%%%%%%%%%%%%%%%%%%%%%%%%%%%%%%%%%%%%%%%%%%%%%%%%%%%%%%%
%%%%%%%%%%%%%%%%%%%%%%%%%%%%%%%%%%%%%%%%%%%%%%%%%%%%%%%%%%%%%%%%%%%%%%

%%%%%%%%%%%%%%%%%%%%%%%%%%%%%%%%%%%%%%%%%%%%%%%%%%%%%%%%%%%%%%%%%%%%%%
\section{DISCUSSION AND CONCLUSIONS}
\label{sec:conclusions}
Any detection or non-detection of fluctuations in NIRB can give us
information on stars forming at high redshifts, 
stars that could have helped to reionize the universe.  The escape fraction
of ionizing photons, the star formation efficiency, and the mass and
metallicity of the stars can affect the amplitude and shape of the angular
power spectrum of fluctuations in NIRB.

We modeled the angular power spectrum from halos and the surrounding IGM by
combining the analytic formulas for the luminosity of halos and the IGM with
$N$-body simulations coupled with radiative transfer for several different
populations of stars.   
Shot noise is a major contributor to the angular power spectrum of halos at small
scales, so it is important to include larger scales in observations to
minimize the component of shot noise.  

The star formation efficiency has a significant effect on the amplitude of
the angular power spectrum, with the amplitude of the angular power
spectrum being proportional to $f_*^2$.  
For a fixed $f_{\gamma}/t_{\rm SF}=f_{\rm esc}f_*N_i/t_{\rm SF}$ (i.e.,
for a given 
reionization history), a combination of parameters that
maximize the star formation efficiency, $f_*$, give the largest NIRB
power spectrum; thus, the stars that are less massive and have
more metals (smaller $N_i$), and are in halos with a lower escape
fraction (smaller $f_{\rm esc}$) will all 
increase the amplitude of the NIRB angular power spectrum.  In general, the amplitude of the angular power spectrum of
the halos (and the mean NIRBE) is mostly dominated by $f_*$, while the
angular power spectrum of the IGM can probe the ionization history though
the factor $f_\gamma/t_{SF}$.  

If we do not fix the reionization history, 
there are other parameters that can change the amplitude of
the angular power spectrum significantly.  The angular power spectrum is
inversely 
proportional to the star formation time scale squared, $C_l\propto
1/t_{\rm SF}^2$.  This uncertainty in the star formation time scale can be
directly related to our uncertainty in the mass to light ratio of
galaxies, for $t_{\rm SF} \propto M_h/L_h$.  
As changes in $t_{\rm SF}$ result in different reionization
histories (for a given $f_\gamma$), the other tracers of reionization,
e.g., the electron-scattering optical depth measured by the WMAP
satellite and the abundance of Lyman-$\alpha$ emitting galaxies, should
help narrow down a range of magnitudes of the NIRB 
fluctuations that are consistent with what we already know about the cosmic
reionization.  

The angular power spectrum of the
IGM is typically a minor contributor to the overall fluctuations; 
however, the IGM contribution can be comparable to the halo
contribution (the stellar contribution as well as the nebular
contribution from within the halo), especially if the escape fraction
of ionizing photons from halos is high.  
In the limit that $f_{\rm esc}$ is close to unity, we expect
$C_l^{\rm IGM}/C_l^{\rm halo}\propto f_{\rm esc}^2$, as $C_l^{\rm halo}$
would be completely dominated by the stellar emission.
One can even make the IGM contribution dominate over the halo contribution by
increasing $t_{\rm SF}$, which  suppresses the halo contribution as
$C_l^{\rm halo}\propto t_{\rm SF}^{-2}$, and leads to a delay in the
reionization. Yet, this would not change the IGM 
power spectrum significantly, as the IGM luminosity power spectrum
saturates when the ionization fraction reaches $X_e\sim 0.5$. 
Of course, we need to make sure that such a model can still complete the
reionization by $z\sim 6$, and  can reproduce the electron-scattering
optical depth measured by WMAP.

The redshift at which the formation of stars contributing to NIRB ends,
$z_{\rm end}$, 
can also change the 
amplitude of the angular power spectrum significantly.  Changing $z_{\rm
end}$ affects not only the amplitude of the
angular power spectrum, but also the shape. 
The attenuation of Lyman-$\alpha$ photons
for the most part, does not affect the angular power spectrum of the halos
greatly, but could affect the amplitude of the IGM by about a factor of 2.

Previous estimates of the angular power spectrum of the NIRB by
\citet{cooray/etal:2004} neglected to account for nonlinear bias.  For our
simulation with the minimum halo mass of $2.2\times 10^9~M_\sun$, the
nonlinear bias 
is large enough to change the prediction for the shape of the angular
power spectrum qualitatively: a turnover of $l(l+1)C_l$ at $l\sim 10^3$
that was predicted by \citet{cooray/etal:2004} is not seen in 
our calculation, and the shape of the clustering component (i.e., minus
the shot noise), is consistent with a pure power law, 
$l(l+1)C_l\propto l^{0.5}$. 

Note that our results for the shape of the angular power spectrum are
valid for the minimum halo mass of $M_{\rm min} = 2.2\times 10^9 \: M_{\sun}$.  
The non-linear bias would be smaller for smaller mass halos.
For 
example, the simulation carried out by \citet{Trac/Cen:2007} resolves
halos down to a smaller mass, $M_{\rm min} =6 \times 10^7~h^{-1}~M_{\sun}$,
and would therefore find a smaller average bias. The halo bias is mass
dependent, more massive halos being more strongly biased. 
The effective
linear bias, $b_{\rm eff,lin}$, is the integral of the linear halo bias
for a given 
mass, $b_1(M_h)$, times the mass function weighted by mass
above a certain minimum mass, i.e.,
$b_{\rm eff,lin}=[\int_{M_{\rm min}}^\infty dM_h~M_h
(dn_h/dM_h)b_1(M_h)]/[\int_{M_{\rm min}}^\infty dM_h~M_h (dn_h/dM_h)]$. 
As there are many more halos at lower masses, lowering $M_{\rm min}$
would result in a lower average bias. As the degree of the non-linear bias
increases as the linear bias increases, one would find a smaller
non-linear bias for a lower $M_{\rm min}$.
Therefore, we might still
see a turnover if these smaller halos are bright enough to contribute to
the NIRB. 

However, the real situation would be more
complex than the above picture. 
In \citet{iliev/etal:2007}, we also performed reionization simulations which
resolve source halos down to this lower minimum mass of $\sim
10^8~M_\sun$.  There, however, unlike \citet{Trac/Cen:2007}, we took account of
the fact that those small-mass ($\lesssim 10^9~M_\sun$) halos are subject to
Jeans filtering, and thus their star formation is suppressed if they
reside within the ionized regions.
The suppression occurs disproportionately on
the low-mass halos clustered around the high-density peaks (which are
the first to be ionized). 
Therefore, the ultimate effect may not be as large as
one might think by just including all halos down to $\sim 10^8~M_\sun$.
While it is plausible that there may be a turn-over, it is also
quite plausible that the location of the turn-over would be on a smaller
scale than what would be predicted by the linear bias model.

This can, in principle, be studied using higher-resolution simulations
like those in \citet{iliev/etal:2007} that include the Jeans filtering
effect, but the simulation box size of $35~h^{-1}~{\rm Mpc}$ there is
not quite large enough to give a reliable statistical measure of the
large-scale structure and angular fluctuations in which we are
interested. Towards that end, we have more recently performed a new set
of large-box, higher-resolution simulations that include the Jeans
filtering effect, reported in \citet{shapiro/etal:prep} and \citet{iliev/etal:prep}.
We shall present results on
the NIRB from these higher-resolution simulations elsewhere.
In any case, the above consideration suggests that the shape of the
angular power spectrum 
gives us important information about the nature of sources
contributing to NIRB as well as the physics of cosmic reionization.

Current observations seem to favor low levels of both the fluctuations
and the mean NIRB due to the high-$z$ (i.e., $z\gtrsim 7$) sources.  
The current observations of the mean 
intensity of the NIRB seem to rule out high levels of
$f_*$, i.e., $f_* \gtrsim 0.2$. 
Most of our models for fluctuations still lie beneath 
the current observations of the fluctuations of the NIRB. The upcoming
CIBER missions will improve the sensitivity of 
observations, but many of our models still lie below their sensitivity limits.
Nevertheless, these new observations should be able to put tighter
constraints on which 
high-$z$ galaxy populations are allowed and which are ruled out.  Given
the lack of direct observational probes of high-$z$ galaxy populations
contributing to the cosmic reionization, 
the NIRB continues to offer invaluable information regarding the physics
of cosmic reionization that is difficult to probe by other means.

We would like to thank Asantha Cooray, Daniel Eisenstein, Donghui
Jeong, Yi Mao, and Rodger Thompson for helpful discussions.
This study was supported by Spitzer Space Telescope theory grant
1310392, NSF grant AST 0708176, NASA grants NNX07AH09G and
NNG04G177G, Chandra grant SAO TM8-9009X, and Swiss National Science Foundation grant 200021-116696/1.
ERF acknowledges support from the University of Colorado
Astrophysical Theory Program through grants from NASA (NNX07AG77G) and
NSF (AST07-07474). EK acknowledges support from an Alfred P. Sloan
Research Fellowship. 
%%%%%%%%%%%%%%%%%%%%%%%%%%%%%%%%%%%%%%%%%%%%%%%%%%%%%%%%%%%%%%%%%%%%%%
\appendix
\section{Derivation of Angular Power Spectrum of NIRB Fluctuations} 
\label{sec:cl_derivation}

The observed intensity (energy received per unit time, unit area, unit
solid angle, and unit frequency) of the NIRB toward a direction on the sky
$\hat{\mathbf n}$, $I_\nu(\hat{\mathbf n})$, is related to the spatial
distribution of the volume emissivity (luminosity emitted per frequency 
per comoving volume) at various redshifts,
$p(\nu,{\mathbf x},z)$, as \citep[][p.91]{peacock:1999}
\begin{equation}
 I_\nu(\hat{\mathbf n})
= \frac{c}{4\pi}\int dz\frac{p[\nu(1+z),\hat{\mathbf n}r(z),z]}{H(z)(1+z)},
\end{equation}
where $r(z)=c\int_0^zdz'/H(z')$ is the comoving distance. The
band-averaged intensity is then  given by
\begin{eqnarray}
 I(\hat{\mathbf n})
&\equiv& \int_{\nu_1}^{\nu_2}d\nu~I_\nu(\hat{\mathbf n})
= \frac{c}{4\pi}\int
dz\frac{\int_{\nu_1}^{\nu_2}d\nu~p[\nu(1+z),\hat{\mathbf
n}r(z),z]}{H(z)(1+z)}\\
&=& 
\frac{c}{4\pi}\int
dz\frac{\int_{\nu_1(1+z)}^{\nu_2(1+z)}d\tilde{\nu}~p[\tilde{\nu},\hat{\mathbf
n}r(z),z]}{H(z)(1+z)^2},
\end{eqnarray}
where $\tilde{\nu}=\nu(1+z)$. Note that the denominator now contains $(1+z)^2$
instead of $(1+z)$.

Now, using the luminosity density integrated over bands,
$\rho_L({\mathbf x},z)\equiv 
\int_{\nu_1(1+z)}^{\nu_2(1+z)}d\nu~p(\nu,{\mathbf x},z)$, we obtain
\begin{equation}
 I(\hat{\mathbf n})
= \frac{c}{4\pi}\int dz\frac{\rho_L[\hat{\mathbf n}r(z),z]}{H(z)(1+z)^2}.
\end{equation}
The spherical harmonic transform of $I(\hat{\mathbf n})$, $a_{lm}=\int
d\hat{\mathbf n}~I(\hat{\mathbf n})Y_{lm}^*(\hat{\mathbf n})$, is
then related to the three-dimensional Fourier transform of 
$\rho_L({\mathbf x},z)$, $\tilde{\rho}_L({\mathbf k},z)=\int d^3{\mathbf
x}~\rho_L({\mathbf x},z)e^{i{\mathbf k}\cdot{\mathbf x}}$ (the inverse transform is
$\rho({\mathbf x},z) = \int d^3{\mathbf k}/(2\pi)^3~\tilde{\rho}_L({\mathbf
k},z)e^{-i{\mathbf k}\cdot{\mathbf x}}$), as,
\begin{equation}
 a_{lm} = 
\frac{c}{4\pi}\int \frac{dz}{H(z)(1+z)^2}
\left[
4\pi (-i)^l 
\int \frac{d^3{\mathbf k}}{(2\pi)^3}
\tilde{\rho}_L({\mathbf k},z)j_l[kr(z)]Y_{lm}^*(\hat{\mathbf k})
\right],
\end{equation}
where we have used Rayleigh's formula:
\begin{equation}
 e^{-i{\mathbf k}\cdot{\hat\mathbf n}r(z)} 
= 4\pi  \sum_{lm} (-i)^lj_l[kr(z)] Y_{lm}^*(\hat{\mathbf k})Y_{lm}(\hat{\mathbf n}).
\end{equation}

The angular power spectrum, $C_l=\langle |a_{lm}|^2\rangle$, is then given
by ($\langle\rangle$ is the statistical ensemble average)
\begin{equation}
 C_l = \left(\frac{c}{4\pi}\right)^2
\int \frac{dz}{H(z)(1+z)^2}\int \frac{dz'}{H(z')(1+z')^2}
\left[
\frac{2}{\pi}
\int k^2dk~P_L(k,z)j_l[kr(z)]j_l[kr(z')]
\right],
\end{equation}
where we have used the definition of the luminosity-density power
spectrum  (Eq.~(\ref{eq:PkL_def})), and the normalization of the
spherical harmonics, $\int d\hat{\mathbf k}Y_{lm}(\hat{\mathbf
k})Y^*_{lm}(\hat{\mathbf k})=1$. 

Now, when $l\gg 1$, the integral within the square bracket can be
approximated as\footnote{The exact integral of a product of two spherical
Bessel functions is given by
$\frac{2}{\pi}\int k^2dk~j_l(kr)j_l(kr')=\delta(r-r')/r^2$.
When a function, $F(k)$, is a slowly-varying function of $k$ compared to 
$j_l(kr)j_l(kr')$, which is a highly oscillating function for $l\gg 1$,
we may obtain
$\frac{2}{\pi}\int k^2dk~F(k)j_l(kr)j_l(kr')\approx F(k=l/r)\delta(r-r')/r^2$. This
is the so-called Limber's approximation.}
\begin{equation}
 \frac{2}{\pi}
\int k^2dk~P_L(k,z)j_l[kr(z)]j_l[kr(z')]
\approx 
\frac{\delta\left[r(z)-r(z')\right]}{r^2(z)}P_L\left(k=\frac{l}{r(z)},z\right).
\end{equation}
Therefore,
\begin{eqnarray}
\nonumber
 C_l &\approx& \left(\frac{c}{4\pi}\right)^2
\int \frac{dz}{H(z)(1+z)^2}\int dr'\frac{dz'/dr'}{H(z')(1+z')^2}
\frac{\delta(r-r')}{r^2(z)}P_L\left(k=\frac{l}{r(z)},z\right)\\
&=&
\frac{c}{(4\pi)^2}
\int \frac{dz}{H(z)r^2(z)(1+z)^4}
P_L\left(k=\frac{l}{r(z)},z\right),
\label{eq:clapp}
\end{eqnarray}
where $r=r(z)$ and $r'=r(z')$, and we have used $dz/dr=H(z)/c$. 
This is Eq.~(\ref{eq:cl}).

On the other hand, if one chooses to calculate $C_l$ from
a pair of $I_\nu(\hat{\mathbf
n})$ and $I_{\nu'}(\hat{\mathbf n})$
 instead
of a pair of the band-averaged intensities $I(\hat{\mathbf
n})$, then $(1+z)^4$ in the denominator of Eq.~(\ref{eq:clapp}) becomes 
$(1+z)^2$: 
\begin{equation}
 C_l^{\nu\nu'}
=
\frac{c}{(4\pi)^2}
\int \frac{dz}{H(z)r^2(z)(1+z)^2}
P_p\left(\nu(1+z),\nu'(1+z); k=\frac{l}{r(z)},z\right),
\label{eq:clcooray}
\end{equation}
where $P_p(\nu,\nu';k,z)$ is the power spectrum of the volume
emissivity, $p(\nu,z)$. Eq.~(\ref{eq:clcooray}) 
agrees with Eq.~(10) of
\citet{cooray/etal:2004}. Note that their $j_\nu$ is $p(\nu)/(4\pi)$ in
our notation, which explains the absence of $1/(4\pi)^2$ in their
Eq.~(10). However, this result does not agree with Eq.~(3) of
\citet{kashlinsky/etal:2004}, which originates from Eq.~(11) of
\citet{kashlinsky/odenwald:2000}. (See also Eq.~(14)
of \citet{kashlinsky:2005}.)
Kashlinsky et al.'s formula has
$(1+z)$ in the denominator of Eq.~(\ref{eq:clcooray}) instead of
$(1+z)^2$, and thus it misses one factor of $1/(1+z)$. We were unable to
trace the cause of this discrepancy. It is therefore possible that
\citet{kashlinsky/etal:2004} over-estimated $C_l$ by a factor of $\sim
10$. As they have not taken into account a short lifetime of massive
stars in their calculations (i.e., they used Eq.~(\ref{eq:lnu2}) instead
of Eq.~(\ref{eq:lnu1})), which yields another factor of $\sim 100$ in
$C_l$, it is possible that they over-estimated $C_l$ by a factor of
$\sim 10^3$.

%%%%%%%%%%%%%%%%%%%%%%%%%%%%%%%%%%%%%%%%%%%%%%%%%%%%%%%%%%%%%%%%%%%%%%%%%%%%%

\end{document}